\documentclass[%
 reprint,
 amsmath,amssymb,
 aps,
]{revtex4-2}

\usepackage{graphicx}
\usepackage{dcolumn}
\usepackage{bm}
\usepackage{xcolor}
\usepackage{amsmath,amssymb}

\usepackage{microtype}
\usepackage{nonfloat}
\usepackage{graphicx}
\usepackage{subfigure}
\usepackage{booktabs} 

\usepackage{amsfonts}
\usepackage{color}

\usepackage{multirow}

\usepackage[colorlinks,linkcolor=black,citecolor=blue,urlcolor=blue ]{hyperref}

\newcommand{\lensfit}{\textsc{Lensfit}}
\newcommand{\bpz}{\texttt{BPZ}}
\newcommand{\healpix}{\texttt{HEALPix}}
\newcommand{\cosmogrid}{\textsc{CosmoGrid}}
\newcommand{\pkdgrav}{\textsc{PkdGrav3}}
\newcommand{\pyccl}{\textsc{PyCCL}}
\newcommand{\hmcode}{\textsc{HMCode}}
\newcommand{\ufalcon}{\textsc{UFalcon}}
\newcommand{\amiga}{\textsc{Amiga}}
\newcommand{\deepsphere}{\textsc{DeepSphere}}
\newcommand{\moped}{\textsc{MOPED}}
\newcommand{\class}{\textsc{Class}}
\newcommand{\concept}{\textsc{CO\textbf{\textit{N}}CEPT}}
\newcommand{\emcee}{\texttt{emcee}}

\defcitealias{kids1000_shear}{A20}
\defcitealias{kids_net}{F19}

\begin{document}

\preprint{APS/123-QED}

\title{A Full $w$CDM Analysis of KiDS-1000 Weak Lensing Maps using Deep Learning}

\author{Janis Fluri$^\mathrm{a,b}$}%
\email{janis.fluri@phys.ethz.ch}
\author{Tomasz Kacprzak$^\mathrm{a,c}$}
\author{Aurelien Lucchi$^\mathrm{d}$}
\author{Aurel Schneider$^\mathrm{e}$}
\author{Alexandre Refregier$^\mathrm{a}$}
\author{Thomas Hofmann$^\mathrm{b}$}
\affiliation{\vspace*{10pt}$^\mathrm{a}$Institute of Particle Physics and Astrophysics, Department of Physics, ETH Zurich, Switzerland}
\affiliation{$^\mathrm{b}$Data Analytics Lab, Department of Computer Science, ETH Zurich, Switzerland}
\affiliation{$^\mathrm{c}$Swiss Data Science Center, Paul Scherrer Institute, Switzerland}
\affiliation{$^\mathrm{d}$Department of Mathematics and Computer Science, University of Basel}
\affiliation{$^\mathrm{e}$Center for Theoretical Astrophysics and Cosmology, Institute for Computational Science, University of Zurich,
Switzerland}
\date{\today}%

\begin{abstract}
We present a full forward-modeled $w$CDM analysis of the KiDS-1000 weak lensing maps using graph-convolutional neural networks (GCNN). Utilizing the \cosmogrid, a novel massive simulation suite spanning six different cosmological parameters, we generate almost one million tomographic mock surveys on the sphere. Due to the large data set size and survey area, we perform a spherical analysis while limiting our map resolution to \healpix~$n_\mathrm{side}=512$. We marginalize over systematics such as photometric redshift errors, multiplicative calibration and additive shear bias. Furthermore, we use a map-level implementation of the non-linear intrinsic alignment model along with a novel treatment of baryonic feedback to incorporate additional astrophysical nuisance parameters. We also perform a spherical power spectrum analysis for comparison. The constraints of the cosmological parameters are generated using a likelihood-free inference method called Gaussian Process Approximate Bayesian Computation (GPABC). Finally, we check that our pipeline is robust against choices of the simulation parameters. We find constraints on the degeneracy parameter of $S_8 \equiv \sigma_8\sqrt{\Omega_M/0.3} = 0.78^{+0.06}_{-0.06}$ for our power spectrum analysis and $S_8 = 0.79^{+0.05}_{-0.05}$ for our GCNN analysis, improving the former by 16\%. This is consistent with earlier analyses of the 2-point function, albeit slightly higher. Baryonic corrections generally broaden the constraints on the degeneracy parameter by about 10\%. These results offer great prospects for full machine learning based analyses of on-going and future weak lensing surveys.
\end{abstract}

\maketitle


\section{Introduction}
\label{sec:intro}

The large scale structures (LSS) of the Universe contains a wealth of information that can be used to test our cosmological models. Weak gravitational lensing (WL) (see e.g. \cite{Schneider2005,Kilbinger2015review} for reviews) utilizes the laws of general relativity along the shape measurements of millions of galaxies to reconstruct the projected matter distribution of the Universe. The unique ability of WL to directly observe the matter contents of the Universe makes it an ideal probe to constrain  cosmological parameters. This has already been demonstrated by WL surveys such as the Canada France Hawaii Telescope Lensing Survey (CFHTLenS)\footnote{\url{cfhtlens.org}} \cite{CFHTLenS2013}, the Kilo-Degree Survey (KiDS)\footnote{\url{kids.strw.leidenuniv.nl}} \cite{Hildebrandt2018viking,kids1000_shear}, the Dark Energy Survey (DES)\footnote{\url{darkenergysurvey.org}} \cite{desy3_fidu,desy3_peaks}, and the Subaru Hyper Suprime-Cam (HSC)\footnote{\url{hsc.mtk.nao.ac.jp/ssp/survey}} \cite{hsc_y1}. Future surveys such as Euclid \cite{Euclid2011}, the Vera C. Rubin Observatory \cite{Chang2013} or the Wide-Field Infrared Survey Telescope (WFIRST) \cite{wfirst} will be able to provide even more precise measurements.

On large scales, the projected matter distribution of the Universe approximately follows the distribution of a Gaussian random field, making statistics based on the two-point correlation function ideal tools to analysis the of data~\cite{kids1000_shear,desy3_fidu, hsc_y1}. However, the increasing quality and quantity of the available data makes it possible to probe scales, where the evolution of the matter distribution is dominated by non-linear effects, leading to information that cannot be fully extracted by the two-point correlation function alone. This motivates the search for novel summary statistics with the ability to extract non-Gaussian information. Such approaches include weak lensing peak statistics (e.g. \cite{jialiu2015,Dietrich2010peaks,Kacprzakpeaks, Fluri2018peak, liu:hal-01439964, KiDspeaks1, KiDspeaks2,dominik_non_gauss,desy3_peaks}), the three-point correlations function (e.g. \cite{threepoint1, threepoint2}) or machine learning based methods (e.g. \cite{Fluri2018,Ribli2019,methodpaper,kids_net,similar_paper}). These higher order statistics have the potential to greatly improve the cosmological parameter constraints and reduce the systematic effects. However, one problem is that analytical predictions are typically not available, making it necessary to predict them with costly simulations. Additionally, the common assumption of a Gaussian likelihood might break down for complicated statistics and likelihood-free inference methods may  have to be used.

In this work we perform a forward-model $w$CDM machine-learning analysis of the KiDS-1000 data~\cite{kids_dr4} using very similar settings as the KiDS-1000 two-point correlation function analysis~{\cite[hereafter \citetalias{kids1000_shear}]{kids1000_shear}}. To achieve this we  build upon our previous analysis of the KiDS-450 data~{\cite[hereafter \citetalias{kids_net}]{kids_net}}, but significantly improve the pipeline by increasing the number of cosmological and nuisance parameters, as well as using a different type of neural networks along with the likelihood-free inference method Gaussian Process Approximate Bayesian Computation (GPABC)~\cite{methodpaper} to constrain the cosmological parameters. Similar to~\citetalias{kids_net} we perform an analysis based on neural networks as well as a standard power spectrum analysis for comparison. In contrast to~\citetalias{kids_net}, we do not need to cut the observed data in independent patches, and instead treat the whole survey at once on the sphere. We create the necessary mock surveys for the evaluation of the power spectra and the training of the networks  using the \ufalcon~\cite{ufalcon_1,ufalcon_2} software and a novel simulations suite, called the \cosmogrid~\cite{cosmogrid}. The \cosmogrid~is an extensive simulation suite that contains $\sim$20'000 independent simulations of 2'500 distinct parameter combinations of the $w$CDM model, varying 6 cosmological parameters, namely the total matter density $\Omega_M$, the fluctuation amplitude $\sigma_8$, the spectral index $n_s$, the Hubble parameter $H_0 \equiv 100h$, the baryon density $\Omega_b$, and the dark energy equation of state parameter $w_0$. The large size of the \cosmogrid~poses computational challenges and we therefore limit the resoltion of our mock surveys to a \healpix~$n_\mathrm{side}=512$. Our generated mock surveys incorporate systematics such as photometric redshift errors or the multiplicative and additive shear biases in a similar way as~\citetalias{kids1000_shear}. Furthermore, we utilize the map-level implementation of the non-linear intrinsic alignment model (NLA)~\cite{nla1,nla2,nla3} presented in~\citetalias{kids_net} to constrain the intrinsic alignment amplitude $A_\mathrm{IA}$. Additionally, we develop a new method to treat baryonic feedback for the \cosmogrid~simulations based on~\cite{aurel_bary1,aurel_bray2,aurel_bary3} that allows us to marginalize over baryonic effects. We evaluate the power spectrum of the generated, spherical mock surveys and further analyse them with graph-convolutional neural networks utilizing \deepsphere~\cite{deepsphere} and an information maximizing loss~\cite{Charnock2018,methodpaper}. Finally, we perform a likelihood-free parameter inference using GPABC~\cite{methodpaper}.

This paper is structured as follows. In section~\ref{sec:data} we give an overview of the used data. In section~\ref{sec:blinding} we describe our blinding scheme. Section~\ref{sec:methods} contains a detailed description of our methods, including the mock survey generation, treatment of all systematics, the used networks and our inference pipeline. We discuss our tests of the whole pipeline with respect to simulation settings in section~\ref{sec:mock_tests} and present our results in section~\ref{sec:results}, which is followed by our conclusions in section~\ref{sec:conclusion}. Afterwards, appendix~\ref{ap:priors} includes more details about certain prior choices. In appendix~\ref{ap:sim_acc} we present tests for the accuracy of our mock surveys and in appendix~\ref{ap:shell_bary} we describe our baryon model in more detail. More information and tests for our examined networks are presented in appendix~\ref{ap:more_nets} and finally, our full parameter constraint can be found in appendix~\ref{ap:full_cons}.

\section{KiDS-1000 Data}
\label{sec:data}

In this work, we analyse the fourth data release \cite{kids_dr4} of the Kilo-Degree Survey (KiDS~\footnote{\url{http://kids.strw.leidenuniv.nl/}}) which is a public survey by the European Southern Observatory (ESO). This release contains $\sim1000$ deg$^2$ of images giving it the name KiDS-1000. KiDS is observing in four band (\textit{ugri}) using the OmegaCAM CCD mosaic camera mounted at the Cassegrain focus of the VLT Survey Telescope (VST). This combination was designed to have a well-behaved and almost round point spread function (PSF), as well as a small camera shear, making it ideal for weak lensing measurements. The combination with its partner survey, VIKING (VISTA Kilo-degree INfrared Galaxy survey~\cite{viking_2013}), increases the number of observed optical and near infrared bands of the galaxies to nine (\textit{ugriZYJHK$_s$}), greatly improving their photometric redshift estimates. 
The observed data is processed with the two pipelines \textsc{Theli}~\cite{theli_2013} and \textsc{Astro-Wise}~\cite{astro_wise_2013} and the galaxy shear estimates are obtained with \lensfit~\cite{lensfit_1,lensfit_2}. \lensfit~uses a likelihood-based method to fit a disc and bulge model to each of the observed galaxies to extract their shear values. Each galaxy is also assigned a weight based on this likelihood, with large weight indicating a high confidence of the fit. Additionally, photometric redshifts of the galaxies are obtained using the \bpz~code~\cite{bpz_code}. A detailed description of the KiDS-1000 methodology can be found in \cite{kids1000_cata}.

\subsection{Data Preparation}

\begin{figure}
    \centering
    \includegraphics[width=0.5\textwidth]{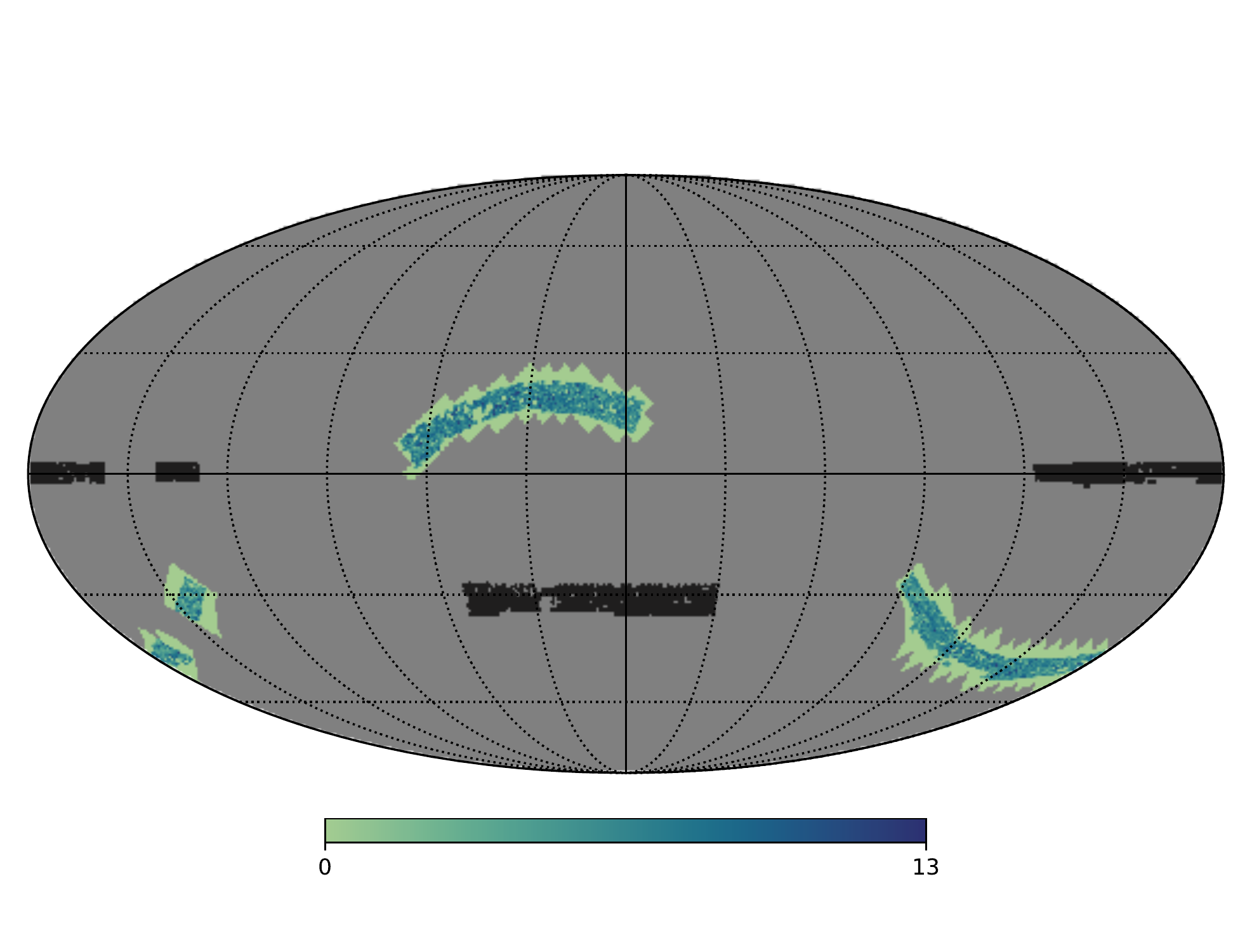}
    \caption{The original KiDS-1000 survey footprint is shown in dark grey and the galaxy density [galaxies/arcmin$^2$] of the rotated footprint in blue. The pixels with zero galaxies are part of the padding necessary to process the maps with the graph convolutional neural network (see section~\ref{sec:network}) and not part of the original survey. \label{fig:footprint_ori}}
\end{figure}

We split the galaxies into five tomographic bins using their \bpz~estimated in the same way as in~\citetalias{kids1000_shear}. The most important quantities of these bins can be found in table 1 of~\citetalias{kids1000_shear}. To create shear maps from the observed galaxies, we project them onto \healpix~\cite{Gorski_2005} maps with a resolution of $n_\mathrm{side} = 512$. However, before the projection, we rotate the positions of the galaxies, as well as their shear values. This rotation is performed to increase the number of cutouts that can be obtained from simulated maps as described in section~\ref{sec:mocks} and is visualized in Figure~\ref{fig:footprint_ori}.
After the rotation, the shear value of each pixel is calculated as the weighted mean of the shear values $e_i$ of the galaxies inside the pixel
\begin{equation}
    \gamma_\mathrm{pix} = \frac{\sum_{i\in\mathrm{pix}}w_i(e_i - c_z)}{\sum_{i\in\mathrm{pix}}w_i},
\end{equation}
where the $w_i$ are the \lensfit~weights and the constants $c_z$ are chosen such that the weighted mean shear of each redshift bin is zero (see section \ref{sec:m_and_c_bias}).

\section{Blinding}
\label{sec:blinding}

The analysis presented in this work was performed in a blinded fashion to avoid a confirmation bias. The shear values of the observed galaxies were only used to generate noise footprints in the analysis pipeline (see section~\ref{sec:noise}). We only unblinded the actual observed data after two conditions were met. First, the power spectra analysis (see section~\ref{sec:powerspec_ana_method}) had to be consistent with the network analysis (see section~\ref{sec:network_ana_method}). Second, the analysis had to be robust against the simulation settings used for the data generation (see section~\ref{sec:mock_tests}). The observed data was only analyzed after passing both of these requirements. After the unblinding we did not change our analysis pipeline. However, we did run an additional analysis, where we adapted a hyperparameter of our inference method. This is explained in more details in section~\ref{sec:results}.

\section{Methodology}
\label{sec:methods}

\begin{figure}
    \centering
    \includegraphics[width=0.5\textwidth]{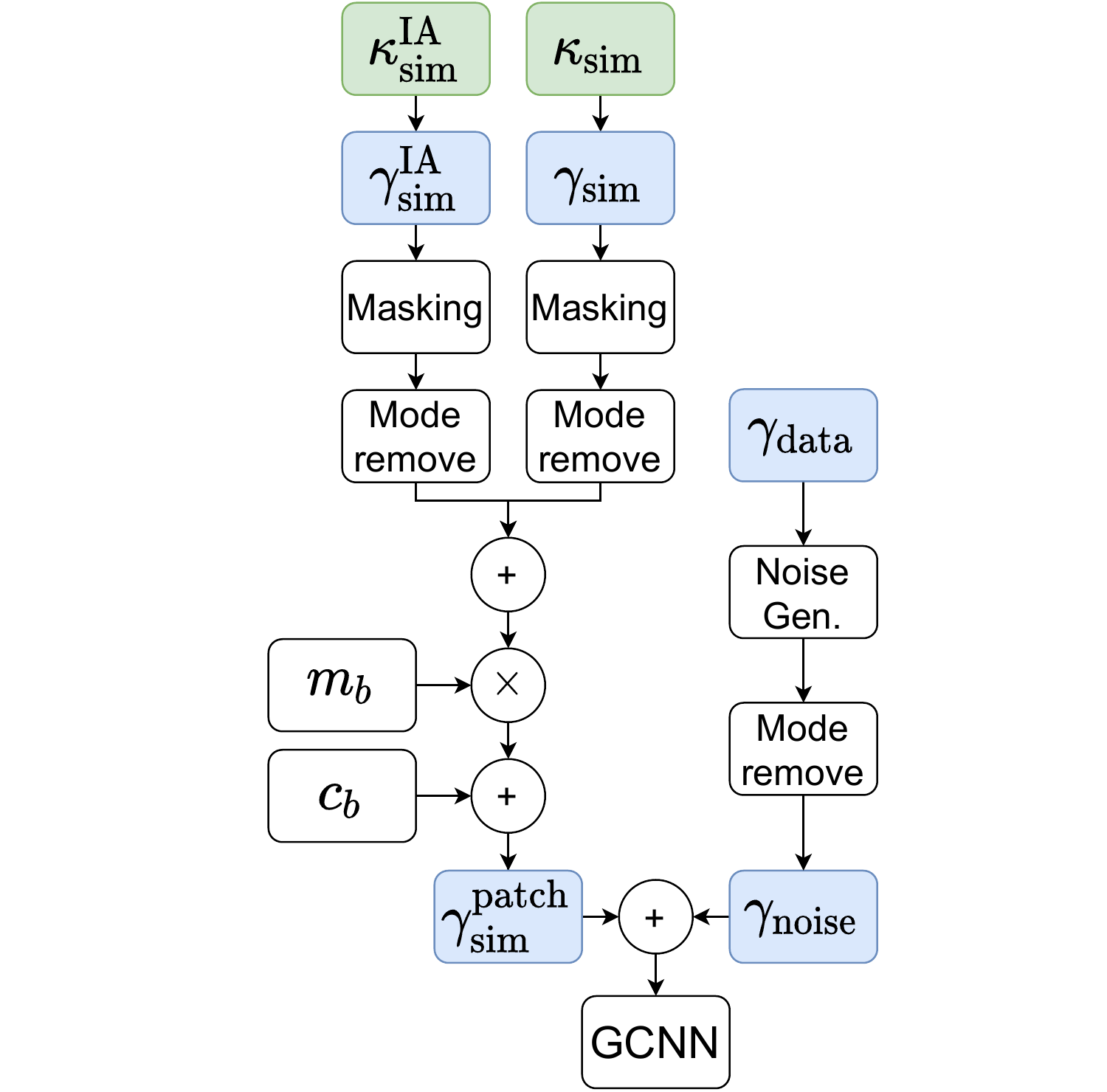}
    \caption{Overview of the data flow used in the analysis. Section~\ref{sec:mocks} explains the generation of full sky convergence $\kappa_\mathrm{sim}$ and intrinsic alignment maps $\kappa_\mathrm{sim}^\mathrm{IA}$, as well as, the spherical KS inversion to transform the convergence maps to shear maps $\gamma_\mathrm{sim}$ \& $\gamma_\mathrm{sim}^\mathrm{IA}$ and the subsequent masking. The mode removal strategy is covered in section~\ref{sec:mode_remove}. Section~\ref{sec:noise} describes the generation of noise maps $\gamma_\mathrm{noise}$ from the actual observed data $\gamma_\mathrm{data}$. Finally, the handling of the multiplicative shear bias $m_b$ and additive shear bias $c_b$ is described in section~\ref{sec:m_and_c_bias}. \label{fig:pipeline}}.
\end{figure}

An overview of the data flow is given by Figure~\ref{fig:pipeline}. 
The depicted flowchart explains most of the crucial parts of the data generation procedure. Additional parts of the pipeline that are not shown include our handling of the photometric redshift error (see section~\ref{sec:redshift}), our treatment of baryonic effects (see section~\ref{sec:barys}), details about the used networks (see sections~\ref{sec:network} and~\ref{sec:sumnet}) and, finally, our inference method (see section~\ref{sec:inference}).

\subsection{Simulations}
\label{sec:simulations}

\subsubsection{Parameters}

\begin{figure}
    \centering
    \includegraphics[width=0.5\textwidth]{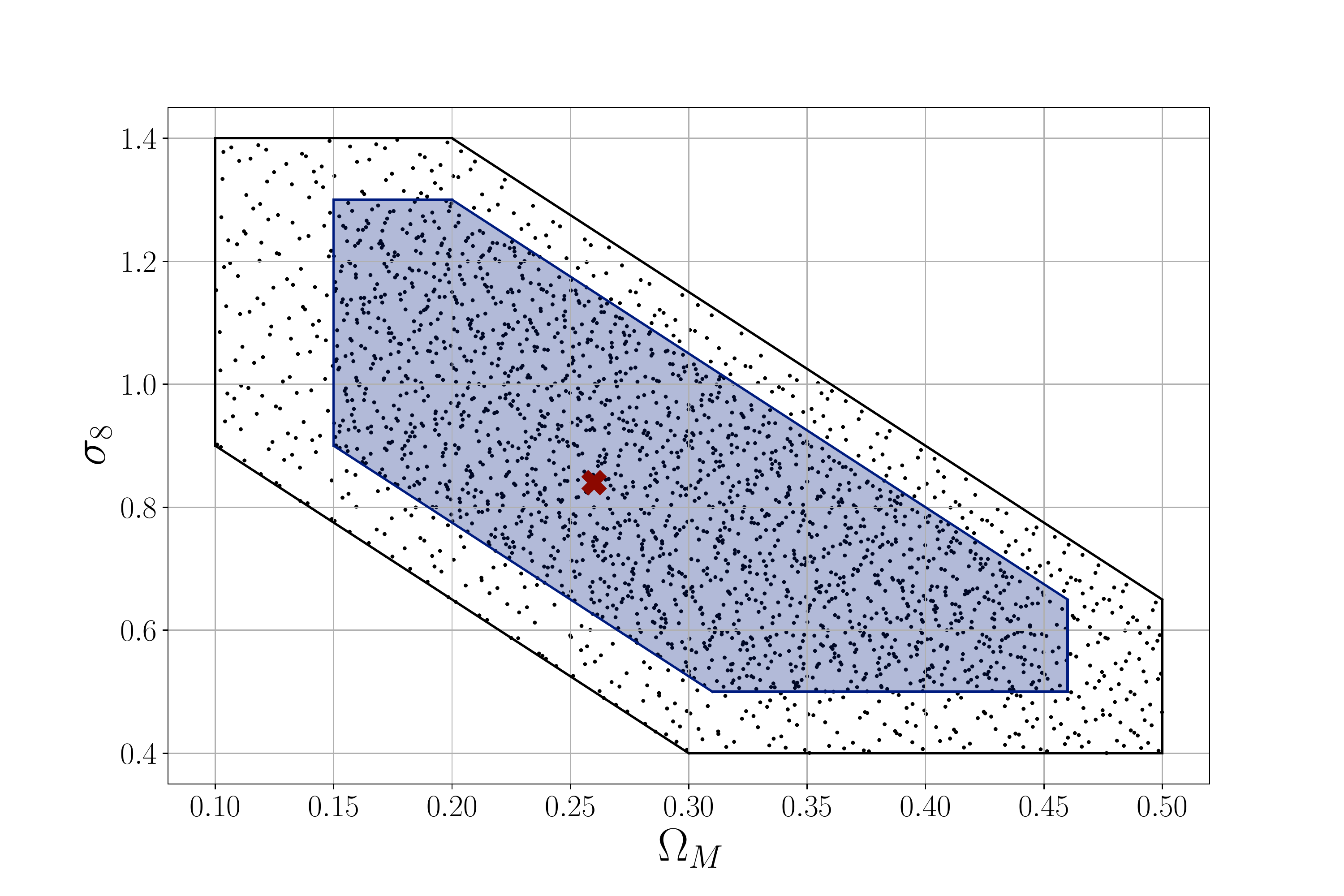}
    \caption{The 2'500 grid points projected onto the $\Omega_M-\sigma_8$ plane. The outer boundary shows the additional prior and the shaded region corresponds to the tight prior. The red cross in the middle indicates the fiducial cosmology. \label{fig:grid_plot}}
\end{figure}

The training of the GCNN as well as the inference of the cosmological parameters requires a tremendous amount of simulations. In this work, we use the \cosmogrid~\cite{cosmogrid} to generate the necessary weak lensing maps. The \cosmogrid~is a simulation suite generated with the fast and efficient \pkdgrav~code~\cite{Stadel2001}. It contains seven independent simulations for each of the 2'500 grid cosmologies varying six different cosmological parameters, as well as, 200 distinct simulations of a fiducial cosmology and perturbations of it. The perturbed fiducial simulations are simulations for which a single cosmological parameter $\theta$ was changed by a small shift $\pm\Delta\theta$, but the seeds of the initial conditions remained unchanged. These simulations can be used to estimate derivatives of quantities with respect to the cosmological parameters via finite difference methods (see e.g. section~\ref{sec:training}). The six varying cosmological parameters are listed in table~\ref{tab:grid}. The dark energy density $\Omega_\Lambda$ was adapted for each simulation to achieve a flat $w$CDM universe and the neutrino mass was fixed to three degenerate neutrinos each having a mass of $m_\nu = 0.02$ eV. The 2'500 grid points are split into two sub-grids: 1'250 grid points follow a Sobol sequence inside the priors listed in table~\ref{tab:grid} and 1'250 grid points follow a Sobol sequence with tighter priors around the fiducial parameters (see e.g. Figure~\ref{fig:grid_plot}), focusing on the region of interest. A Sobol sequence is a quasi-random low-discrepancy sequence that uniformly samples the space. It is similar to Latin hypercube sampling. However, it has the advantage that the sample size does not have to be set in advance, allowing the addition of new samples to the sequence. Another important property is that the number of dimensions of the samples can also be extended, making it easy to add new parameters. One should note that the tight prior is not relevant for this analysis as we only consider the full range of the parameters. The fiducial parameters and their perturbations are also listed in table~\ref{tab:grid}. 

There are two additional priors not stated in table~\ref{tab:grid}. The first one is in the $\Omega_M-\sigma_8$ plane, shown in Figure~\ref{fig:grid_plot}, and motivated by the degeneracy region of the two parameters. The second one is in the $\Omega_M-w_0$ plane and caused by the effective nature of the $w$CDM model and the relativistic fields in the simulations. More details about this can be found in appendix~\ref{ap:priors}.

\subsubsection{Settings}

The grid and fiducial simulations were all run with 832$^3$ particles in a box with a side length of 900 Mpc/$h$. The simulations also include three degenerate massive neutrinos each having a mass $m_\nu = 0.02$ eV and general relativistic effects, which are modeled as described in~\cite{neutrinos}. This model requires a lookup table of highly accurate transfer functions. This table was generated using \class~\cite{class_2011} and then transformed into the proper N-body gauge using \concept~\cite{concept_2019}. This lookup table was also used to generate the initial conditions at $z = 99$. The remaining precision parameters were left to their default values as in~\citetalias{kids_net}.

After the initial condition generation, the simulations used 140 base time steps~\cite{trillion_parts} to evolve the particles to $z = 0$. The first 70 time steps were equally spaced in proper time and evolved the particles to $z = 4$ and the remaining steps were equally spaced in proper time between $z = 4$ and $z = 0$. This split in the evolution was done to reduce the number of steps at low redshifts to reduce the amount of generated data. The simulations were run in lightcone mode and after crossing $z=3.5$ a shell of particles with a resolution of $n_\mathrm{side} = 2048$ was generated for each time step, containing the particles that left the observed lightcone. The simulated box was replicated to cover the relevant range if necessary. 
\pkdgrav~also includes a friends-of-friends (FoF) halo finder. For each time step, a halo catalog was generated using a linking length of one-fifth of the mean particle separation. This catalog was later used to include baryonic effects into the generated weak lensing maps (see section~\ref{sec:barys}).

\subsubsection{Benchmarks}
\label{sec:benchmarks}

Additionally to the fiducial and grid simulations, the \cosmogrid~contains different types of benchmark simulations. These simulations were all run using the fiducial cosmology. Each benchmark contains five independent simulations and the random seeds for the initial conditions were the same across the different settings. This was done to reduce the cosmic variance between the different benchmarks, which makes it easier to use them to test various systematics. The first benchmark simulations were run with exactly the same parameters as the fiducial simulations. This benchmark can be used to generate mock observations that are equivalent to the other benchmarks in terms of cosmic variance, but run with the normal simulation settings. For the second benchmark the particle count was increased to 2048$^3$. This can be used to check the convergence of the simulations and with respect to the particle count. For the third benchmark the boxsize was increased to 2250 Mpc/$h$, while keeping the particle density the same, which can be used to investigate the effect of modes that are larger than the fiducial setup. The last benchmark was generated to check the redshift resolution of the fiducial setup. For this run, the number of base timesteps was increased to 500, of which 100 were equally spaced in proper time to $z = 4$ and the remaining 400 were equally spaced between $z = 4$ and $z = 0$. All these benchmarks were used to check the robustness of our pipeline with respect to simulation settings and the results are presented in section~\ref{sec:mock_tests}.

\begin{table}
    \centering
    \begin{tabular}{lrrl}
    \hline 
        $\theta$ & $\theta_\mathrm{fid}$ & $\Delta\theta$ & Prior \\
    \hline
    \hline
        $\Omega_M$ & 0.26 & 0.01 & [0.1, 0.5] \\
        $\sigma_8$ & 0.84 & 0.015 & [0.4, 1.4] \\
        $h$ & 0.6736 & 0.02 & [0.64, 0.82] \\
        $\Omega_b$ & 0.0493 & 0.001 & [0.03, 0.06] \\
        $n_s$ & 0.9649 & 0.02 & [0.87, 1.07] \\
        $w_0$ & -1.0 & 0.05 & [-2.0, -0.333] \\
        $m_\nu$ & 0.02 & N/A & fixed \\
    \hline
        $A_\mathrm{IA}$ & 0.0 & 0.1 & [-3.0, 3.0] \\
        $\lg(M^0_c)$ & 13.82 & 0.1 & [12.0, 15.0] \\
        $\nu$ & 0.0 & 0.1 & [-2.0, 2.0] \\
    \hline
        $\delta_z$ & N/A & N/A & $\mathcal{N}(\mathbf{\mu}_z,\mathbf{C}_z)$ \\
        $m$ & N/A & N/A & $\mathcal{N}(\mathbf{\mu}_m,\mathbf{C}_m)$ \\
        $\delta_c$ & N/A & N/A & $\mathcal{N}(0,2.3\times10^{-4})$ \\
    \end{tabular}
    \caption{Cosmological and nuisance parameters used in this analysis. We provide the fiducial values $\theta_\mathrm{fid}$ along with their perturbations $\delta\theta$ used for the training of the network and the \moped~compression (see sections~\ref{sec:network} and~\ref{sec:powerspec_ana_method}). We vary six cosmological parameters in flat priors and also constrain three astrophysical nuisance parameters: the instrinsic alignment amplitude $A_\mathrm{IA}$ (section~\ref{sec:IA}) and the baryon parameters $M^0_c$ and $\nu$ (section~\ref{sec:barys}). The priors of the mean shift of the redshift distributions $\delta_z$ (section~\ref{sec:redshift}) and the multiplicative bias $m$  are modeled as multivariate Gaussians with means $\mathbf{\mu}_z$ and $\mathbf{\mu}_m$, and covariance matrices $\mathbf{C}_z$ and $\mathbf{C}_m$. The uncertainty of the additive bias $\delta_c$ (section~\ref{sec:m_and_c_bias}) is modeled with an univariate Gaussian prior. The parameters $\delta_z$, $m$ and $\delta_c$ are marginalized in the analysis and not constrained. \label{tab:grid}}
\end{table}

\subsection{Mock surveys}
\label{sec:mocks}

Similarly to \citetalias{kids_net} and \cite{desy3_peaks}, we generate full sky convergence maps with the simulated shells using \ufalcon~\cite{ufalcon_1,ufalcon_2}. However, we downgrade the simulated shells to an $n_\mathrm{side}$ of 512 to alleviate the computational costs and reduce the memory consumption of the generated surveys. \ufalcon~uses the Born approximation to efficiently generate convergence and intrinsic alignment (see section~\ref{sec:IA}) maps. The convergence of any given pixel $\theta_\mathrm{pix}$ is calculated in the following way
\begin{equation}
\kappa(\theta_\mathrm{pix}) \approx \frac{3}{2}\Omega_\mathrm{m}\sum_bW_b\frac{H_0}{c}\int_{\Delta z_b}\frac{c\mathrm{d}z}{H_0E(z)}\delta\left(\Vec{v}_\mathrm{pix},z\right), \nonumber
\end{equation}
where $\mathcal{D}(z)$ is the dimensionless comoving distance and the vector $\Vec{v}_\mathrm{pix}$ is given by
\begin{equation*}
    \Vec{v}_\mathrm{pix} = \frac{c}{H_0}\mathcal{D}(z)\hat{n}_\mathrm{pix}.
\end{equation*}
The vector $\hat{n}_\mathrm{pix}$ is a unit vector pointing to the pixels center and $E(z)$ is given by
\begin{equation}
\mathrm{d}\mathcal{D} = \frac{\mathrm{d}z}{E(z)}.
\end{equation}
The sum runs over all redshift shells and $\Delta z_b$ is the thickness of shell $b$.
Each shell gets the additional weight $W_b$ which depends on the redshift distribution of the source galaxies. We refer the interested reader to~\cite{ufalcon_1,ufalcon_2} for a detailed description. The generation of the intrinsic alignment maps works in a similar way, but uses different weights for the shells, which is described in~\citetalias{kids_net}. The Born approximation used to generate the maps might have an impact on the constraints. However,~\cite{born_is_fine} showed that the impact of the approximation is negligible for stage-III surveys like KiDS, which was also verified by the systematics test in~\citetalias{kids_net}. 

We generate the maps using the same redshift distributions as~\citetalias{kids1000_shear}. A detailed description of the construction of the redshift distributions can be found in~\cite{kids1000_redshift} and our treatment of the photometric redshift error is described in section~\ref{sec:redshift}. It is important to note that these redshift distributions span a range of $0 \leq z \leq 6$ which is significantly larger than the range of our simulations. However, the lensing contribution of the matter between $3.5 \leq z \leq 6$ is negligible for KiDS-1000. We therefore used the whole range of the redshift distributions, but ignored the contribution of the matter in the high redshift regime outside of the range of our simulations. We compare the power spectra of the generated maps to predictions obtained with \pyccl~\footnote{\url{https://github.com/LSSTDESC/CCL}} in appendix~\ref{ap:sim_acc}.

After generating full sky convergence maps, we used the spherical Kaiser-Squires (KS) inversion~\cite{spherical_KS} to transform them into weak lensing shear maps. This transformation has two unwanted side effects that have to be addressed in the analysis. First, such a transformation acts as a low-pass filter on the maps, because the highest $\ell$-mode in the transformation is directly linked to the resolution of the maps and a perfect reconstruction of the original convergence map is not possible. Second, the generated map contains only E-modes. We address these issues after the masking with a ``mode removal'' (see section~\ref{sec:mode_remove}) strategy similar to the one used in~\citetalias{kids_net}.

\begin{figure}
    \centering
    \includegraphics[width=0.5\textwidth]{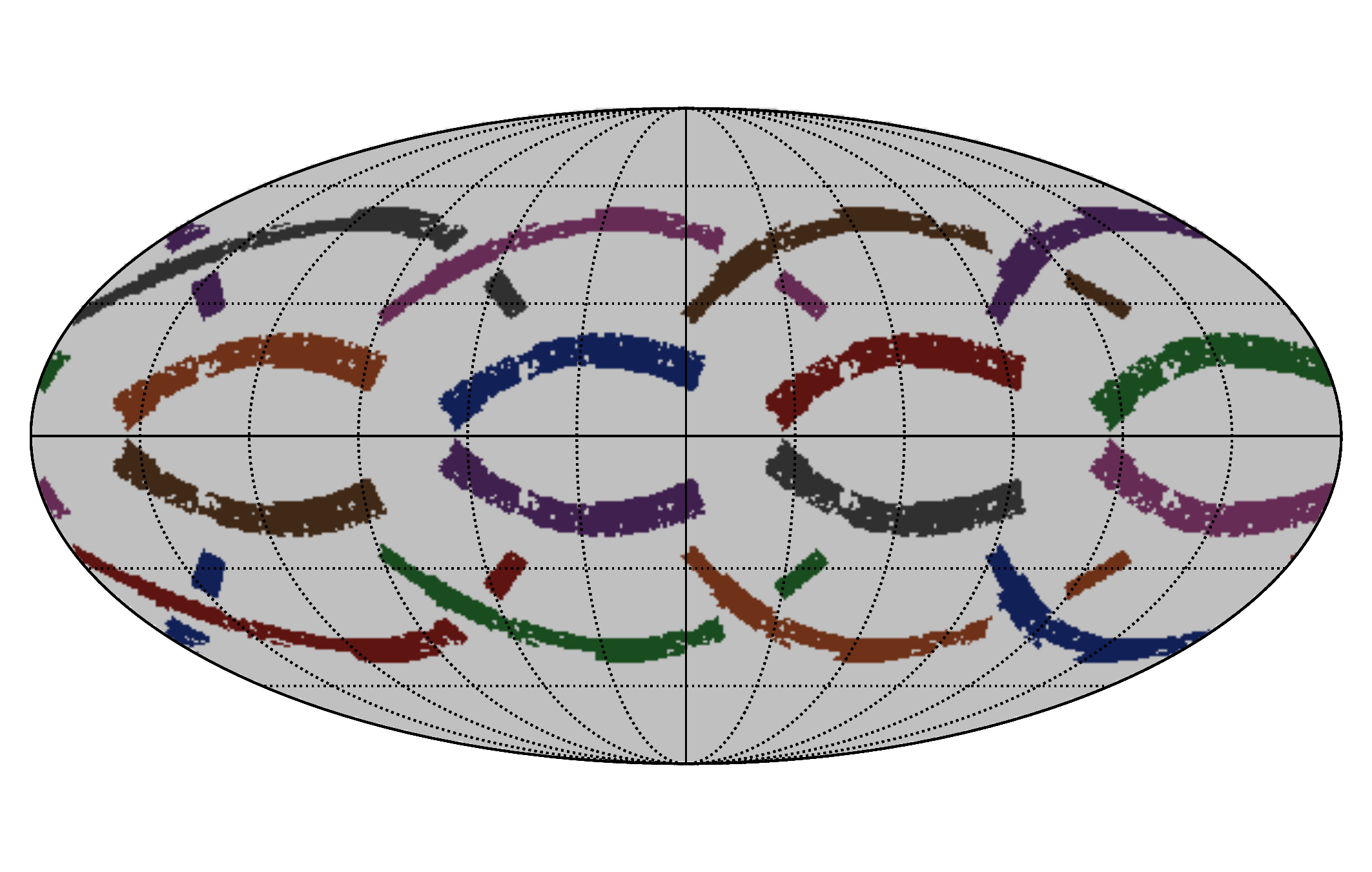}
    \caption{The eight patches that were to make better use of the full sky simulations. The patches respect the \healpix~octahedral symmetry, meaning that one can perfectly map every pixel of each patch onto any other patch without causing unwanted artefacts. \label{fig:patches}}
\end{figure}

After the first spherical KS inversion, we mask the full sky cosmic shear maps to generate mock surveys. To make better use of the simulated area, we cut out eight survey footprints out of a single simulation. These footprints are shown in Figure~\ref{fig:patches}. Each of the shown footprints can be obtained from the original (rotated) survey footprint (see Figure~\ref{fig:footprint_ori}) with a transformation that respects the \healpix~octahedral symmetry, meaning that each pixel can be mapped to exactly one pixel without overlapping with any other pixels. For the grid cosmologies, we generated five full sky convergence maps per patch with a different redshift distribution (see section~\ref{sec:redshift}), resulting in 280 (seven simulations $\times$ eight patches $\times$ five realizations) mock surveys per grid point after the KS inversion. This results in a total of 700'000 mock surveys for all grid cosmologies combined. For the fiducial cosmology, including its perturbations and benchmarks, we generated a total of ten full sky convergence maps per patch, resulting in 16'000 mock surveys for the fiducial simulations and its perturbations and 400 mock surveys for the benchmarks.

The combined size of all generated mock surveys with a resolution of $n_\mathrm{side} = 512$ is approximately $\sim 11$TB, showing the immense size of the data set. Increasing the resolution to $n_\mathrm{side} = 1024$ would increase the memory usage of the mock surveys by a factor of four. Additionally, it would significantly increase the time to train the neural networks and to evaluate the power spectra. We therefore leave the analysis of smaller scales that could be probed with a higher resolution to future work. 

At this point the generated mock surveys do not contain any noise and still suffer from the side effects of the KS inversion. Also, they no not include effects of the multiplicative and additive shear bias yet. We will cover all these aspects in the following sections. 

\subsection{Shape and Measurement Noise}
\label{sec:noise}

The simulation of realistic noise maps is crucial for the analysis. To do this, we use the same technique as in~\citetalias{kids_net} and generate noise maps using random rotations of the observed shear values
\begin{equation}
    \gamma_\mathrm{pix}^\mathrm{noise} = \frac{\sum_{j \in \mathrm{pix}}\exp\left(\vartheta_ji\right)w_j(\gamma_j - c_z)}{\sum_{j \in \mathrm{pix}}w_j},
\end{equation}
where the $\vartheta_j$ are drawn uniformly from $[0, 2\pi)$. The rotation and projection of the galaxies is computationally very efficient and could theoretically be done on the fly during the training of the networks. However, noise maps that are generated this way are not band limited and also contain B-modes. Therefore, we have to apply our mode removal (see section \ref{sec:mode_remove}) which requires a spherical harmonics decomposition. Such a transformation cannot be done on-the-fly during the training or evaluation of the network and we decided to create a static dataset of noise footprints. We produced one noise map for each generated mock survey of the grid cosmologies and two for each mock survey of the fiducial cosmology, one for the training and one for the evaluation. The generation of two independent noise maps for each mock survey of the fiducial simulations was necessary to avoid overfitting during the training of the network (see section~\ref{sec:training}).

\subsection{Mode Removal}
\label{sec:mode_remove}

Projecting the shear of the observed galaxies onto a spherical grid leads to shear maps that contain B-modes and are not band-limited. We define a band-limited map as a map that can be decomposed into spherical harmonics without losing information. Theoretically, this is always possible, but it can be numerically unstable. Because of this, \healpix~limits the resolution of the spherical harmonics decomposition to modes with $\ell \leq 3n_\mathrm{side} - 1$ per default. Therefore, the simulated shear maps do not have modes above the cutoff scale. However, the projected galaxies from the survey and the noise realizations still contain these modes, which can potentially lead to biases in the results. Additionally, the simulated convergence maps will produce shear maps that only contain E-modes, as opposed to the observed maps and noise maps, which can contain B-modes. These missing modes can also lead to biases. We mitigate these potential biases by performing an additional spherical harmonic decomposition on all masked maps. During this decomposition we remove all B-modes from the shear maps. This way, all generated simulations, noise maps and the true data are band limited, contain the same masking effects and are E-modes only. We dub this additional decomposition ``model removal'' (see Figure~\ref{fig:pipeline}).

\subsection{Systematics}

In this section we describe our treatment of the systematic effects that we consider in this work. In our analysis we marginalize over photometric redshift errors (section~\ref{sec:redshift}), intrinsic alignment (section~\ref{sec:IA}), the multiplicative and additive shear bias (section~\ref{sec:m_and_c_bias}), as well as baryonic feedback (section~\ref{sec:barys}). Furthermore, we check that our results are robust to the simulation settings of the \cosmogrid~(section~\ref{sec:sim_settings}).

\subsubsection{Photometric Redshift}
\label{sec:redshift}

The redshift distributions of each tomographic bin of the KiDS-1000 dataset have been calibrated using the self-organising map (SOM) method of~\cite{som1} with additional requirements described in~\cite{som2}. The resulting distributions have a reliable mean redshift with a high accuracy coming from the nine-band photometry of the galaxies. The expected uncertainty on the mean redshift was estimated using mock simulations. Following~\citetalias{kids1000_shear}, we use the same means and covariance matrix to model the uncertainty on the mean redshifts of the five tomographic distributions as multivariate Gaussians. Each generated mock survey (see section~\ref{sec:mocks}), either for the training or evaluation, has been produced with a redshift distribution that is shifted by a random draw from this multivariate Gaussian to marginalize over the photometric redshift errors.

\subsubsection{Multiplicative and Additive Shear Bias}
\label{sec:m_and_c_bias}

We use the common model of an additive and multiplicative shear bias correction to model the calibration uncertainty
\begin{equation}
    \gamma^\mathrm{obs} = (1 + m_z)\gamma + c_z,
\end{equation}
where $m_z$ and $c_z$ are the multiplicative and additive correction, respectively. We treat the multiplicative bias in the same way as the ``free $m$ correlated'' model of~\citetalias{kids1000_shear}, using the same mean and covariance matrix for the Gaussian prior (see Table~\ref{tab:grid}). Each redshift bin is corrected by a multiplicative bias term that is drawn from a Gaussian distribution and the terms of the different bins are fully correlated. It was shown in~\citetalias{kids1000_shear} that treating the multiplicative biases as correlated or uncorrelated does not significantly affect the resulting constraints. We marginalize over this uncertainty by correcting each of the generated mock surveys (see section~\ref{sec:mocks}) with a randomly drawn bias term after the mode removal (see section~\ref{sec:mode_remove} and Figure~\ref{fig:pipeline}).

We also treat the additive shear biases for each redshift bin. We remove the mean shear from all shear maps, simulated and observed alike, and then model the remaining uncertainty similar to the treatment of the real space correlation function $\xi_+$ in~\citetalias{kids1000_shear}. The mode removal (see section~\ref{sec:mode_remove}) removes all B-modes from the shear maps and hence removes any additive correction from the imaginary part of the shear maps. Therefore, we only need to model the additive correction of the real part. We expect this correction to have a negligible impact on the results. Nevertheless, we use the same uncertainty $\delta_c$ to model this additive correction as~\citetalias{kids1000_shear} used for their treatment of the correction in their $\xi_1$ analysis. This correction is applied to each simulated survey after the mode removal (see section~\ref{sec:mode_remove} and Figure~\ref{fig:pipeline}).

\subsubsection{Intrinsic Alignment}
\label{sec:IA}

The intrinsic alignment of galaxies around massive objects is one of the most important systematic effects in weak gravitational lensing. We use the same model as in~\citetalias{kids_net} which is based on the non-linear intrinsic alignment model (NLA)~\cite{nla1,nla2,nla3}. We refer the reader to~\citetalias{kids_net} and~\cite{dominik_non_gauss} for details about the concrete implementation. Similar to~\citetalias{kids_net} and~\citetalias{kids1000_shear}, we ignore the redshift and luminosity dependence of the NLA model and only consider the intrinsic alignment amplitude $A_\mathrm{IA}$ as a free parameter. We create an intrinsic alignment map along each mock survey containing the cosmological signal. This intrinsic alignment map, scaled by the amplitude $A_\mathrm{IA}$, can then be added to the actual signal map (see Figure~\ref{fig:pipeline}). We also check the accuracy of our IA only maps by comparing their power spectra to \pyccl~predictions. This comparison is presented in appendix~\ref{ap:sim_acc}. 

To constrain the intrinsic alignment amplitude, we extent the Sobol sequence of the grid cosmolgies by an additional dimension and scale it to our prior on $A_\mathrm{IA}$ (see Table~\ref{tab:grid}). This way, each simulated cosmology has a unique amplitude which can be used for the parameter inference (see section~\ref{sec:inference}). 

\subsubsection{Baryonic Feedback}
\label{sec:barys}

Baryonic effects can have a significant impact on the small scale clustering in large cosmological simulations~\cite[see e.g.][]{vandaalen_2011,McCarthy:2016mry,Chisari:2019tus}. However, modeling these effects is a challenging task as they depend on hydrodynamical interactions which include sub-grid stellar and active galactic nuclei (AGN) feedback phenomena. In~\citetalias{kids1000_shear} these corrections are modelled at the level of the matter power spectrum by using the \hmcode~\cite{hmcode}. The \hmcode~introduces the baryonic feedback parameter $A_\mathrm{bary}$ that controls the strength of the corrections, with $A_\mathrm{bary} = 3.13$ corresponding to the dark-matter only case. Such a model is, however, not applicable for our map-based analysis. A possible way to introduce baryonic feedback into our simulations would be to run large-scale hydrodynamic simulations. However, these simulations are computationally very expensive, which is why multiple approaches to incorporate baryonic feedback into dark-matter-only simulations have been proposed \cite{aurel_bary1,painting_barys,arico_2020,lu_2021}. For example, \cite{painting_barys} applied a deep learning based approach to directly ``paint'' baryonic feedback onto weak lensing maps. In this work we use a modified version of the \textit{baryonification model} of~\cite{aurel_bary1,aurel_bray2,aurel_bary3} to incorporate baryonic feedback into our simulations. A different modified version of this method has already been used by \cite{2021arXiv210911060L} to perform a neural network analysis of flat weak lensing maps, showing that neural networks are capable to learn the relevant model parameters. The method works by adjusting the particle positions of dark-matter-only simulations such that the resulting density field includes the effects of gas and stars at cosmological scales. It has been shown by~\cite{aurel_bray2} that the corrected density field is in good agreement with full hydrodynamical simulations. We will only cover the most relevant aspects of the baryonification model here, and provide more details and benchmarks in appendix~\ref{ap:shell_bary}. We refer the interested reader to~\cite{aurel_bray2} for a detailed description of the original model. 

The baryonification model of~\cite{aurel_bray2} works by displacing $N$-body particles around dark matter halos, such that the resulting profiles include effects from stars and gas that are influenced by feedback processes. The original model has several free parameters that control the various effects. In our model we only vary the parameter $M_c$ which dictates the mass dependence of the gas profile and has been shown to be the most important parameter for cosmology \cite{Giri:2021qin}. In contrast to the original model of~\cite{aurel_bray2}, we do not directly use $M_c$ as a baryonic parameter, but we instead assume
\begin{equation}\label{eq:Mc_of_z}
    M_c = M_c^0(1 + z)^\nu,
\end{equation}
where $M_c^0$ and $\nu$ correspond to our new model parameters. Equation~\eqref{eq:Mc_of_z} corresponds to an extension of the model in~\cite{aurel_bray2}, allowing for an additional redshift dependence of the baryonic feedback effects. We fix the remaining parameters of the baryonification model to the best-guess parameters (model B-avrg), which can be found in table 2 of~\cite{aurel_bray2}. The values of the fixed parameters are motivated by observed X-ray gas fractions~\cite{Sun2009, Vikhlinin2009, Gonzalez2013} including current uncertainties of the hydrostatic mass bias~\cite{Eckert2016}. The fiducial values and priors of the two varying parameters $(M_c^0, \nu)$ are listed in table~\ref{tab:grid}. 

In the original model, the particles are moved inside snapshots at a fixed redshift. However, the simulations of the \cosmogrid~were run in lighcone mode, meaning that we only have access to the particle shells instead of full snapshots. Therefore, we use the friends-of-friends (FoF) halo catalogue of the \cosmogrid~simulations to adapt the particle shells instead of the positions of individual particles inside snapshots. We provide more details about this procedure in appendix~\ref{ap:shell_bary}. Similar to the treatment of the intrinsic alignment amplitude, we extent the Sobol sequence of the grid cosmologies by two dimension to assign each simulated cosmology a unique set of parameters $(M_c^0, \nu)$. We then generate a second set of simulated shells that contain the baryonic corrections for all simulations. For the fiducial simulations, their perturbations and the benchmarks, we chose our fiducial baryon correction parameters and their corresponding perturbations according to table~\ref{tab:grid}. One should note that the fiducial baryon correction parameters are not equivalent to the dark-matter-only case, but reflect the best-guess parameters of~\cite{aurel_bray2}. Using this second set of shells, we repeat the previously described procedure to generate mock surveys. This allows us to perform an analysis that fully incorporates baryonic effects at the level of the model described above.

\subsubsection{Simulation Settings}
\label{sec:sim_settings}

The results of the analysis should be robust with respect to the simulation settings. We check this robustness using mock observations of the benchmark simulations (see section~\ref{sec:benchmarks}). In this way we can test the influence of the number of particles, the box size, and the redshift resolution of the shells on the cosmological constraints. The number of particles is relevant for the resolution at small scales. Using a larger box size enables us to measure the impact of potential super survey modes. Finally, the benchmark with the higher redshift resolution makes it possible to check if the number of base time steps of the simulations is sufficiently large. The results of these tests are presented in section~\ref{sec:mock_tests}.

\subsection{Graph Convolutional Neural Network}
\label{sec:network}

Due to the large survey area and footprint of the KiDS-1000 data, the curvature of the sky needs to be taken into account. In~\citetalias{kids_net}, this problem was solved by projecting parts of the survey onto flat patches. However, in this work, we choose to treat the survey as a whole. There are multiple approaches to deal with data on the sphere in machine learning. Following~\cite{methodpaper} we chose to use 
\deepsphere~\footnote{\url{https://github.com/deepsphere/deepsphere-cosmo-tf2}}, described in~\cite{deepsphere}, which treats the pixel of the maps as nodes in a graph. \deepsphere~has two main types of layers that are described in~\cite{deepsphere,methodpaper}. First, pseudo convolutions make use of the \healpix~pixelization and are used to reduce the $n_\mathrm{side}$ of the maps with trainable weights. These layers can only be applied if the input maps are padded correctly (see~\cite{methodpaper}). This padding, visualized in Figure~\ref{fig:footprint_ori}, is necessary such that the lower resolution maps have a valid number of pixels. The other type of layers are graph convolutions, which are expressed in terms of Chebyshev polynomials and applied to the data via the graph Laplacian. These layers can be used to construct composite layers like residual layers. In this work we use the same settings and types of layers as in~\cite{methodpaper}. Our fiducial architecture is listed in table~\ref{tab:architecture}. All layers of networks used in this work, if not mentioned otherwise, use the linear rectified unit (ReLU) as activation function. We trained a total of three networks with the same fiducial architecture (see section~\ref{sec:training}). Additionally, we trained three similar networks to check the robustness of the results regarding certain choices of hyperparameters. The first network was trained with a larger batch size (see section~\ref{sec:training}). The second was trained with more residual layers and the third one was trained without graph convolutional layers. These networks are described in more detail in appendix~\ref{ap:more_nets}, and we will refer to them as ``benchmark networks''. 

\begin{table}[t]
    \centering
    \caption{Fiducial architecture of the used GCNN. We report layer type, output shape ($N_b$ being the batch size) and number of trainable parameters. The residual layer (see Figure~5 of~\cite{methodpaper}) is repeated five times. The output of the network is only four dimensional since we decided to ignore certain parameters in the loss (see section~\ref{sec:training}). 
    \label{tab:architecture}}
    \begin{tabular}{llr}
         \textbf{Layer Type} & \textbf{Output Shape} & \textbf{\# Parameter} \\
         \hline\vspace{-5pt} \\ 
         Input & ($N_b$,\ 149504,\ 10) & 0 \\
         Pseudo Convolution & ($N_b$, 37376, 32) & 1312 \\
         Pseudo Convolution & ($N_b$, 9344, 64)  & 8256 \\
         Pseudo Convolution & ($N_b$, 2336, 128) & 32896 \\
         Chebyshev Convolution & ($N_b$, 2336, 256) & 163840 \\
         Layer-Normalization & ($N_b$, 2336, 256) & 512 \\
         Pseudo Convolution & ($N_b$, 584, 256) & 262400 \\
         Chebyshev Convolution & ($N_b$, 584, 256) & 327680 \\
         Layer-Normalization & ($N_b$, 584, 256) & 512 \\
         Pseudo Convolution & ($N_b$, 146, 256) & 262400 \\
         Residual Layer & ($N_b$, 146, 256) & 656896 \\
         \multicolumn{3}{c}{$\vdots$} \\
         Residual Layer & ($N_b$, 1672, 128) & 170528 \\
         Flatten & ($N_b$, 37376) & 0 \\
         Layer-Normalization & ($N_b$, 37376) & 74752 \\
         Fully Connected &  ($N_b$, 4) &  149508
    \end{tabular}
\end{table}

\subsubsection{Training}
\label{sec:training}

We used both shear components of all five redshift bins as the input of the graph convolutional neural network (GCNN), resulting in a total of ten input channels (see table~\ref{tab:architecture}). We trained the GCNNs using the fiducial simulations of the \cosmogrid~and the information maximizing loss presented in~\cite{methodpaper} that is based on~\cite{Charnock2018} and optimizes the Cram\'er-Rao bound. The loss function is given by
\begin{equation}
    L = \log\det(\mathrm{Cov}_\theta(\hat{s})) 
                            - 2\log\left\vert\det\left(\frac{\partial \Psi_\theta(\hat{s})}{\partial \theta}\right)\right\vert, \label{eq:loss_1}
\end{equation}
where $\hat{s}$ is the output of the network (summary), and $\Psi_\theta(\hat{s})$ is the expected value of $\hat{s}$ for input maps that were generated with the cosmological and nuisance parameters $\theta$. We found that using all parameters leads to instabilities during the training because the signal originating from some parameters is fairly weak. We therefore decided to only use the four parameters for the training to which weak lensing is most sensitive to $\theta = (\Omega_M, \sigma_8, w_0, A_\mathrm{IA})$. One should note that this does not correspond to a marginalization of the remaining parameters. The network will still output different results if for example $\Omega_b$ is changed. However, the network is not aware of the missing parameters during the training. Additionally, to make the training more stable we used the regularizer proposed in~\cite{methodpaper}
\begin{equation}
    L_\mathrm{regu} = \lambda\left\Vert \frac{\partial \Psi_\theta(\hat{s})}{\partial \theta} - \mathbf{I} \right\Vert, \label{eq:regu_loss}
\end{equation}
where we set $\lambda = 100$ and $\mathbf{I}$ is the identity matrix.
The training is performed on the mock surveys without baryon corrections. We found that simply evaluating the normally trained network on the mock surveys containing baryon corrections leads to similar results as retraining the network with the new maps. The reason for this is most likely the high noise level of the maps and the low signal originating from the baryon nuisance parameters. The covariance $\mathrm{Cov}_\theta(\hat{s}))$ is estimated empirically during the training and the derivatives of the expected value are calculated empirically and with a finite difference method using the fiducial simulations with perturbed parameters (see table~\ref{tab:grid}). Each network, fiducial architecture or benchmark, is trained for 100'000 steps on eight GPUs in a data-parallel fashion on the supercomputer Piz Daint~\footnote{\url{https://www.cscs.ch/computers/piz-daint/}}. Each GPU had a local batch size of eight, leading to a final batch size of 64, with the exception of the benchmark model that used a larger total batch size (see appendix~\ref{ap:more_nets}), which was trained on 16 GPUs. The weight optimization is done using the Adam optimizer~\citep{Adamopt} with an initial learning rate of 0.0001 and moments $\beta_1=0.9$ and $\beta_2=0.999$. As in~\cite{methodpaper}, we clip the gradient, reducing the global norm to 5.0 if necessary to avoid large weight updates. One should note that the actual input of the network is a batch of 576 because of the perturbed maps that have to be evaluated to estimate the derivative used in the loss (see equation~\eqref{eq:loss_1}). During every training step each fiducial map in the current batch is augmented with random multiplicative and additive shear biases and paired with a random noise realization. We observed that the networks are overfitting by learning the 16'000 noise maps of our fiducial set, if we trained the networks for long enough. This overfitting could potentially be avoided by using more regularization techniques like dropout~\cite{dropout}. We leave such applications to future work. To mitigate the overfitting during the inference, we use new noise realizations (see section~\ref{sec:noise}) for the evaluation. We evaluate the whole dataset, fiducial and grid, every 10'000 steps.

\subsection{Summary Networks}
\label{sec:sumnet}

Additionally to the GCNN, we train standard fully connected networks on pre-calculated summary statistics. To train the network we evaluate the summary statistics of 128'000 combinations of signal and noise maps for the fiducial simulations and perturbations using only the noise generated for the training. We also evaluate the summary statistics using the maps of the grid cosmologies and the evaluation set of the fiducial simulations. For each generated survey we evaluate the auto- and cross-spectra in 30 linearly spaced bins for $\ell \in [0, 1500]$, covering almost all resolvable scales. Further, we perform a peak count analysis with auto- and cross-peaks similar to~\cite{desy3_peaks}. However, we found that including the peak counts in the input of the networks did not significantly increase the performance. This is most likely because of the low resolution of the maps and the small survey area, making it hard for peak counts to extract a strong signal. Therefore, we did not use any additional summaries besides the auto- and cross-correlations. The architecture of the networks is presented in table~\ref{tab:architecture_sumnet}.

\begin{table}
    \centering
    \caption{Architecture of the used summary networks. We report layer type, output shape ($N_b$ being the batch size) and number of trainable parameters. The fully connected layers in the middle are repeated nine times. The output of the network is only four dimensional since we decided to ignore certain parameters in the loss (see section~\ref{sec:training_sumnet}) 
    \label{tab:architecture_sumnet}}
    \begin{tabular}{llr}
         \textbf{Layer Type} & \textbf{Output Shape} & \textbf{\# Parameter} \\
         \hline\vspace{-5pt} \\ 
         Input & ($N_b$,\ 450) & 0 \\
         Layer-Normalization & ($N_b$, 450) & 900 \\
         Fully Connected &  ($N_b$, 512) & 230912 \\
         Fully Connected &  ($N_b$, 256) & 131328 \\
         Fully Connected &  ($N_b$, 256) & 65792 \\
         \multicolumn{3}{c}{$\vdots$} \\
         Fully Connected &  ($N_b$, 256) & 65792 \\
         Fully Connected &  ($N_b$, 3) &  1028
    \end{tabular}
\end{table}

\begin{figure*}[t!]
    \centering
    \includegraphics[width=1.0\textwidth]{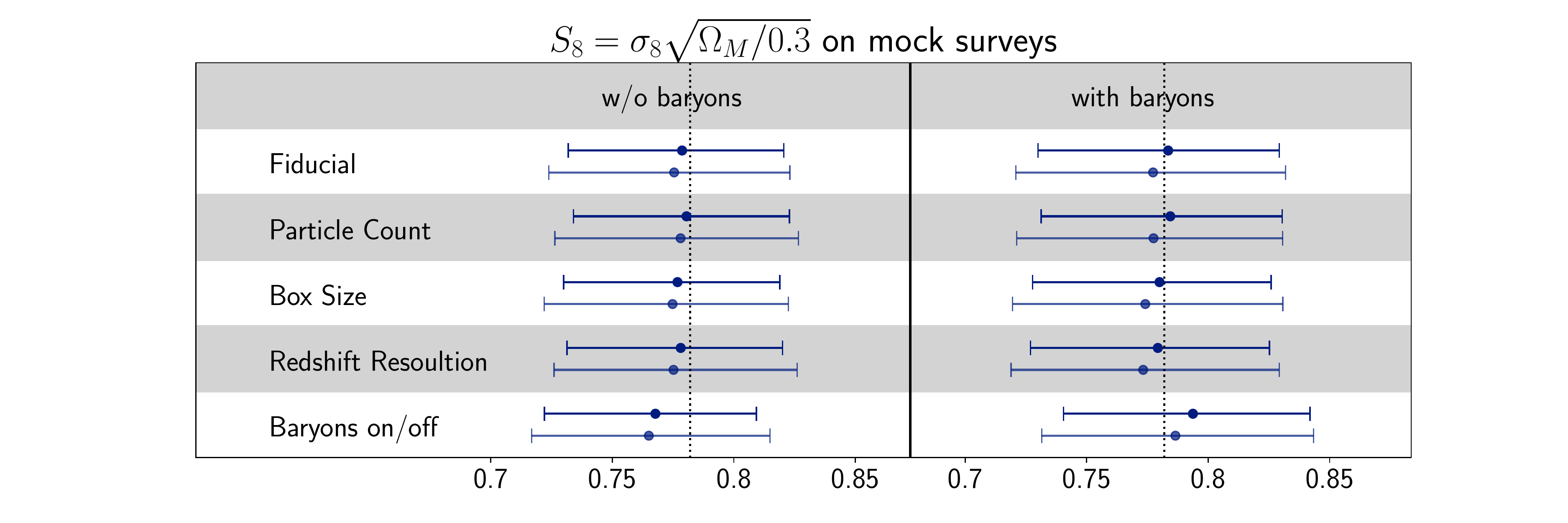}
    \caption{Constraints of the degeneracy parameter $S_8$ using mock observations generated from the \cosmogrid~benchmark simulations. The error bars correspond to the 68\% confidence intervals and the dots indicate the expected value. The upper constraints on each stripe were generated using the GCNNs and summary networks while the lower ones stem from our power spectra analysis. The left column of the plot shows the constraints without including Baryonic corrections in the simulated maps used for the inference, i.e. the maps used for the evaluation of the GNCC and power spectra, while the constraints in the right column includes them. Besides the last row, the mock observations used in the left column also ignore Baryon feedback. The last row shows the constraints for the fiducial mock observation (first row) again, but including Baryonic correction in the left column and ignoring them in the right column.  \label{fig:s8_mocks}}
\end{figure*}

\subsubsection{Training}
\label{sec:training_sumnet}

We train three instances of the summary networks presented in table~\ref{tab:architecture_sumnet} for 100'000 steps using the same optimizer settings and gradient clipping as for the GCNN. However, since the input data is significantly smaller, we could perform the training on a single GPU with a batch size of 1'024. As for the GCNN we reduce the relevant parameters for the training to $\theta = (\Omega_M, \sigma_8, w_0, A_\mathrm{IA})$. The input of the network is the logarithm of the absolute value of the auto- and cross-spectra. This transformation is performed to avoid the usually very small values of the spectra. Training the network this way can lead to significant overfitting, as only 128'000 different samples are available. Therefore, we added a standard L2 regularization term with a weight of 0.25 to the loss. Again, we evaluate the whole evaluation dataset, fiducial and grid, every 10'000 steps.

\subsection{Parameter Inference}
\label{sec:inference}

We perform the parameter inference in the same way as~\cite{methodpaper}, by using Gaussian Process Approximate Bayesian Computation (GPABC). We estimate the ABC log-posterior of a grid point $\theta$ by using the summaries of the $n_x = 280$ mock surveys via
\begin{equation}
    \hat{\mathcal{L}}(\theta) = \log\left(\frac{p(\theta)}{n_x}\sum_{i=1}^{n_x}K_h\left(\left\Vert \hat{s}_i - \hat{s}_\mathrm{obs} \right\Vert_M\right)\right),
\end{equation}
where $p(\theta)$ is the (flat) prior, $K_h$ is a kernel with scale $h$ and $\Vert \cdot \Vert_M$ represents the Mahalanobis distance using the fiducial covariance matrix
\begin{equation}
    \Vert y \Vert^2_M = y^T\mathrm{Cov}_\theta(\hat{s})^{-1}y.
\end{equation}
Additionally, the variance of this estimate can be obtained via
\begin{equation}
    \sigma^2(\theta) = \frac{p(\theta)^2}{n_x}\frac{\mathrm{Var}\left[K_h\left(\left\Vert \hat{s}_i - \hat{s}_\mathrm{obs} \right\Vert_M\right)\right]}{\hat{p}_\mathrm{ABC}(\theta\vert\hat{s}_\mathrm{obs})},
\end{equation}
where the ABC posterior is defined via
\begin{equation}
    \hat{\mathcal{L}}(\theta) = \log\left( \hat{p}_\mathrm{ABC}(\theta\vert\hat{s}_\mathrm{obs}) \right).
\end{equation}
In this work, we exclusively use the sigmoid kernel
\begin{equation}
    K_h(x) = \frac{2}{\pi h}\frac{1}{\exp\left(\frac{x}{h}\right) + \exp\left(-\frac{x}{h}\right)}.
\end{equation}
The scale $h$ is a hyperparameter and generally depends on the observation. These log-posterior estimates along with their uncertainty can then be interpolated using Gaussian process (GP) regression. The GP can then be evaluated to obtain confidence intervals for the model parameters with standard techniques like Markov Chain Monte Carlo (MCMC). In this work, we use the \emcee~algorithm described in~\cite{emcee} with 1'024 walkers each sampling 5'000 samples. With these settings, sampling the emulated posterior takes roughly three minutes on a single GPU on Piz Daint.

\begin{figure*}[t!]
    \centering
    \includegraphics[width=0.49\textwidth]{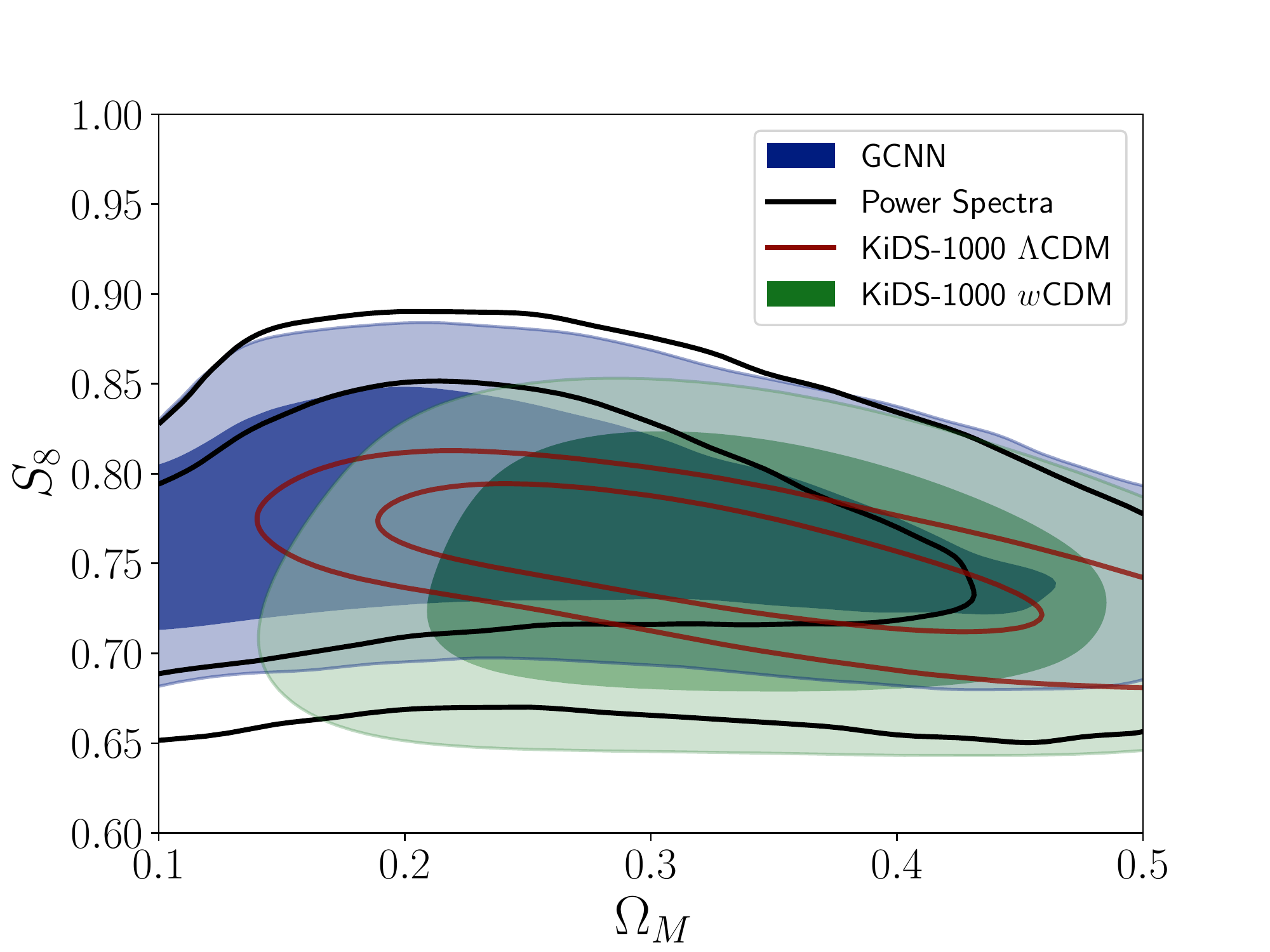}
    \includegraphics[width=0.49\textwidth]{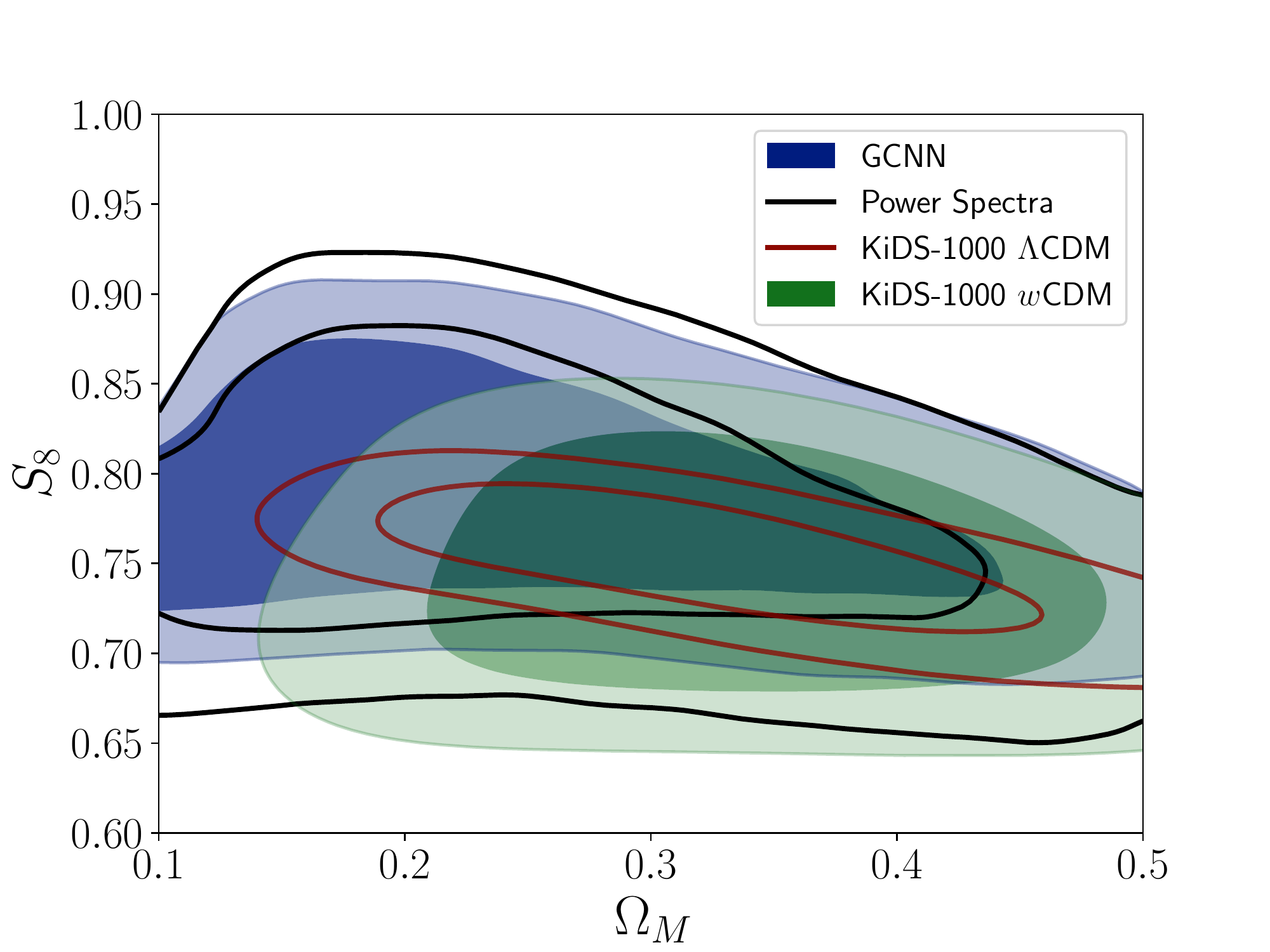}
    \caption{The fiducial results of our analysis compared to the band power constraints of~\citetalias{kids_net} and the KiDS-1000 $w$CDM analysis~\cite{kids_wcdm}. The left plot shows the results without the treatment of baryons, while the right plot includes baryonic feedback. \label{fig:res_OM_S8}}
\end{figure*}

\subsubsection{Power Spectra Analysis}
\label{sec:powerspec_ana_method}

To check the consistency of our network analysis, we first perform a power spectrum analysis using a standard compression scheme. The compression is necessary as the ABC log-posterior estimates generally become less noisy if the summary vector has fewer dimensions, which is known as the curse of dimensionality. The dimensional reduction is performed using the standard \moped~compression~\cite{moped}. The \moped~algorithm performs a linear transformation that aims to preserve as much information about the relevant model parameter as possible. Given any summary $\hat{s}$, the $i$-th entry of the compressed vectors is obtained via
\begin{equation}
    \hat{y}_i = b^T_i\hat{s},
\end{equation}
where the linear transformation is defined via 
\begin{equation}
    b_i = \frac{\mathrm{Cov}_\theta(\hat{s})^{-1}\frac{\partial \Psi_\theta(\hat{s})}{\partial \theta_i} - \sum_{q=1}^{i-1}\left(\frac{\partial \Psi_\theta(\hat{s})}{\partial \theta_i}^Tb_q\right)b_q}{\sqrt{\frac{\partial \Psi_\theta(\hat{s})}{\partial \theta_i}\mathrm{Cov}_\theta(\hat{s})^{-1}\frac{\partial \Psi_\theta(\hat{s})}{\partial \theta_i} - \sum_{q=1}^{i-1}\left(\frac{\partial \Psi_\theta(\hat{s})}{\partial \theta_i}^Tb_q\right)^2}}. \label{eq:moped}
\end{equation}
We compress the auto- and cross-correlation of all mock surveys using the compression matrix generated with the auto- and cross-spectra of the fiducial simulations and their perturbations. The output vector $\hat{y}$ contains one entry for each model parameter, which is seven for mock surveys without baryonic corrections and nine if baryonic corrections are included. The compressed vectors can then be used for parameter inference with GPABC. 

\subsubsection{Network Analysis}
\label{sec:network_ana_method}

Similar to~\citetalias{kids_net}, we decided to combine multiple networks to improve the results. We therefore use all predictions from all fiducial GCNN and summary networks. All networks were evaluated every 10'000 training steps, meaning that the output of a single network for a single mock observation is 40 dimensional (four outputs $\times$ ten evaluations). All six networks combined would therefore have an output dimension of 240. This high dimensionality would deteriorate the performance of the GPABC significantly. In~\cite{methodpaper} it was therefore proposed to perform as a second compression using a fully connected network and the previously introduced information maximizing loss (see equation~\eqref{eq:loss_1}). However, at this point, we only have access to 16'000 evaluations of the fiducial mock surveys to optimize the weights of the networks. This low number might lead to severe overfitting in the optimization. Therefore, we decided to use the standard \moped~compression instead. To perform a \moped~compression, one has to calculate the inverse of the covariance matrix of the fiducial predictions (see equation~\eqref{eq:moped}). This inversion can be numerically unstable if the covariance matrix is not well conditioned, e.g. because the vectors are very correlated. To avoid this instability, we performed a \moped~compression for all predictions of the GCNN and summary networks individually. This reduced the dimensionality of the outputs from the three GCNN and summary networks from 120 to the four, one for each parameter that we trained the networks on (see sections~\ref{sec:training} and~\ref{sec:training_sumnet}). Afterward, we took the mean of the two compressed vectors as the final summary. Theoretically, it would be better to perform a single \moped~compression on the full 240 dimensional output, as it would also consider the correlations between the predictions of the GCNNs and summary networks. However, we found that the numerical instability caused by the ill-conditioned matrix actually degrades the performance. This could be solved by increasing the number of fiducial mock surveys, which we will leave to future work.

All results of our network analysis presented in the main paper use this combination of six networks. We present some constraints with different settings, including constraints from our benchmark networks, in appendix~\ref{ap:more_nets}.

\begin{figure*}
    \centering
    \includegraphics[width=0.49\textwidth]{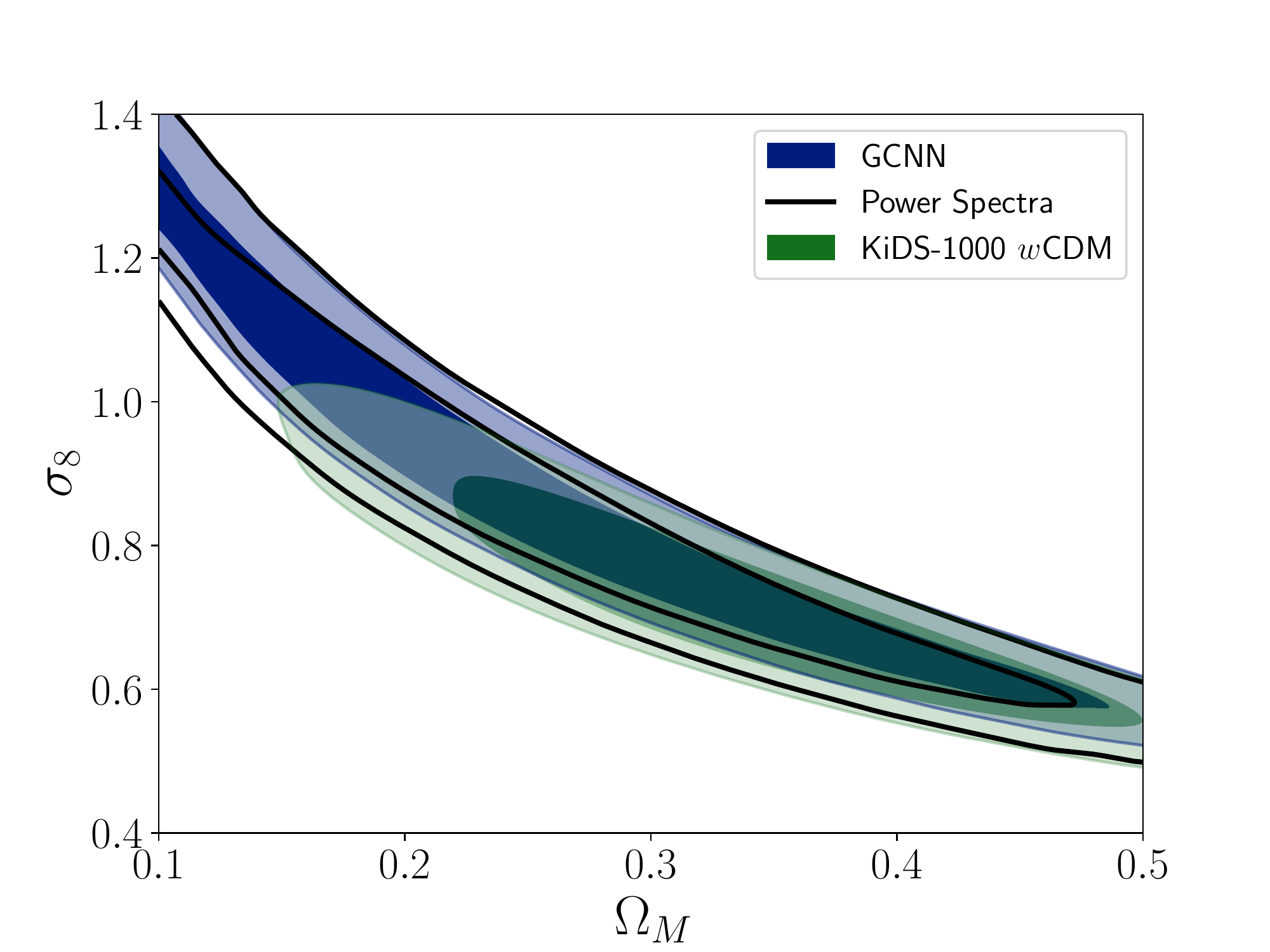}
    \includegraphics[width=0.49\textwidth]{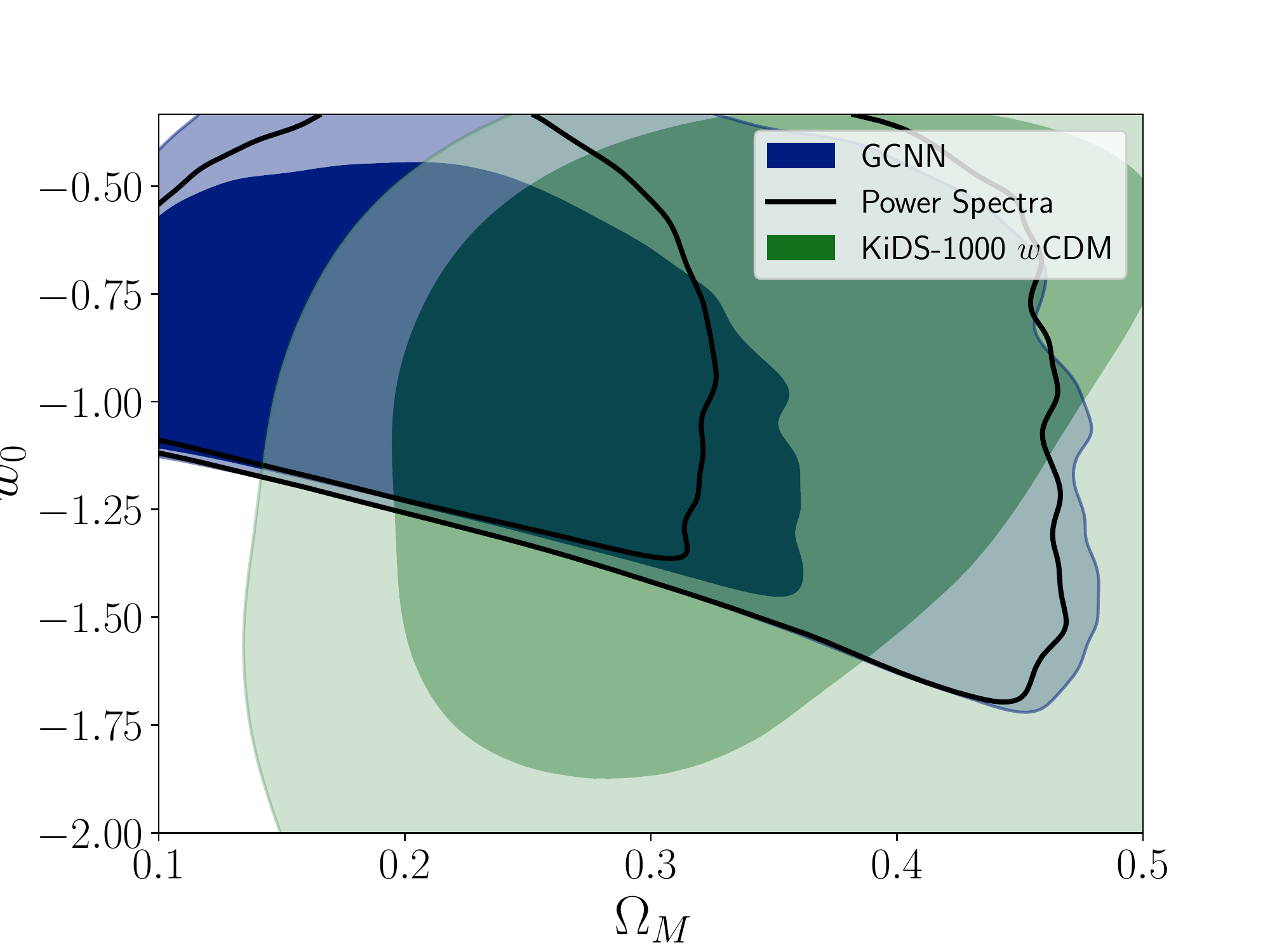}
    \caption{The fiducial constraints of our analysis including baryon feedback compared to the KiDS-1000 $w$CDM analysis~\cite{kids_wcdm}. The left plot shows the degeneracy of the $\Omega_M - \sigma_8$ plane, while the right plot shows the constraints of the dark energy equation of state parameter $w_0$. The lower bound of $w_0$ of our constraints purely dominated by our prior (see appendix~\ref{ap:priors}). \label{fig:res_more}}
\end{figure*}

\section{Tests on Mock Observations}
\label{sec:mock_tests}

In this section we present cosmological constraints that are generated with mock observations from the \cosmogrid~benchmark simulations. These tests are performed for three reasons. First, to check if our power spectrum analysis is consistent with the network analysis. Second, to see if our pipeline can correctly recover the input parameters of the mock observations. And third, to examine the effects of the different simulation settings (including baryonic feedback effects) on the constraints. The inference is done as explained in section~\ref{sec:inference} with the scale parameter set to $h = 0.4$. As mock observations, we choose the mean predictions of all 400 mock surveys from the \cosmogrid~benchmark simulations. We choose the mean predictions to decrease the spread of the cosmic variance. We present the 68\% confidence regions and mean predictions of the degeneracy parameter
\begin{equation}
    S_8 \equiv \sigma_8\sqrt{\frac{\Omega_M}{0.3}}
\end{equation}
in Figure~\ref{fig:s8_mocks}.
It can be seen that all constraints are consistent with each other and also with the input parameter. The impact of the different simulation settings is small and similar to the findings of~\citetalias{kids_net}. However, the impact of the baryonic corrections is significantly smaller. This can be seen in the last row of the plot, where we use the fiducial mock observation including Baryonic feedback in the analysis that ignores Baryons or ignore Baryonic feedback in the analysis that uses the baryonic correction model respectively. The smaller impact is mainly caused by the chosen scale cuts. The resolution of the maps in~\citetalias{kids_net} is much higher, leading to more non-Gaussian information, but also increasing the impact of baryonic physics. The scale cuts in this work ($\ell_\mathrm{max} \sim 1500$) are similar to the ones in~\citetalias{kids1000_shear} and the measured shifts caused by either including the baryonic corrections in the mock observation, but not in the data or vice versa are consistent with the findings in~\citetalias{kids1000_shear}. This shows that the implemented baryonification model is consistent with the \hmcode~implementation used in~\citetalias{kids1000_shear}, at least at the level of the cosmological constraints. Generally, including baryonic feedback into the analysis increases the confidence intervals by $\sim10 \%$, which is expected.

Another important difference to the results of~\citetalias{kids_net} is that the constraints of the networks analysis are on average only $\sim10\%$ tighter than the constraints of the power spectrum analysis, as opposed to the $\sim30\%$ improvement reported in~\citetalias{kids_net}. There are multiple reasons for this difference. The most important two being the different scale cuts and the choice of the scale parameter $h$ in the GPABC. The high resolution maps used in~\citetalias{kids_net} include much more non-Gaussian information, increasing the performance of the networks. We expected the gain of the networks to drop, considering the low resolution maps used in this analysis. Further, all constraints generated with GPABC require the choice of the scale parameter $h$. This parameter acts like a smoothing scale on the constraints and makes it possible to obtain constraints even if the simulated grid is very coarse. Increasing $h$ leads to broader constraints, but the expected value remains largely unchanged. Theoretically, one should optimize $h$ for all constraints individually. However, since here, we are only interested in the consistency of the constraints as opposed to the actual size, we chose a fairly conservative $h = 0.4$ for all constraints. We decided to only optimize $h$ for the final constraints after the unblinding (see sections~\ref{sec:blinding} and~\ref{sec:results}).

All constraints presented in this section are generated using the fiducial setting described in section~\ref{sec:inference}. Further tests can be found in appendix~\ref{ap:more_nets}.

\section{Results}
\label{sec:results}

As mentioned in section~\ref{sec:blinding}, we unblinded our results after making sure that our pipeline successfully recovers the parameters of the mock observations, that the results of the network and power spectra analysis are consistent, and that the pipeline is robust against simulation settings. Our fiducial constraints are presented in Figure~\ref{fig:res_OM_S8}. The results were generated in the same way as the constraints in section~\ref{sec:mock_tests} and generally consistent with the results of~\citetalias{kids1000_shear} and the KiDS-1000 $w$CDM analysis~\cite{kids_wcdm}. The results including the baryon corrections prefer a slightly higher value of the degeneracy parameter $S_8$. This is consistent with our checks from section~\ref{sec:mock_tests}. A comparison of $S_8$ constraints of various settings and external analyses is shown in Figure~\ref{fig:res_S8_comp}. We show the same constraints in the $\Omega_M - \sigma_8$ and $\Omega_M - w_0$ planes in Figure~\ref{fig:res_more}. Additionally, we present the full parameter constraints in appendix~\ref{ap:full_cons}. We constrain the degeneracy parameter to $S_8 = 0.76^{+0.05}_{-0.05}$ with our power spectrum analysis ignoring including baryon feedback and to $S_8 = 0.78^{+0.06}_{-0.06}$ if we include our baryon treatment. Including the baryon correction increases the constraints by 14\%, which is roughly consistent with the findings of~\citetalias{kids1000_shear} and our tests on mock observations. Our fiducial analysis with the GCNN constrains the degeneracy parameter to $S_8 = 0.77^{+0.04}_{-0.05}$ when ignoring baryon corrections and to $S_8 = 0.79^{+0.05}_{-0.05}$ when including baryon feedback. This is an 14\% and 16\% improvement respectively when compared to our power spectrum analysis, which is also consistent with our tests on mock observations. 
\begin{figure}[b]
    \centering
    \includegraphics[width=0.49\textwidth]{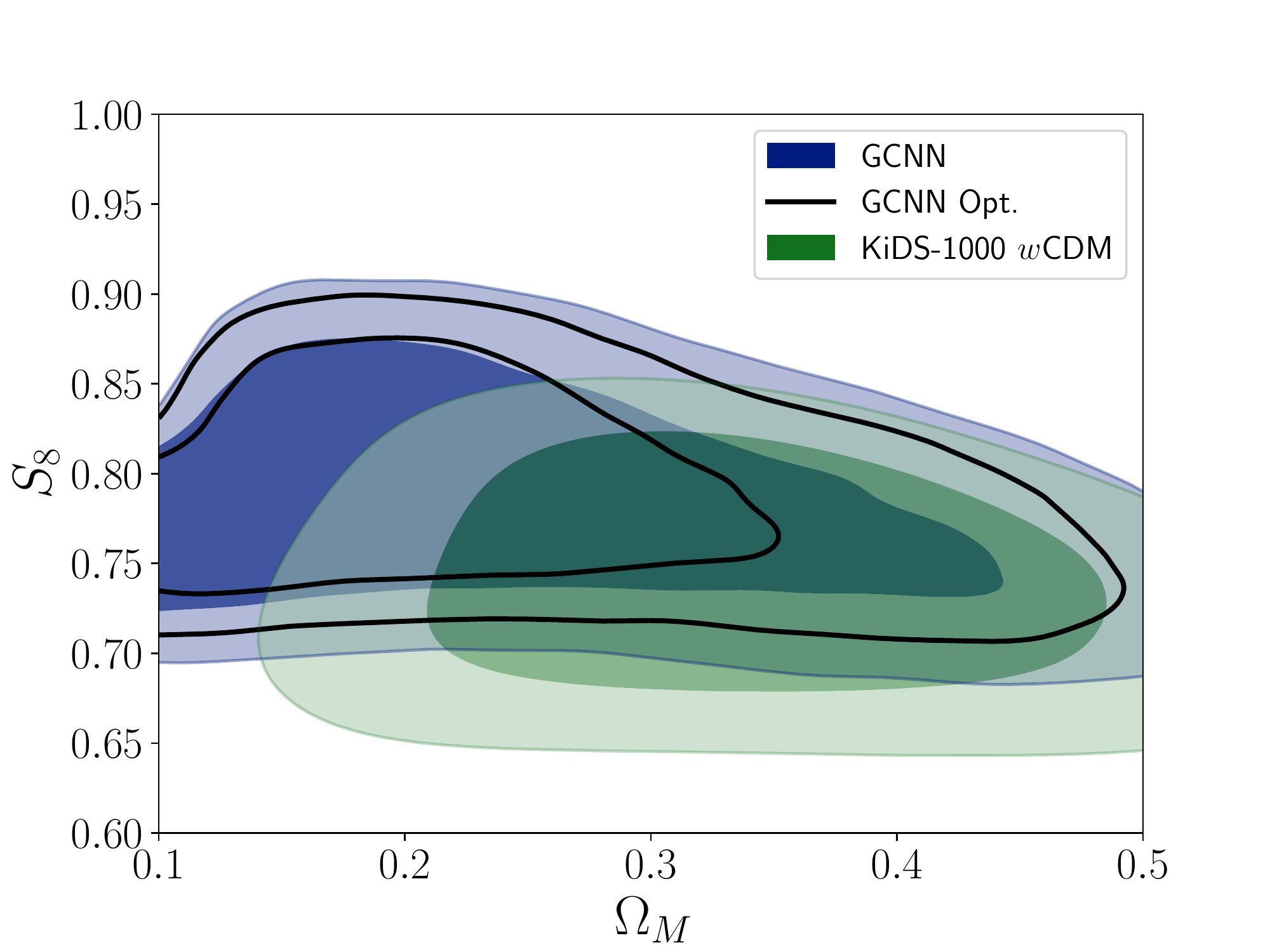}
    \caption{The fiducial GCNN constraints of this work compared to the optimized version that was obtained by reducing the scale parameter of the GPABC. \label{fig:res_opt}}
\end{figure}
\begin{figure*}[t!]
    \centering
    \includegraphics[width=1.0\textwidth]{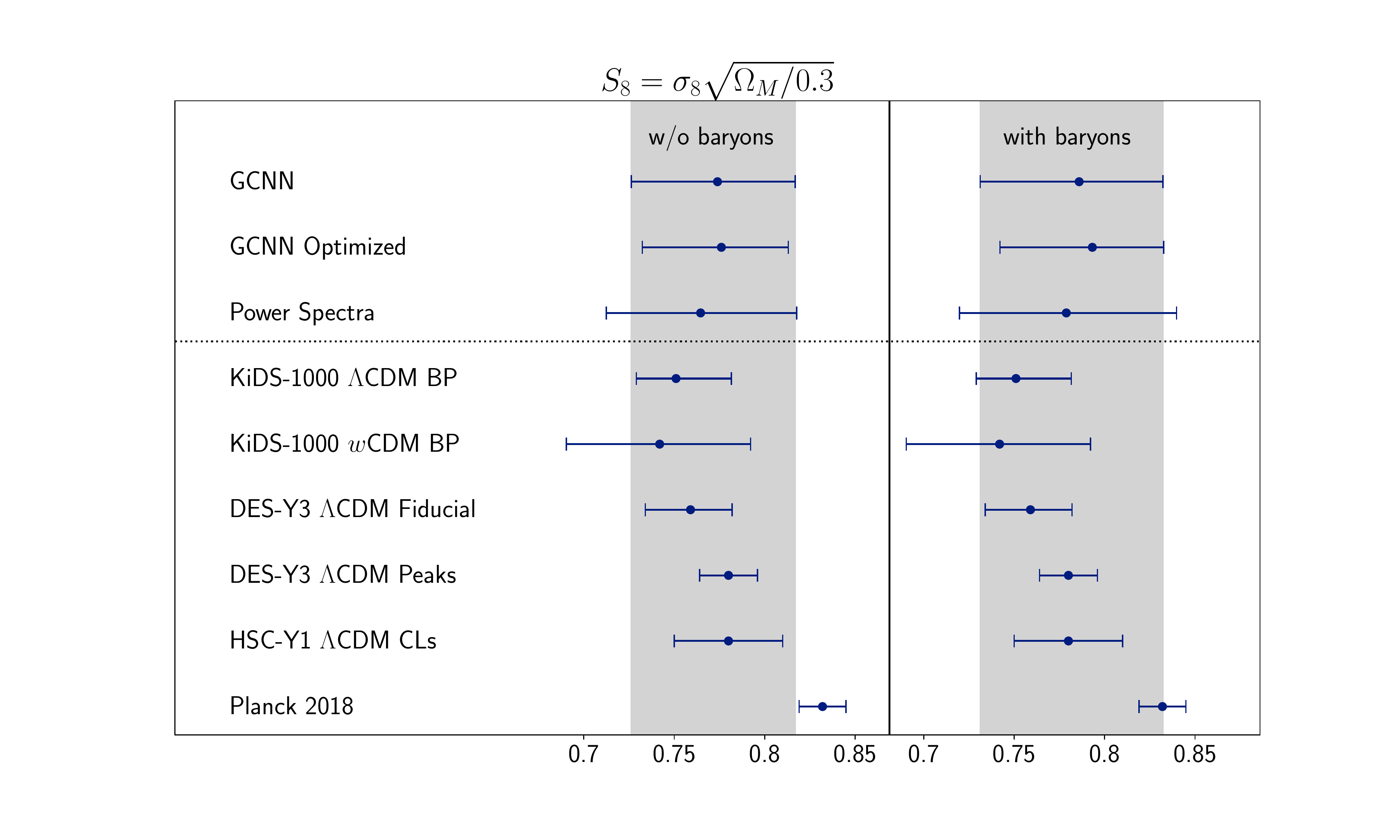}
    \caption{A comparison of our constraints of the degeneracy parameter $S_8$ with different external analyses. The first three results are from this work, where the left column shows our results without the treatment of baryon correction and the results in the right column include baryon corrections. We compare our results to the band power constraints of~\citetalias{kids1000_shear}, the KiDS-1000 $w$CDM analysis~\cite{kids_wcdm}, the fiducial DES-Y3 results~\cite{desy3_fidu}, the DES-Y3 peaks analysis~\cite{desy3_peaks}, the HSC-Y1 results~\cite{hsc_y1} and the Planck 2018 (TT,TE,EE + lowE + lensing)~\cite{planck_18} constraints. \label{fig:res_S8_comp}}
\end{figure*}
Both, our power spectrum and GCNN analysis, show the typical degeneracy in the $\Omega_M-\sigma_8$ plane, which is shown in Figure~\ref{fig:res_more}. Our constraints on the dark energy equation of state parameter are mostly dominated by our prior and consistent with the $\Lambda$CDM model. However, our fiducial GCNN analysis is able to put an upper bound on $w_0 = -0.93^{+0.32}_{-0.29}$ at the 68\% confidence level, while the lower bound is determined by our prior distribution. Finally, we are also able to constrain the intrinsic alignment amplitude. The fiducial results of our power spectrum analysis constrain the intrinsic alignment amplitude to $A_\mathrm{IA} = 0.77^{+0.86}_{-0.70}$ without baryon corrections and to $A_\mathrm{IA} = 0.73^{+0.88}_{-0.70}$ when including baryons. The GCNN analysis yields the constraints $A_\mathrm{IA} = 0.51^{+0.67}_{-0.66}$ and $A_\mathrm{IA} = 0.46^{+0.67}_{-0.67}$, respectively, leading to an improvement of $\sim15$\%. These constraints are consistent with the GCNN analysis having the tendency to prefer slightly lower intrinsic alignment amplitudes as the power spectrum analysis.

In Figure~\ref{fig:res_opt}, we show the constraints of our GCNN analysis, where we decreased the smoothing scale of the GPABC to $h = 0.2$, which we dub ``GCNN optimized". As mentioned in section~\ref{sec:mock_tests}, the smoothing scale of the GPABC is a hyperparameter that can be optimized for each observation. For our fiducial constraints, we chose the same, conservative scale $h = 0.4$ as for our tests on mock observations. Decreasing this parameter can lead to tighter constraints, however, choosing a value that is too small will lead to a noisy posterior similar to choosing a threshold that is to small in a standard rejection ABC analysis. A demonstration of the impact of the smoothing parameter is given in the appendix of~\cite{methodpaper}. Our results are generally robust to small changes to this parameter, indicating that the parameter choice did not lead to a noisy posterior. Decreasing it further only leads to marginal improvements of the constraints. Using this new scale, the constraints of the degeneracy parameter improve to $S_8 = 0.78^{+0.04}_{-0.04}$ without the treatment of baryons and to $S_8 = 0.79^{+0.04}_{-0.05}$ if baryon corrections are included. This corresponds to a $\sim11$\% improvement when compared to the fiducial GCNN results. 

Our results are also consistent with the results of~\citetalias{kids1000_shear} and the KiDS-1000 $w$CDM analysis~\cite{kids_wcdm}. Our analysis results in slightly higher values for the degeneracy parameter $S_8$, but the difference is less than one standard deviation. The $S_8$ constraints of our fiducial power spectrum analysis including baryon corrections are 18\% larger than the results of the KiDS-1000 $w$CDM analysis~\cite{kids_wcdm}. Looking at Figures~\ref{fig:res_S8_comp} and~\ref{fig:res_more}, one can see that the difference is mostly coming from the low $\Omega_M$ regime that the KiDS-1000 $w$CDM analysis is able to exclude as compared to our analysis. These differences are most likely caused by the different modeling choices. The KiDS-1000 $w$CDM analysis and our power spectrum analysis are fairly similar, however, there are some important differences. The most important being the different weighing schemes of the $\ell$-modes, the baryon treatment, and the different prior choices. While the scale cuts are comparable, it is worth noting that our map-based approach with a resolution of $n_\mathrm{side} = 512$ significantly down-weights the high $\ell$-modes because of the pixel window function. Furthermore, as we described in section~\ref{sec:barys}, the baryon treatment is based on two entirely different methods. And lastly, the KiDS-1000 $w$CDM analysis has a flat prior in $\omega_\mathrm{cdm} \equiv \Omega_\mathrm{cdm}h^2$ and $\omega_b \equiv \Omega_bh^2$ which does not translate into a flat prior in $\Omega_M$ and $\Omega_b$ that our analysis is using. These modeling differences can potentially explain the differences in the results. 

Our fiducial $S_8$ constraints including baryon correction of the GCNN analysis have the same size as the constraints of the KiDS-1000 $w$CDM analysis~\cite{kids_wcdm}, but are $\sim1\sigma$ higher, while our optimized results are $~11$\% smaller. Due to the preference of higher $S_8$-values in our analysis, we find a lower tension with the results of Planck~\cite{planck_18}. Our constraints of the intrinsic alignment amplitude are generally consistent with the results from~\citetalias{kids1000_shear} and KiDS-1000 $w$CDM analysis~\cite{kids_wcdm}, with a tendency towards slightly smaller values.

\section{Conclusion}
\label{sec:conclusion}

In this work, we presented a fully forward-modeled $w$CDM analysis of the KiDS-1000 data that is consistent with previous results and shows the potential of neural networks in cosmological parameter inference. We used the simulations of the \cosmogrid~to generate almost one million mock surveys of the KiDS-1000 data, including the treatment of systematic effects like the photometric redshift bias or the multiplicative and additive shear biases. Additionally, we treat the effect of intrinsic galaxy alignment and baryon corrections on map level to incorporate them into our analysis. We then performed a standard power spectrum analysis and we trained GCNNs using \deepsphere~with an information maximizing loss. Finally, we applied the method described in~\cite{methodpaper} to perform a likelihood-free cosmological parameter inference. 

Out results of the power spectrum and network analysis are generally consistent with each other and also with the results from~\citetalias{kids1000_shear} and the KiDS-1000 $w$CDM analysis~\cite{kids_wcdm}. We find a preference for slightly higher $S_8$ values which can potentially be explained by the different modeling choices. Similar to the results of~\citetalias{kids_net} we find that the networks generally perform better than the power spectrum analysis, leading to $\sim$15\% smaller constraints in $S_8$ and $A_\mathrm{IA}$, with our GCNN analysis giving similar constraints as the KiDS-1000 $w$CDM analysis~\cite{kids_wcdm}. Optimizing the hyperparameter of the GPABC leads to a further decrease of the $S_8$ constraints by $\sim$11\%.

Including baryon corrections into our analysis generally broadens the constraints by $\sim 10$\%, affecting the network and power spectrum analysis to a similar degree. Additionally, we find that including baryons into the analysis leads to slightly higher values of $S_8$, which agrees with the results of~\citetalias{kids1000_shear}, showing that the implemented baryon model is consistent with other treatments at the level of the cosmological parameter constraints. 

Potential extensions of this work include a more advanced intrinsic alignment model, less conservative scale cuts and the analysis of other surveys such as the dark energy survey (DES). We believe that using a finer resolution of the mock surveys will, most likely, greatly improve the performance of the networks, similar to gains presented in~\citetalias{kids_net}. A finer resolution of the mock surveys would require additional tests of the simulations and the baryon model. At the same time it would also drastically increase the storage requirements along with the computational resources necessary to analyse the data. For example, increasing the resolution to a \healpix~$n_\mathrm{side} = 1024$ would increase the storage necessary for the mock surveys by a factor of four and similarly affect the training times of the networks. However, larger data sets might also be able to reduce the prior ranges of the cosmological parameters, reducing the necessary number of simulations of the \cosmogrid. Overall this offers great prospects for machine learning inference for current and weak lensing surveys.

\begin{acknowledgments}

JF would like to thank Jeppe Mosgaard Dakin for helpful discussions and especially his contribution to the $w_0-\Omega_M$ prior of the simulation grid. We thank Joachim Stadel and Douglas Potter for helpful discussions.

This work was supported by a grant from the Swiss National Supercomputing Centre (CSCS) under project ID s998, entitled ``Measuring dark energy with deep learning''.

We would like to thank the technical support teams of the Euler and Piz Daint computing clusters.

AS acknowledges support from the Swiss National Science Foundation via the grant PCEFP2\_181157.

This work is based on observations made with ESO Telescopes at the La Silla Paranal Observatory under programme IDs 177.A-3016, 177.A-3017, 177.A-3018 and 179.A-2004, and on data products produced by the KiDS consortium. The KiDS production team acknowledges support from: Deutsche Forschungsgemeinschaft, ERC, NOVA and NWO-M grants; Target; the University of Padova, and the University Federico II (Naples).

\end{acknowledgments}


\bibliographystyle{ieeetr_short}
\bibliography{library}

\begin{thebibliography}{10}

\bibitem{Schneider2005}
P.~{Schneider}, ``{Weak Gravitational Lensing},'' {\em arXiv e-prints},
  pp.~astro--ph/0509252, Sep 2005.

\bibitem{Kilbinger2015review}
M.~{Kilbinger}, ``{Cosmology with cosmic shear observations: a review},'' {\em
  Reports on Progress in Physics}, vol.~78, p.~086901, Jul 2015.

\bibitem{Note1}
\protect \url {cfhtlens.org}.

\bibitem{CFHTLenS2013}
E.~Grocutt, F.~Simpson, C.~Heymans, {\em et~al.}, ``{CFHTLenS tomographic weak
  lensing cosmological parameter constraints: Mitigating the impact of
  intrinsic galaxy alignments},'' {\em Monthly Notices of the Royal
  Astronomical Society}, vol.~432, pp.~2433--2453, 05 2013.

\bibitem{Note2}
\protect \url {kids.strw.leidenuniv.nl}.

\bibitem{Hildebrandt2018viking}
H.~{Hildebrandt}, F.~{K{\"o}hlinger}, J.~L. {van den Busch}, {\em et~al.},
  ``{KiDS+VIKING-450: Cosmic shear tomography with optical+infrared data},''
  {\em arXiv e-prints}, p.~arXiv:1812.06076, Dec 2018.

\bibitem{kids1000_shear}
{Asgari, Marika}, {Lin, Chieh-An}, {Joachimi, Benjamin}, {\em et~al.},
  ``Kids-1000 cosmology: Cosmic shear constraints and comparison between two
  point statistics,'' {\em A\&A}, vol.~645, p.~A104, 2021.

\bibitem{Note3}
\protect \url {darkenergysurvey.org}.

\bibitem{desy3_fidu}
L.~F. {Secco}, S.~{Samuroff}, E.~{Krause}, {\em et~al.}, ``{Dark Energy Survey
  Year 3 Results: Cosmology from Cosmic Shear and Robustness to Modeling
  Uncertainty},'' {\em arXiv e-prints}, p.~arXiv:2105.13544, May 2021.

\bibitem{desy3_peaks}
D.~{Z{\"u}rcher}, J.~{Fluri}, R.~{Sgier}, {\em et~al.}, ``{Dark Energy Survey
  Year 3 results: Cosmology with peaks using an emulator approach},'' {\em
  arXiv e-prints}, p.~arXiv:2110.10135, Oct. 2021.

\bibitem{Note4}
\protect \url {hsc.mtk.nao.ac.jp/ssp/survey}.

\bibitem{hsc_y1}
C.~Hikage, M.~Oguri, T.~Hamana, {\em et~al.}, ``{Cosmology from cosmic shear
  power spectra with Subaru Hyper Suprime-Cam first-year data},'' {\em
  Publications of the Astronomical Society of Japan}, vol.~71, 03 2019.
\newblock 43.

\bibitem{Euclid2011}
R.~{Laureijs}, J.~{Amiaux}, S.~{Arduini}, {\em et~al.}, ``{Euclid Definition
  Study Report},'' {\em arXiv e-prints}, p.~arXiv:1110.3193, Oct 2011.

\bibitem{Chang2013}
C.~{Chang}, M.~{Jarvis}, B.~{Jain}, {\em et~al.}, ``{The effective number
  density of galaxies for weak lensing measurements in the LSST project},''
  {\em Monthly Notices of the Royal Astronomical Society}, vol.~434,
  pp.~2121--2135, Sep 2013.

\bibitem{wfirst}
D.~{Spergel}, N.~{Gehrels}, C.~{Baltay}, {\em et~al.}, ``{Wide-Field InfrarRed
  Survey Telescope-Astrophysics Focused Telescope Assets WFIRST-AFTA 2015
  Report},'' {\em arXiv e-prints}, p.~arXiv:1503.03757, Mar. 2015.

\bibitem{jialiu2015}
J.~Liu, A.~Petri, Z.~Haiman, {\em et~al.}, ``Cosmology constraints from the
  weak lensing peak counts and the power spectrum in cfhtlens data,'' {\em
  Phys. Rev. D}, vol.~91, p.~063507, Mar 2015.

\bibitem{Dietrich2010peaks}
J.~P. {Dietrich} and J.~{Hartlap}, ``{Cosmology with the shear-peak
  statistics},'' {\em Monthly Notices of the Royal Astronomical Society},
  vol.~402, pp.~1049--1058, Feb 2010.

\bibitem{Kacprzakpeaks}
T.~{Kacprzak}, D.~{Kirk}, O.~{Friedrich}, {\em et~al.}, ``{Cosmology
  constraints from shear peak statistics in Dark Energy Survey Science
  Verification data},'' {\em Monthly Notices of the Royal Astronomical
  Society}, vol.~463, pp.~3653--3673, Dec 2016.

\bibitem{Fluri2018peak}
J.~{Fluri}, T.~{Kacprzak}, R.~{Sgier}, {\em et~al.}, ``{Weak lensing peak
  statistics in the era of large scale cosmological surveys},'' {\em Journal of
  Cosmology and Astro-Particle Physics}, vol.~2018, p.~051, Oct 2018.

\bibitem{liu:hal-01439964}
X.~Liu, C.~Pan, R.~Li, {\em et~al.}, ``{Cosmological constraints from weak
  lensing peak statistics with Canada-France-Hawaii Telescope Stripe 82
  Survey},'' {\em {Monthly Notices of the Royal Astronomical Society}},
  vol.~450, pp.~2888--2902, July 2015.

\bibitem{KiDspeaks1}
H.~{Shan}, X.~{Liu}, H.~{Hildebrandt}, {\em et~al.}, ``{KiDS-450: cosmological
  constraints from weak lensing peak statistics - I. Inference from analytical
  prediction of high signal-to-noise ratio convergence peaks},'' {\em Monthly
  Notices of the Royal Astronomical Society}, vol.~474, pp.~1116--1134, Feb
  2018.

\bibitem{KiDspeaks2}
N.~{Martinet}, P.~{Schneider}, H.~{Hildebrandt}, {\em et~al.}, ``{KiDS-450:
  cosmological constraints from weak-lensing peak statistics - II: Inference
  from shear peaks using N-body simulations},'' {\em Monthly Notices of the
  Royal Astronomical Society}, vol.~474, pp.~712--730, Feb 2018.

\bibitem{dominik_non_gauss}
D.~Zürcher, J.~Fluri, R.~Sgier, {\em et~al.}, ``Cosmological forecast for
  non-gaussian statistics in large-scale weak lensing surveys,'' {\em Journal
  of Cosmology and Astroparticle Physics}, vol.~2021, pp.~028--028, jan 2021.

\bibitem{threepoint1}
E.~{Semboloni}, T.~{Schrabback}, L.~{van Waerbeke}, {\em et~al.}, ``{Weak
  lensing from space: first cosmological constraints from three-point shear
  statistics},'' {\em Monthly Notices of the Royal Astronomical Society},
  vol.~410, pp.~143--160, Jan 2011.

\bibitem{threepoint2}
L.~{Fu}, M.~{Kilbinger}, T.~{Erben}, {\em et~al.}, ``{CFHTLenS: cosmological
  constraints from a combination of cosmic shear two-point and three-point
  correlations},'' {\em Monthly Notices of the Royal Astronomical Society},
  vol.~441, pp.~2725--2743, Jul 2014.

\bibitem{Fluri2018}
J.~Fluri, T.~Kacprzak, A.~Refregier, {\em et~al.}, ``Cosmological constraints
  from noisy convergence maps through deep learning,'' {\em Phys. Rev. D},
  vol.~98, p.~123518, Dec 2018.

\bibitem{Ribli2019}
D.~{Ribli}, B.~{{\'A}rmin Pataki}, J.~M. {Zorrilla Matilla}, {\em et~al.},
  ``{Weak lensing cosmology with convolutional neural networks on noisy
  data},'' {\em arXiv e-prints}, p.~arXiv:1902.03663, Feb 2019.

\bibitem{methodpaper}
J.~{Fluri}, A.~{Lucchi}, T.~{Kacprzak}, {\em et~al.}, ``{Cosmological Parameter
  Estimation and Inference using Deep Summaries},'' {\em arXiv e-prints},
  p.~arXiv:2107.09002, July 2021.

\bibitem{kids_net}
J.~Fluri, T.~Kacprzak, A.~Lucchi, {\em et~al.}, ``Cosmological constraints with
  deep learning from kids-450 weak lensing maps,'' {\em Phys. Rev. D},
  vol.~100, p.~063514, Sep 2019.

\bibitem{similar_paper}
T.~L. {Makinen}, T.~{Charnock}, J.~{Alsing}, and B.~D. {Wandelt}, ``{Lossless,
  Scalable Implicit Likelihood Inference for Cosmological Fields},'' {\em arXiv
  e-prints}, p.~arXiv:2107.07405, July 2021.

\bibitem{kids_dr4}
{Kuijken, K.}, {Heymans, C.}, {Dvornik, A.}, {\em et~al.}, ``The fourth data
  release of the kilo-degree survey: ugri imaging and nine-band optical-ir
  photometry over 1000 square degrees,'' {\em A\&A}, vol.~625, p.~A2, 2019.

\bibitem{ufalcon_1}
R.~J. {Sgier}, A.~{R{\'e}fr{\'e}gier}, A.~{Amara}, and A.~{Nicola}, ``{Fast
  generation of covariance matrices for weak lensing},'' {\em Journal of
  Cosmology and Astroparticle Physics}, vol.~2019, p.~044, Jan. 2019.

\bibitem{ufalcon_2}
R.~{Sgier}, J.~{Fluri}, J.~{Herbel}, {\em et~al.}, ``{Fast lightcones for
  combined cosmological probes},'' {\em Journal of Cosmology and Astroparticle
  Physics}, vol.~2021, p.~047, Feb. 2021.

\bibitem{cosmogrid}
T.~{Kacprzak}, J.~{Fluri}, A.~{Schneider}, {\em et~al.}, ``{{\rmfamily
  \textsc{CosmoGrid}}: a fully numerical $w\rm{CDM}$ theory prediction for
  large scale structure cosmology},'' {\em To be submitted to Journal of
  Cosmology and Astroparticle Physics}, 2022.

\bibitem{nla1}
C.~M. Hirata and U.~c.~v. Seljak, ``Intrinsic alignment-lensing interference as
  a contaminant of cosmic shear,'' {\em Phys. Rev. D}, vol.~70, p.~063526, Sep
  2004.

\bibitem{nla2}
S.~Bridle and L.~King, ``Dark energy constraints from cosmic shear power
  spectra: impact of intrinsic alignments on photometric redshift
  requirements,'' {\em New Journal of Physics}, vol.~9, pp.~444--444, dec 2007.

\bibitem{nla3}
{Joachimi, B.}, {Mandelbaum, R.}, {Abdalla, F. B.}, and {Bridle, S. L.},
  ``Constraints on intrinsic alignment contamination of weak lensing surveys
  using the megaz-lrg sample,'' {\em A\&A}, vol.~527, p.~A26, 2011.

\bibitem{aurel_bary1}
A.~{Schneider} and R.~{Teyssier}, ``{A new method to quantify the effects of
  baryons on the matter power spectrum},'' {\em Journal of Cosmology and
  Astroparticle Physics}, vol.~2015, p.~049, Dec. 2015.

\bibitem{aurel_bray2}
A.~{Schneider}, R.~{Teyssier}, J.~{Stadel}, {\em et~al.}, ``{Quantifying baryon
  effects on the matter power spectrum and the weak lensing shear
  correlation},'' {\em Journal of Cosmology and Astroparticle Physics},
  vol.~2019, p.~020, Mar. 2019.

\bibitem{aurel_bary3}
A.~{Schneider}, S.~K. {Giri}, S.~{Amodeo}, and A.~{Refregier}, ``{Constraining
  baryonic feedback and cosmology with weak-lensing, X-ray, and kinematic
  Sunyaev-Zeldovich observations},'' {\em arXiv e-prints}, p.~arXiv:2110.02228,
  Oct. 2021.

\bibitem{deepsphere}
N.~{Perraudin}, M.~{Defferrard}, T.~{Kacprzak}, and R.~{Sgier}, ``{DeepSphere:
  Efficient spherical convolutional neural network with HEALPix sampling for
  cosmological applications},'' {\em Astronomy and Computing}, vol.~27, p.~130,
  Apr. 2019.

\bibitem{Charnock2018}
T.~Charnock, G.~Lavaux, and B.~D. Wandelt, ``Automatic physical inference with
  information maximizing neural networks,'' {\em Phys. Rev. D}, vol.~97,
  p.~083004, Apr 2018.

\bibitem{Note5}
\protect \url {http://kids.strw.leidenuniv.nl/}.

\bibitem{viking_2013}
A.~{Edge}, W.~{Sutherland}, K.~{Kuijken}, {\em et~al.}, ``{The VISTA
  Kilo-degree Infrared Galaxy (VIKING) Survey: Bridging the Gap between Low and
  High Redshift},'' {\em The Messenger}, vol.~154, pp.~32--34, Dec. 2013.

\bibitem{theli_2013}
T.~{Erben}, H.~{Hildebrandt}, L.~{Miller}, {\em et~al.}, ``{CFHTLenS: the
  Canada-France-Hawaii Telescope Lensing Survey - imaging data and catalogue
  products},'' {\em Monthly Notices of the Royal Astronomical Society},
  vol.~433, pp.~2545--2563, Aug. 2013.

\bibitem{astro_wise_2013}
K.~{Begeman}, A.~N. {Belikov}, D.~R. {Boxhoorn}, and E.~A. {Valentijn}, ``{The
  Astro-WISE datacentric information system},'' {\em Experimental Astronomy},
  vol.~35, pp.~1--23, Jan. 2013.

\bibitem{lensfit_1}
L.~{Miller}, C.~{Heymans}, T.~D. {Kitching}, {\em et~al.}, ``{Bayesian galaxy
  shape measurement for weak lensing surveys - III. Application to the
  Canada-France-Hawaii Telescope Lensing Survey},'' {\em Monthly Notices of the
  Royal Astronomical Society}, vol.~429, pp.~2858--2880, Mar. 2013.

\bibitem{lensfit_2}
I.~Fenech~Conti, R.~Herbonnet, H.~Hoekstra, {\em et~al.}, ``{Calibration of
  weak-lensing shear in the Kilo-Degree Survey},'' {\em Monthly Notices of the
  Royal Astronomical Society}, vol.~467, pp.~1627--1651, 01 2017.

\bibitem{bpz_code}
N.~{Ben{\'\i}tez}, ``{Bayesian Photometric Redshift Estimation},'' {\em The
  Astrophysical Journal}, vol.~536, pp.~571--583, June 2000.

\bibitem{kids1000_cata}
{Giblin, Benjamin}, {Heymans, Catherine}, {Asgari, Marika}, {\em et~al.},
  ``Kids-1000 catalogue: Weak gravitational lensing shear measurements,'' {\em
  A\&A}, vol.~645, p.~A105, 2021.

\bibitem{Gorski_2005}
K.~M. Gorski, E.~Hivon, A.~J. Banday, {\em et~al.}, ``{HEALPix}: A framework
  for high-resolution discretization and fast analysis of data distributed on
  the sphere,'' {\em The Astrophysical Journal}, vol.~622, pp.~759--771, apr
  2005.

\bibitem{Stadel2001}
J.~G. {Stadel}, {\em {Cosmological N-body simulations and their analysis}}.
\newblock PhD thesis, UNIVERSITY OF WASHINGTON, 2001.

\bibitem{neutrinos}
T.~{Tram}, J.~{Brandbyge}, J.~{Dakin}, and S.~{Hannestad}, ``{Fully
  relativistic treatment of light neutrinos in N-body simulations},'' {\em
  Journal of Cosmology and Astroparticle Physics}, vol.~2019, p.~022, Mar.
  2019.

\bibitem{class_2011}
J.~{Lesgourgues}, ``{The Cosmic Linear Anisotropy Solving System (CLASS) I:
  Overview},'' {\em arXiv e-prints}, p.~arXiv:1104.2932, Apr. 2011.

\bibitem{concept_2019}
J.~{Dakin}, J.~{Brandbyge}, S.~{Hannestad}, {\em et~al.},
  ``{{\ensuremath{\nu}}CONCEPT: cosmological neutrino simulations from the
  non-linear Boltzmann hierarchy},'' {\em Journal of Cosmology and
  Astroparticle Physics}, vol.~2019, p.~052, Feb. 2019.

\bibitem{trillion_parts}
D.~{Potter}, J.~{Stadel}, and R.~{Teyssier}, ``{PKDGRAV3: beyond trillion
  particle cosmological simulations for the next era of galaxy surveys},'' {\em
  Computational Astrophysics and Cosmology}, vol.~4, p.~2, May 2017.

\bibitem{born_is_fine}
A.~Petri, Z.~Haiman, and M.~May, ``Validity of the born approximation for
  beyond gaussian weak lensing observables,'' {\em Phys. Rev. D}, vol.~95,
  p.~123503, Jun 2017.

\bibitem{kids1000_redshift}
H.~Hildebrandt, J.~L. van~den Busch, A.~H. Wright, {\em et~al.}, ``Kids-1000
  catalogue: Redshift distributions and their calibration,'' {\em Astronomy \&
  Astrophysics}, vol.~647, p.~A124, Mar 2021.

\bibitem{Note6}
\protect \url {https://github.com/LSSTDESC/CCL}.

\bibitem{spherical_KS}
C.~G.~R. {Wallis}, M.~A. {Price}, J.~D. {McEwen}, {\em et~al.}, ``{Mapping dark
  matter on the celestial sphere with weak gravitational lensing},'' {\em arXiv
  e-prints}, p.~arXiv:1703.09233, Mar. 2017.

\bibitem{som1}
{Wright, Angus H.}, {Hildebrandt, Hendrik}, {van den Busch, Jan Luca}, and
  {Heymans, Catherine}, ``Photometric redshift calibration with self-organising
  maps,'' {\em A\&A}, vol.~637, p.~A100, 2020.

\bibitem{som2}
A.~H. {Wright}, H.~{Hildebrandt}, J.~L. {van den Busch}, {\em et~al.},
  ``{KiDS+VIKING-450: Improved cosmological parameter constraints from redshift
  calibration with self-organising maps},'' {\em A\&A}, vol.~640, p.~L14, Aug.
  2020.

\bibitem{vandaalen_2011}
M.~P. van Daalen, J.~Schaye, C.~M. Booth, and C.~D. Vecchia, ``{The effects of
  galaxy formation on the matter power spectrum: A challenge for precision
  cosmology},'' {\em Mon. Not. Roy. Astron. Soc.}, vol.~415, pp.~3649--3665,
  2011.

\bibitem{McCarthy:2016mry}
I.~G. McCarthy, J.~Schaye, S.~Bird, and A.~M.~C. Le~Brun, ``{The BAHAMAS
  project: Calibrated hydrodynamical simulations for large-scale structure
  cosmology},'' {\em Mon. Not. Roy. Astron. Soc.}, vol.~465, no.~3,
  pp.~2936--2965, 2017.

\bibitem{Chisari:2019tus}
N.~E. Chisari {\em et~al.}, ``{Modelling baryonic feedback for survey
  cosmology},'' {\em Open J. Astrophys.}, 2019.

\bibitem{hmcode}
A.~J. {Mead}, J.~A. {Peacock}, C.~{Heymans}, {\em et~al.}, ``{An accurate halo
  model for fitting non-linear cosmological power spectra and baryonic feedback
  models},'' {\em Monthly Notices of the Royal Astronomical Society}, vol.~454,
  pp.~1958--1975, Dec. 2015.

\bibitem{painting_barys}
T.~{Tr{\"o}ster}, C.~{Ferguson}, J.~{Harnois-D{\'e}raps}, and I.~G. {McCarthy},
  ``{Painting with baryons: augmenting N-body simulations with gas using deep
  generative models},'' {\em Monthly Notices of the Royal Astronomical
  Society}, vol.~487, pp.~L24--L29, July 2019.

\bibitem{arico_2020}
G.~Aric{\`o}, R.~E. Angulo, C.~Hern{\'a}ndez-Monteagudo, {\em et~al.},
  ``Modelling the large-scale mass density field of the universe as a function
  of cosmology and baryonic physics,'' {\em Monthly Notices of the Royal
  Astronomical Society}, vol.~495, no.~4, pp.~4800--4819, 2020.

\bibitem{lu_2021}
T.~{Lu} and Z.~{Haiman}, ``{The impact of baryons on cosmological inference
  from weak lensing statistics},'' {\em Monthly Notices of the Royal
  Astronomical Society}, vol.~506, pp.~3406--3417, Sept. 2021.

\bibitem{2021arXiv210911060L}
T.~{Lu}, Z.~{Haiman}, and J.~M. {Zorrilla Matilla}, ``{Simultaneously
  constraining cosmology and baryonic physics via deep learning from weak
  lensing},'' {\em arXiv e-prints}, p.~arXiv:2109.11060, Sept. 2021.

\bibitem{Giri:2021qin}
S.~K. {Giri} and A.~{Schneider}, ``{Emulation of baryonic effects on the matter
  power spectrum and constraints from galaxy cluster data},'' {\em arXiv
  e-prints}, p.~arXiv:2108.08863, Aug. 2021.

\bibitem{Sun2009}
M.~Sun, G.~M. Voit, M.~Donahue, {\em et~al.}, ``{Chandra studies of the X-ray
  gas properties of galaxy groups},'' {\em Astrophys. J.}, vol.~693,
  pp.~1142--1172, 2009.

\bibitem{Vikhlinin2009}
A.~Vikhlinin {\em et~al.}, ``{Chandra Cluster Cosmology Project II: Samples and
  X-ray Data Reduction},'' {\em Astrophys. J.}, vol.~692, pp.~1033--1059, 2009.

\bibitem{Gonzalez2013}
A.~H. Gonzalez, S.~Sivanandam, A.~I. Zabludoff, and D.~Zaritsky, ``{Galaxy
  Cluster Baryon Fractions Revisited},'' {\em Astrophys. J.}, vol.~778, p.~14,
  2013.

\bibitem{Eckert2016}
D.~Eckert {\em et~al.}, ``{The XXL Survey. XIII. Baryon content of the bright
  cluster sample},'' {\em Astron. Astrophys.}, vol.~592, p.~A12, 2016.

\bibitem{Note7}
\protect \url {https://github.com/deepsphere/deepsphere-cosmo-tf2}.

\bibitem{Note8}
\protect \url {https://www.cscs.ch/computers/piz-daint/}.

\bibitem{Adamopt}
D.~P. {Kingma} and J.~{Ba}, ``{Adam: A Method for Stochastic Optimization},''
  {\em arXiv e-prints}, p.~arXiv:1412.6980, Dec. 2014.

\bibitem{dropout}
N.~Srivastava, G.~Hinton, A.~Krizhevsky, {\em et~al.}, ``Dropout: A simple way
  to prevent neural networks from overfitting,'' {\em Journal of Machine
  Learning Research}, vol.~15, no.~56, pp.~1929--1958, 2014.

\bibitem{emcee}
D.~{Foreman-Mackey}, D.~W. {Hogg}, D.~{Lang}, and J.~{Goodman}, ``{emcee: The
  MCMC Hammer},'' {\em PASP}, vol.~125, p.~306, Mar. 2013.

\bibitem{kids_wcdm}
T.~{Tr{\"o}ster}, M.~{Asgari}, C.~{Blake}, {\em et~al.}, ``{KiDS-1000
  Cosmology: Constraints beyond flat {\ensuremath{\Lambda}}CDM},'' {\em
  {A\&A}}, vol.~649, p.~A88, May 2021.

\bibitem{moped}
A.~F. Heavens, R.~Jimenez, and O.~Lahav, ``{Massive lossless data compression
  and multiple parameter estimation from galaxy spectra},'' {\em Monthly
  Notices of the Royal Astronomical Society}, vol.~317, pp.~965--972, 10 2000.

\bibitem{planck_18}
{Planck Collaboration}, N.~{Aghanim}, Y.~{Akrami}, {\em et~al.}, ``{Planck 2018
  results. VI. Cosmological parameters},'' {\em A\&A}, vol.~641, p.~A6, Sept.
  2020.

\bibitem{amiga1}
S.~P.~D. {Gill}, A.~{Knebe}, and B.~K. {Gibson}, ``{The evolution of
  substructure - I. A new identification method},'' {\em Monthly Notices of the
  Royal Astronomical Society}, vol.~351, pp.~399--409, June 2004.

\bibitem{amiga2}
S.~R. {Knollmann} and A.~{Knebe}, ``{AHF: Amiga's Halo Finder},'' {\em The
  Astrophysical Journal}, vol.~182, pp.~608--624, June 2009.

\end{thebibliography}

\clearpage

\appendix
\section{Additional $\Omega_M-w_0$ Priors}
 \label{ap:priors}
 
\begin{figure}
    \centering
    \includegraphics[width=0.5\textwidth]{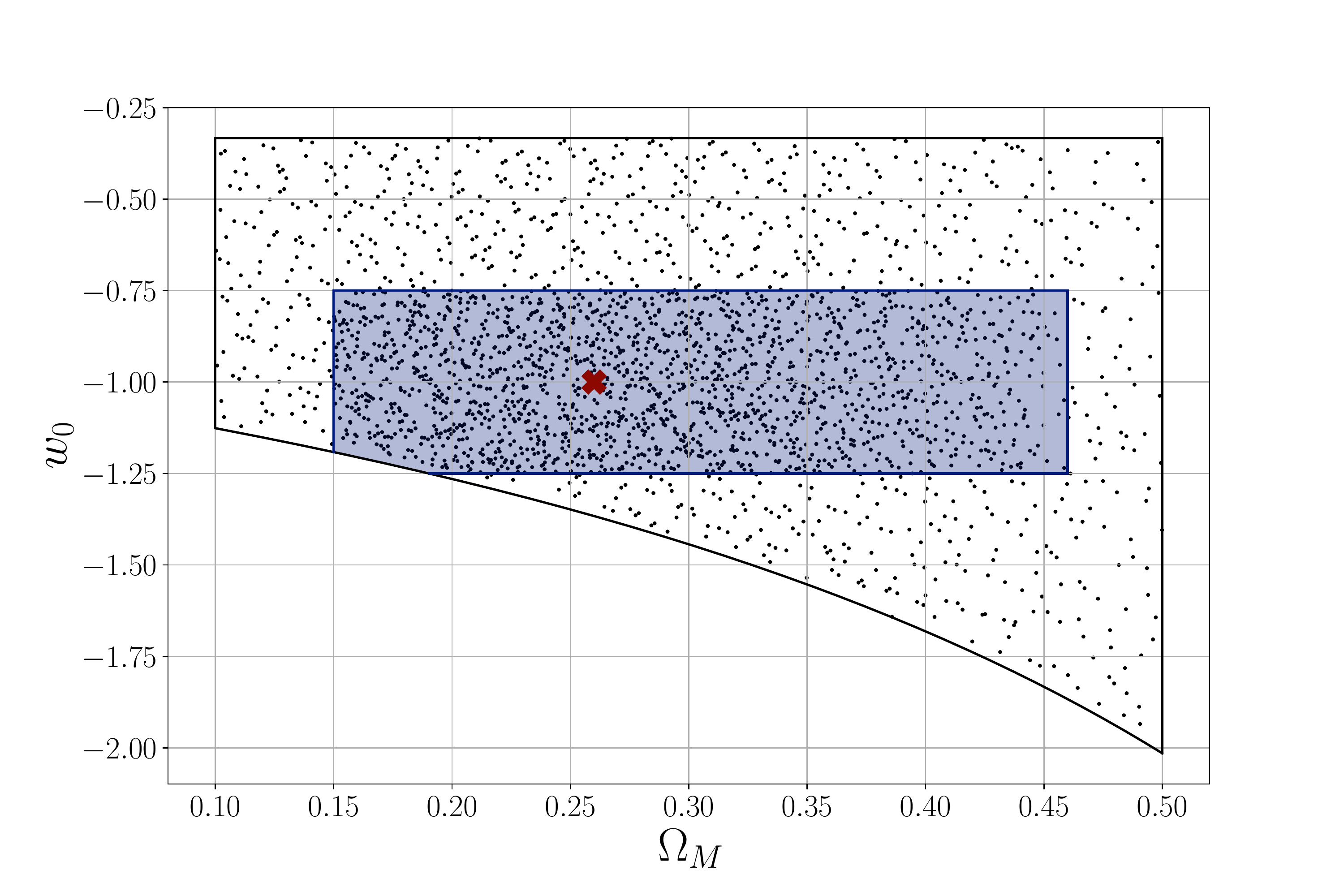}
    \caption{The 2'500 grid points projected onto the $\Omega_M-w_0$ plane. The lower border indicates the additional prior from equation~\ref{eq:phatom_crossing} and the shaded region corresponds to the tight prior mentioned in section~\ref{sec:simulations}. The red cross in the middle indicates the fiducial cosmology. \label{fig:grid_plot_w}}
\end{figure}

 The additional prior in the $\Omega_M-w_0$ plane, mentioned in section~\ref{sec:simulations}, is a consequence of the effective nature of the $w$CDM model along with the modeling of the relativistic fields according to~\cite{neutrinos}. The treatment of the relativistic fields requires a lookup table of transfer functions in the N-body gauge. The \cosmogrid~generated the transfer functions with \class~and then transformed them into the N-body gauge using \concept. This gauge transformation requires the total velocity transfer function $\vartheta_\mathrm{tot}$, which is defined as a weighted sum over the velocity transfer functions $\vartheta_i$ from all relevant relativistic species $i$. The weight of each term is given by the sum of the density and pressure of the given species $(\rho + p)_i$. However, the normalization of the sum
 \begin{equation}
    (\rho + p)_\mathrm{tot} = \sum_i (\rho + p)_i,
 \end{equation}
 will perform a sign flip during the evolution if $w_0 < -1$. When this sign flip occurs, the total velocity transfer function diverges, similar to the phantom crossing for dark energy models, which results in impossibly large linear kicks inside the simulations. This divergence is inevitable for all models with $w_0 < -1$, however, it can occur in the future, i.e. at $z < 0$, for certain parameter choices, which makes it possible to run certain simulations with $w_0 < -1$. For $w_a = 0$ and ignoring photons and neutrinos, the time of the divergence can be calculated via
 \begin{equation}
     a_\mathrm{divergence} = \left(\frac{(\Omega_M - 1)(1 + w_0)}{\Omega_M} \right)^\frac{1}{3w_0}, \label{eq:phatom_crossing}
 \end{equation}
 where $\Omega_M = \Omega_\mathrm{cdm} + \Omega_b$. Using this formula we can select all parameters that are still possible to run with relativistic fields. We show the resulting additional prior in Figure~\ref{fig:grid_plot_w}.

 \section{Accuracy of the Simulations}
 \label{ap:sim_acc}
 
\begin{figure*}
     \centering
     \includegraphics[width=0.47\textwidth]{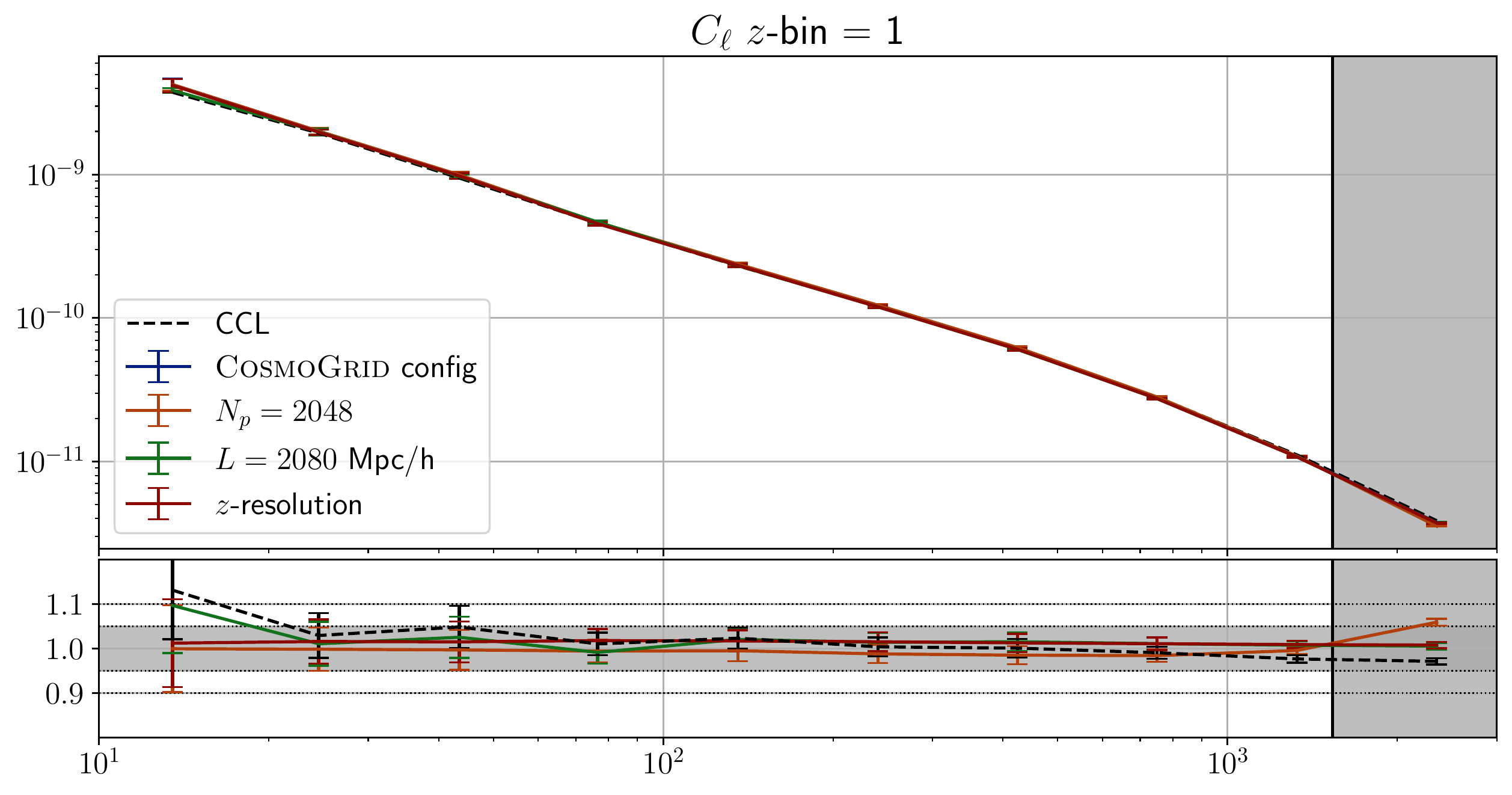}
     \includegraphics[width=0.47\textwidth]{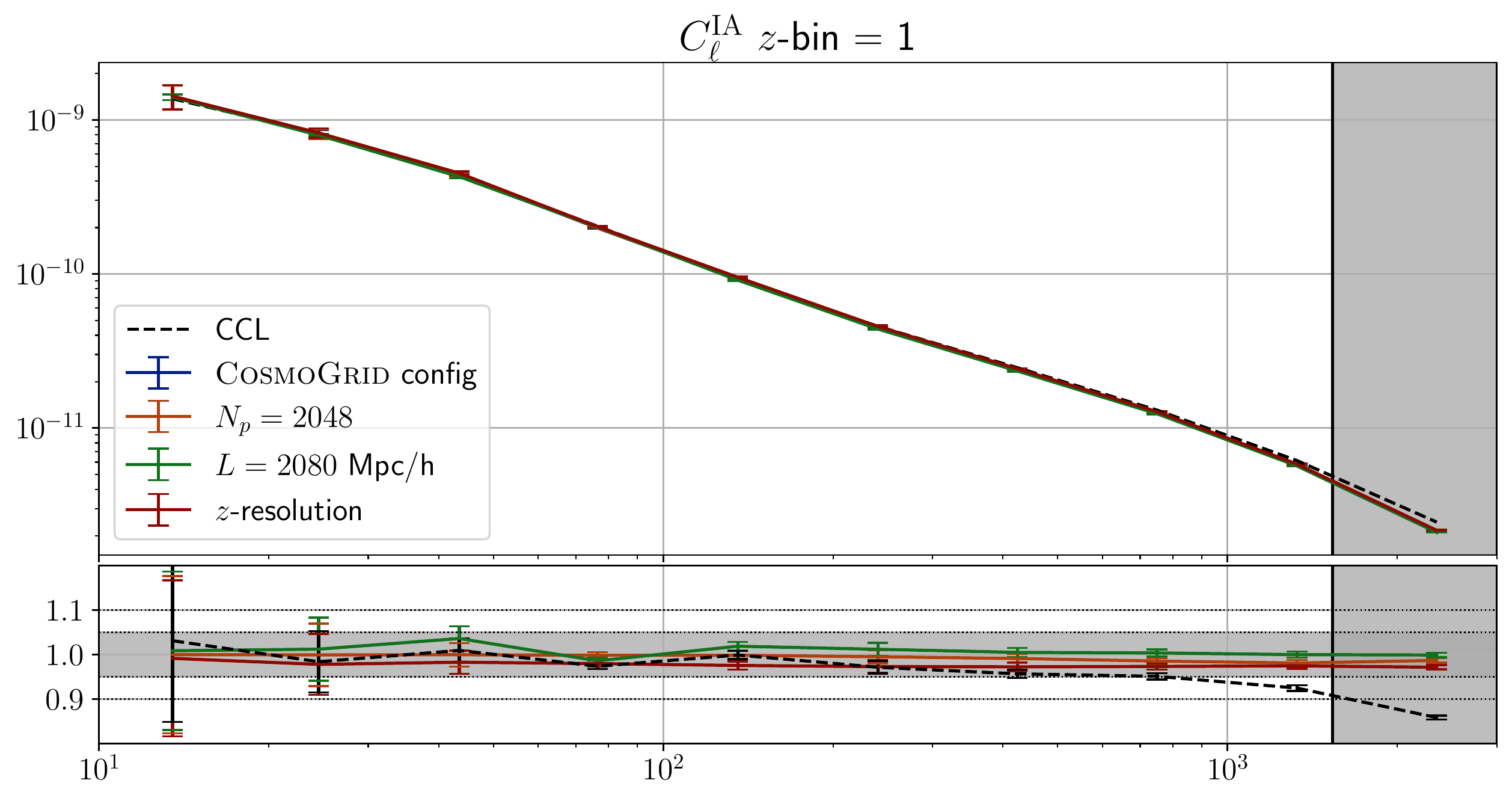}
     \includegraphics[width=0.47\textwidth]{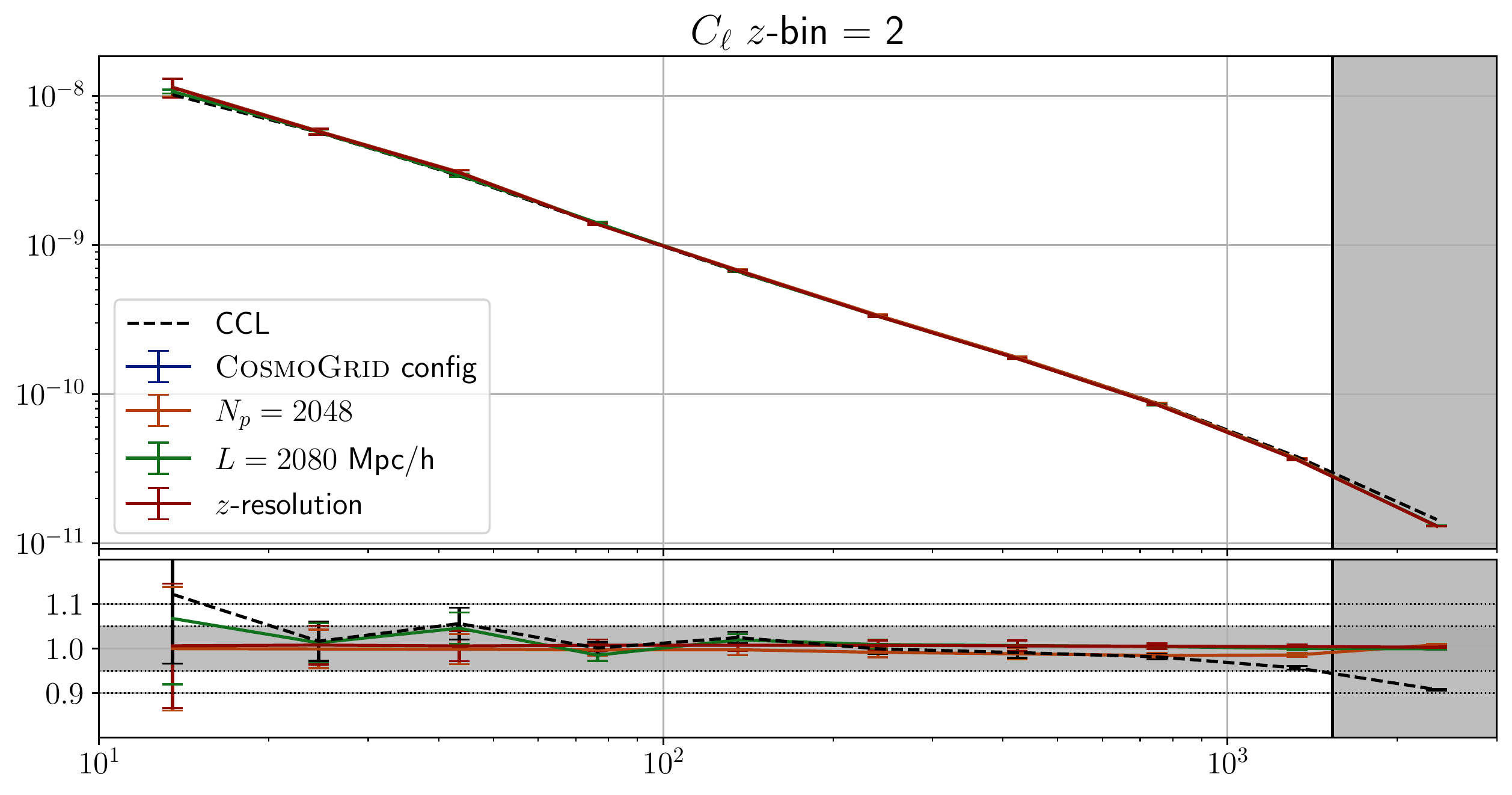}
     \includegraphics[width=0.47\textwidth]{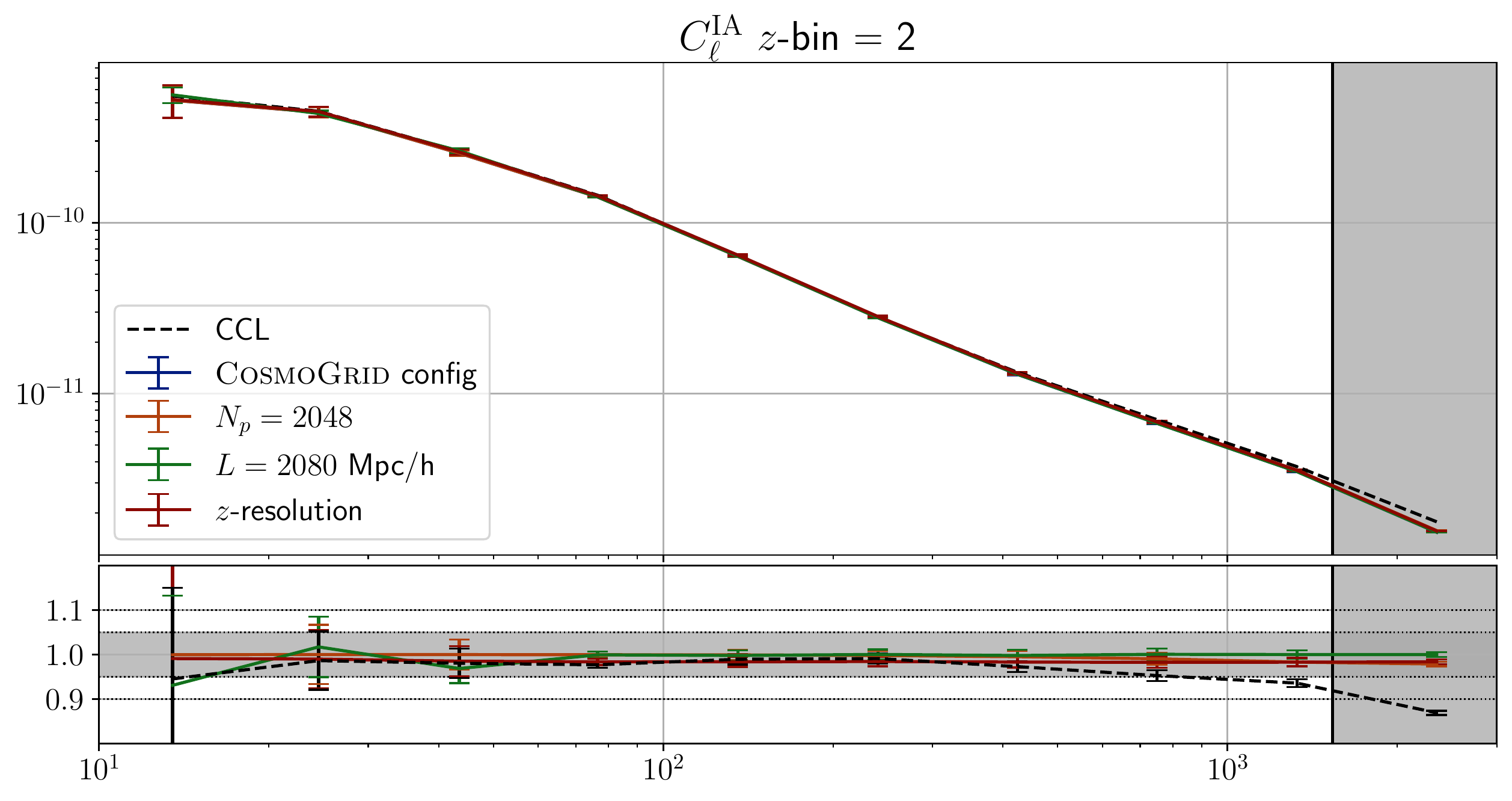}
     \includegraphics[width=0.47\textwidth]{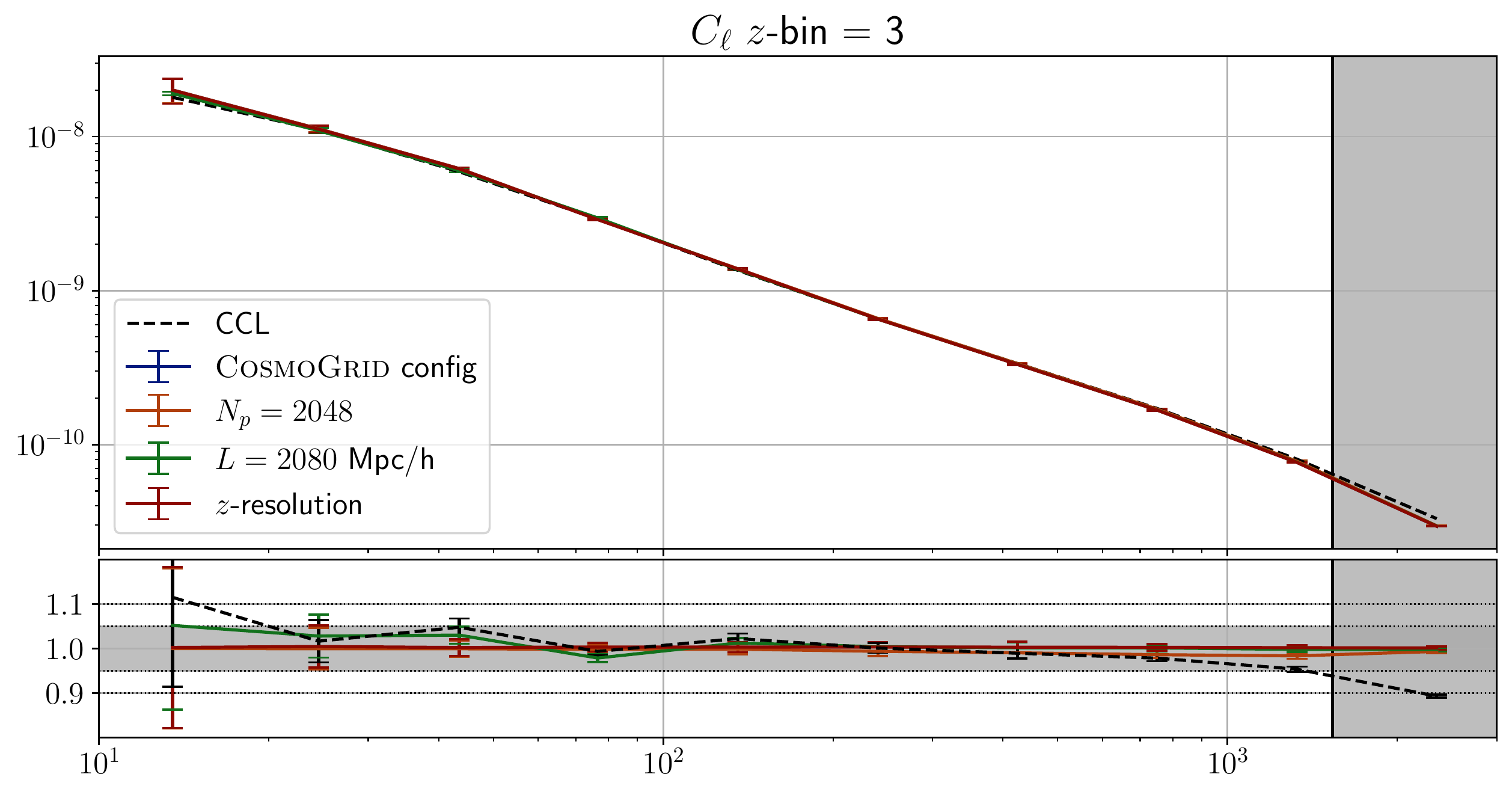}
     \includegraphics[width=0.47\textwidth]{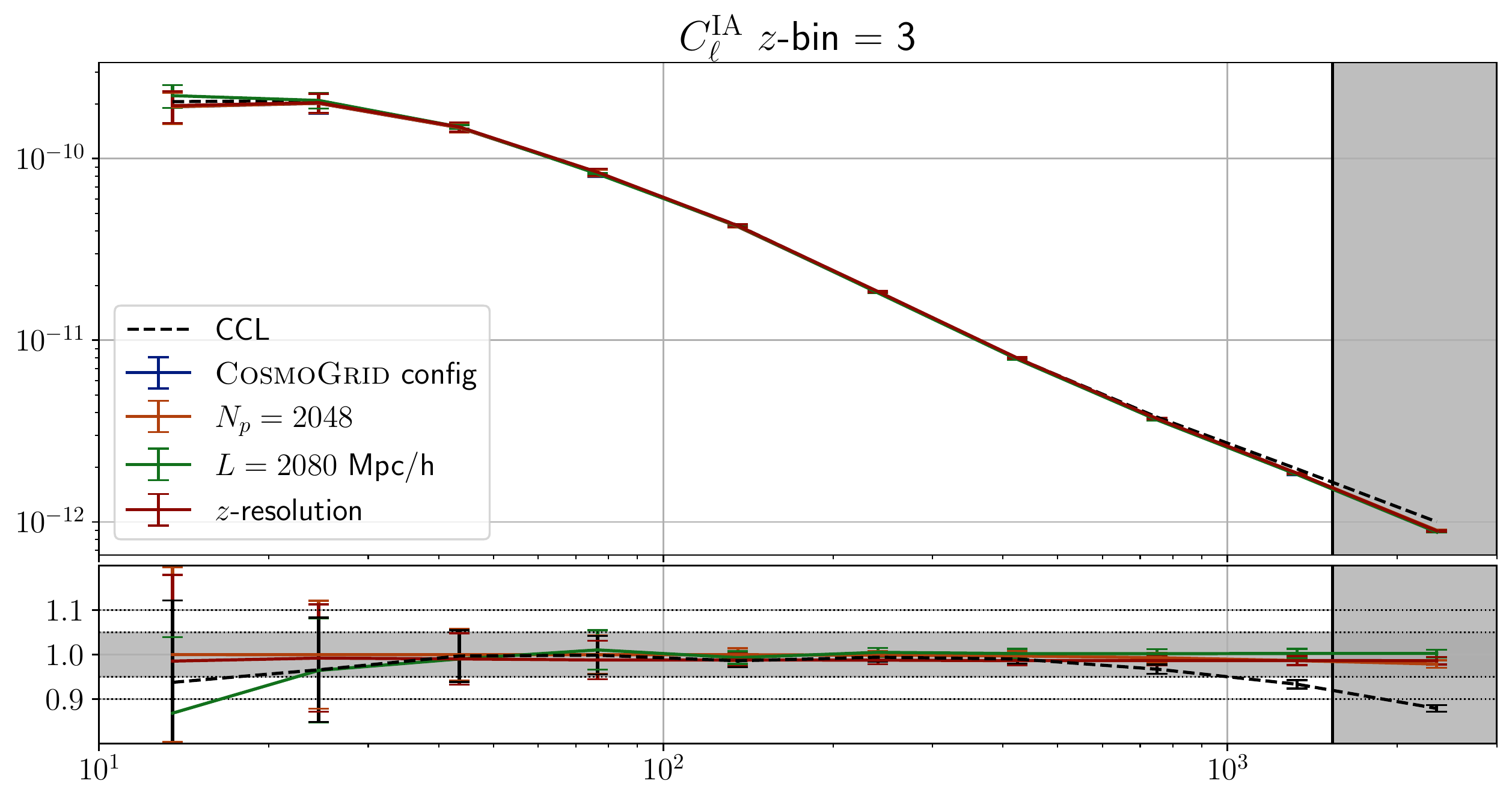}
     \includegraphics[width=0.47\textwidth]{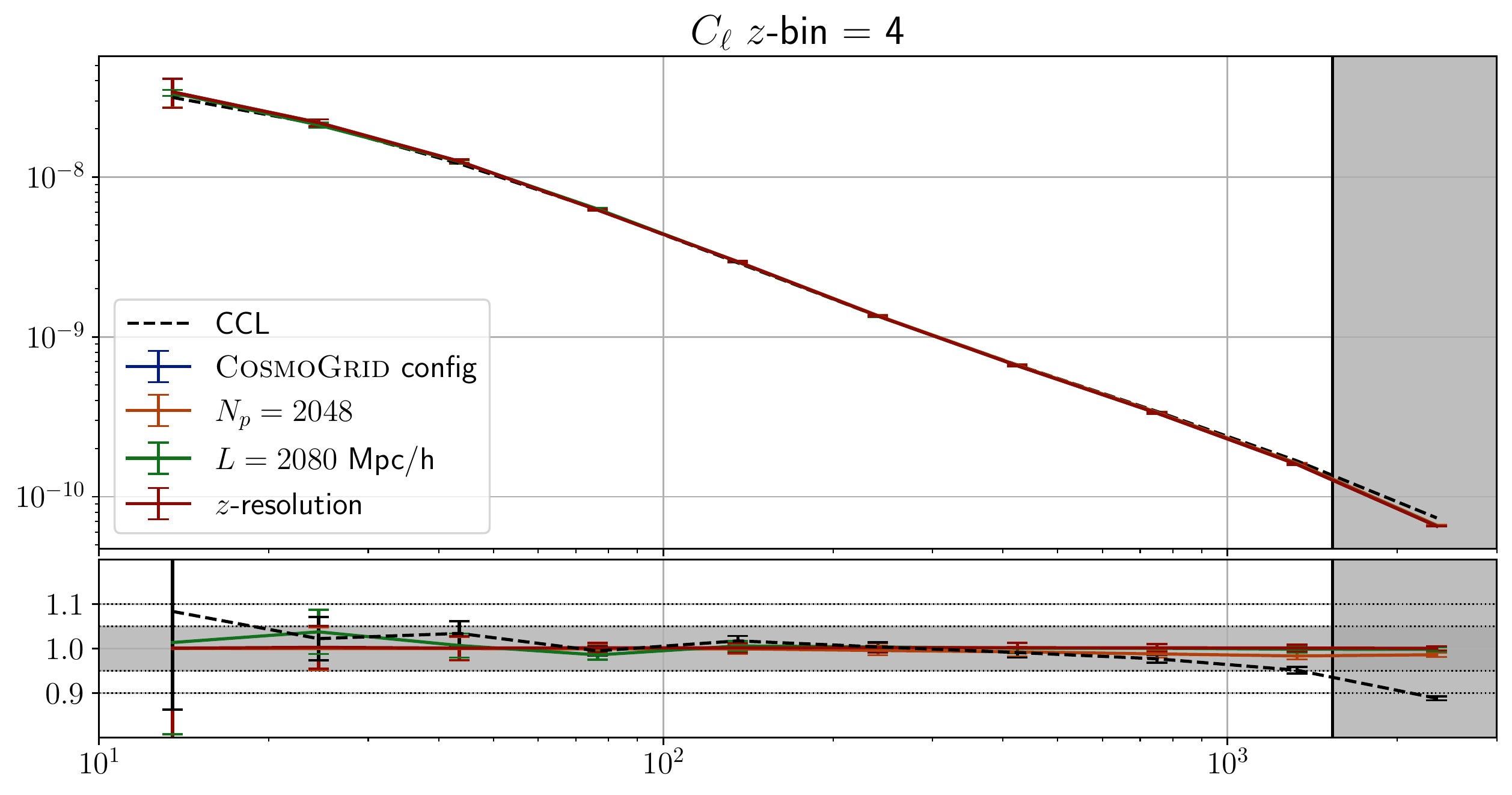}
     \includegraphics[width=0.47\textwidth]{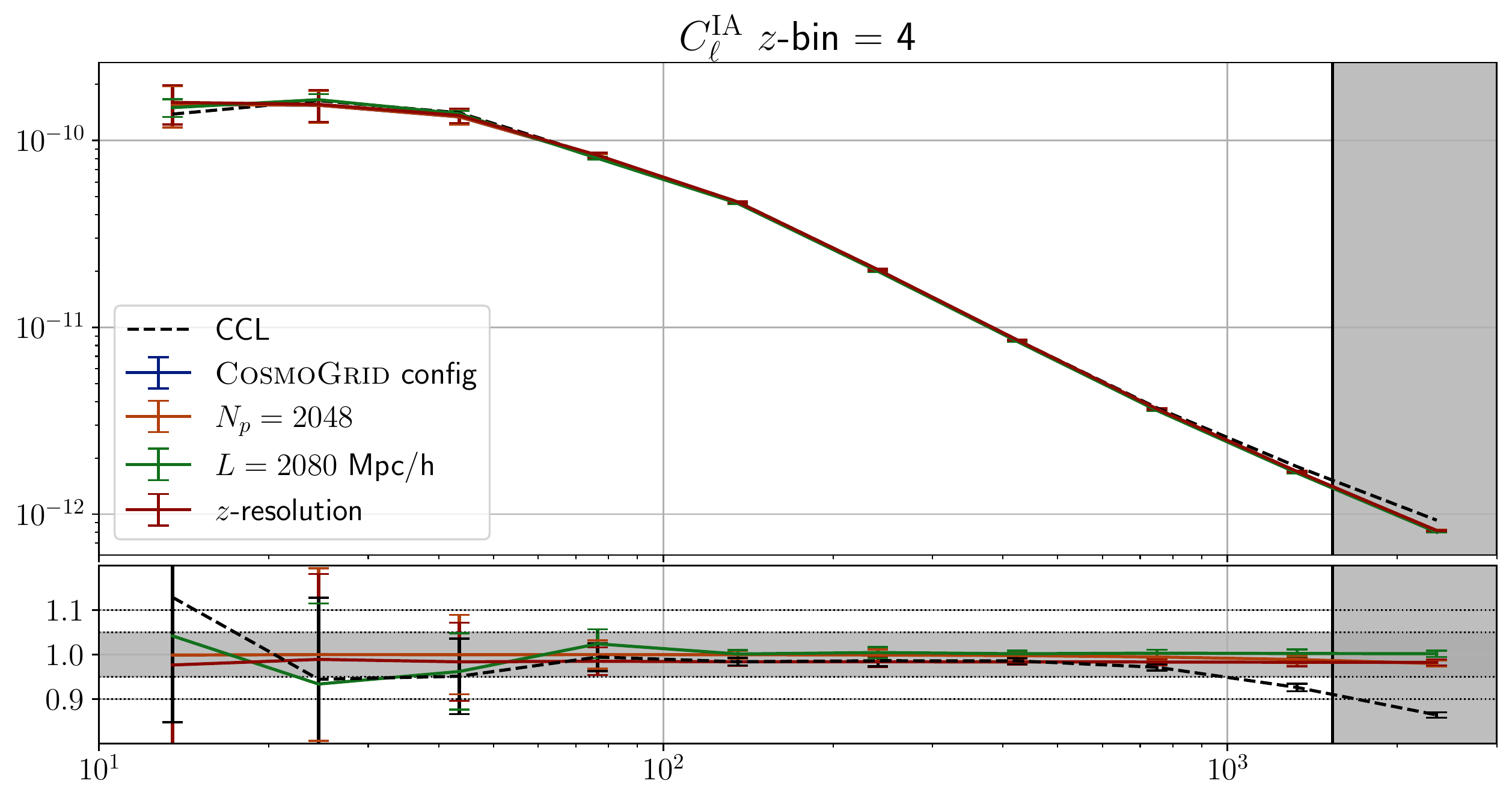}
     \includegraphics[width=0.47\textwidth]{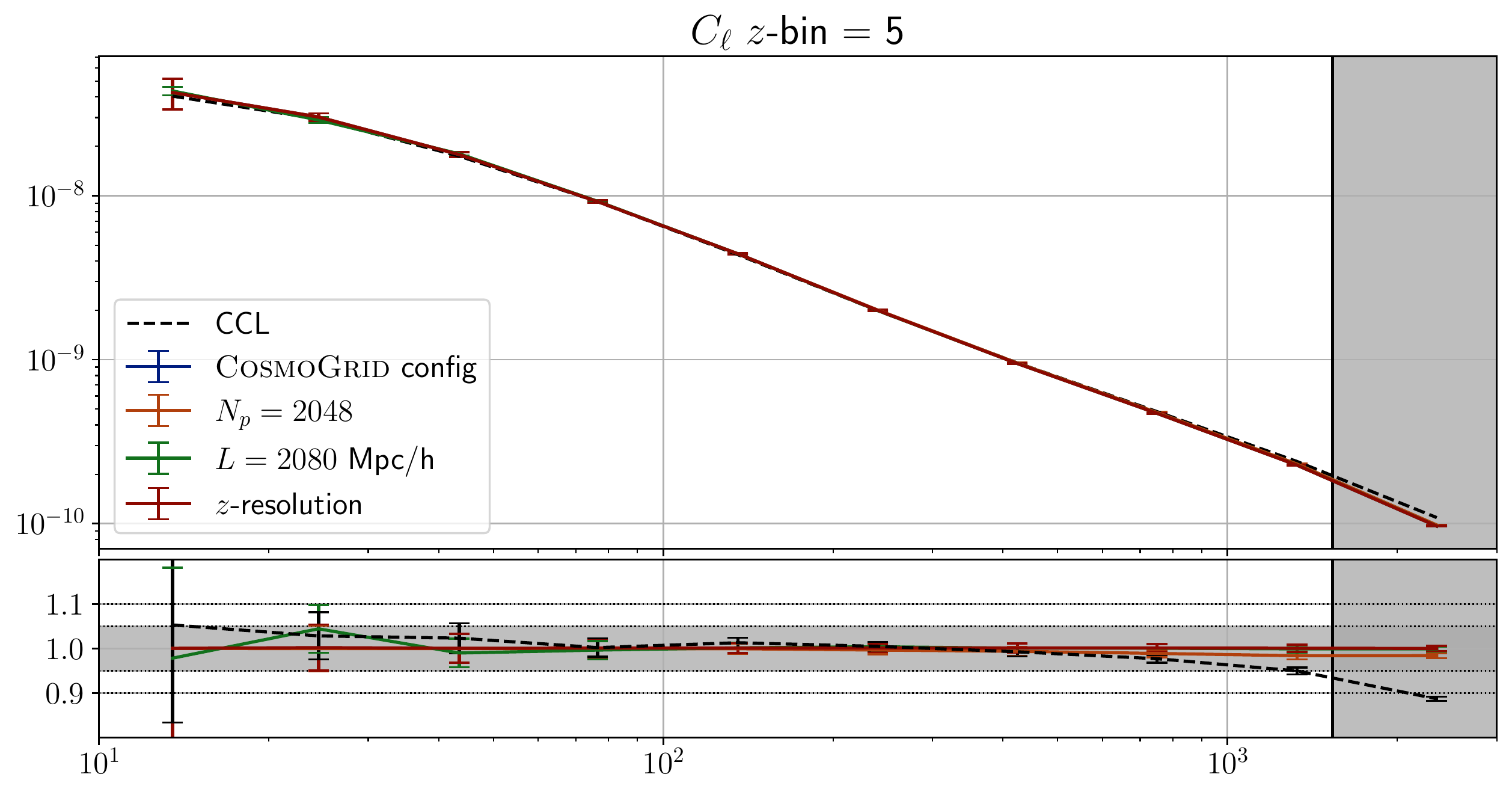}
     \includegraphics[width=0.47\textwidth]{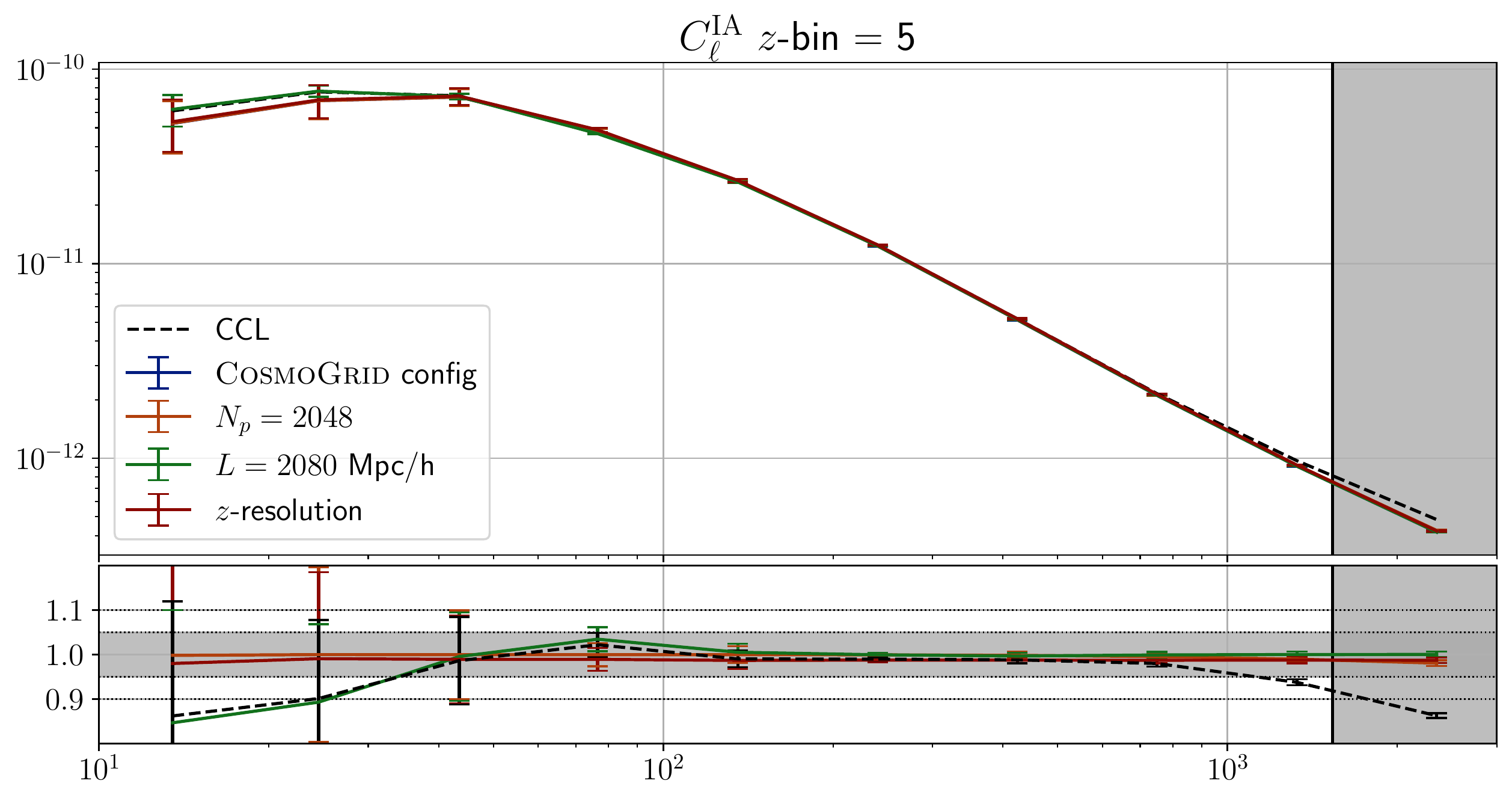}
     \caption{Power spectra comparison of \pyccl~and the benchmark simulations. The spectra are all calculated using maps with an $n_\mathrm{side}$ of 2048 to avoid resolution effects. The left column shows the spectra of the cosmological signal for each redshift bin. The right column shows the same comparison for the intrinsic alignment (II) spectra. The shaded regions on the right side of each plot indicate the scales that can't be resolved with our fiducial resolution of $n_\mathrm{side} = 512$. The lower panels of each plot show the ratio of the \cosmogrid~settings with respect to the other settings. \label{fig:spec_acc}}
     
 \end{figure*}

 We compare the power spectra from all benchmark simulations to check the accuracy of the simulated convergence maps. Further, we also compare our fiducial configuration to spectra predicted with \pyccl~using the whole range of the redshift distributions. The results of this comparison are shown in Figure~\ref{fig:spec_acc}. The power spectra of the benchmarks agree all well with each other and also with the \pyccl~predictions. Meaning that the simulation settings have a negligible impact on the considered scales.
 
 \section{Shell Baryonification}
 \label{ap:shell_bary}

In this section we provide some details about our modified version of the baryon correction model from~\cite{aurel_bray2} which we dub \textit{shell baryonification}. The original correction model starts with the assumption that the dark-matter-only halos are well described by combining the truncated Navarro–Frenk–White (NFW) profile with a 2-halo density component
\begin{equation}
    \rho_\mathrm{dmo}(r) = \rho_\mathrm{nfw}(r) + \rho_\mathrm{2h}(r).
\end{equation}
This profile only depends on the cosmological parameters of the simulations, the viral mass of the halo $M \equiv M_{200}$ and the concentration $c \equiv c_{200}$. However, a more realistic description of the profile would also include gas, the central galaxy and a collisionless matter component
\begin{equation}
    \rho_\mathrm{dmb}(r) = \rho_\mathrm{gas} + \rho_\mathrm{cga}(r) + \rho_\mathrm{clm}(r) + \rho_\mathrm{2h}(r).
\end{equation}
The goal of the correction model is now to transform the dark matter only profile $\rho_\mathrm{dmo}$ from the simulations into a profile that includes this baryonic feedback $\rho_\mathrm{dmb}$. This is done by matching the halo masses
\begin{equation}
    M_\chi(r) = 4\pi\int_0^r s^2\rho_\chi(s) \mathrm{d}s.
\end{equation}
The halo mass functions $M_\mathrm{dmo}(r)$ and $M_\mathrm{dmb}(r)$ are bijective function, making it possible to invert them. The difference between the inverted function gives raise to the displacement function
\begin{equation}
    d(r_\mathrm{dmo}\vert M,c) = r_\mathrm{dmb}(M) - r_\mathrm{dmo}(M),
\end{equation}
for each halo with mass $M$ and concentration $c$. Each particle inside the simulation that is close enough to a halo, is then displaced using this function, transforming the dark matter only profiles $\rho_\mathrm{dmo}$ into $\rho_\mathrm{dmb}$. The model requires that the simulation is run in snapshot mode, such that the position of each particle is known. However, the simulations of the \cosmogrid~were run in lightcone mode and only particle shells are available along with the halo catalogues of each time step, making it impossible to directly apply this correction model. 

The halo catalogues of the \cosmogrid~were generated using a FoF halo finder with a linkage length of 20\% of the mean particle separation and were generated for each time step. For our correction model we used all FoF halos that contain at least 100 particles within the viral radius $r \equiv r_{200}$. We fit a standard NFW profile to each halo using the logarithmically binned mass shells contained in the \cosmogrid~halo catalogues to obtain the concentration $c$ and mass $M$ of each halo. Afterwards, we organized the halos into shells using the same shell boundaries and replication schemes as in the original lightcone mode. Halos that were closer than 20 Mpc to a shell boundary are assigned to both shells of the boundary, making it possible for them to affect the particles in both shells. Following the \ufalcon~convention, we assume that all halos and particles inside a shell lie at the shells mean redshift $z_m$. Next we define the projected halo mass function
\begin{equation}
    M_\chi^p(r) = 2\pi\int_0^r s \int_0^{z_\mathrm{max}} \rho_\chi(s,z) \mathrm{d}z \mathrm{d}s,
\end{equation}
which corresponds to the mass function of a halo that was projected onto a plane. We chose $z_\mathrm{max} = 50r$ as the limit of the integration. Note that the projected mass is actually diverging as $z_\mathrm{max}$ increases, because of the 2-halo component. However, the projected displacement function
\begin{equation}
    d^p(r_\mathrm{dmo}\vert M,c) = r_\mathrm{dmb}(M^p) - r_\mathrm{dmo}(M^p),
\end{equation}
is always finite, because the divergent terms cancel out. The shell baryonification is then performed by displacing the pixels of the high resolution shells ($n_\mathrm{side} = 2048$) of the \cosmogrid~simulations. For each halo we assume a locally flat sky and displace the pixels closer than $50r$ using the projected displacement function. Afterwards, we assign new values to the \healpix~pixel positions using a linear interpolation to the displaced pixels. In a last step, we decrease the resolution of the shells to our final resolution of $n_\mathrm{side} = 512$ which acts as a smoothing kernel.

\begin{figure*}
     \centering
     \includegraphics[width=0.47\textwidth]{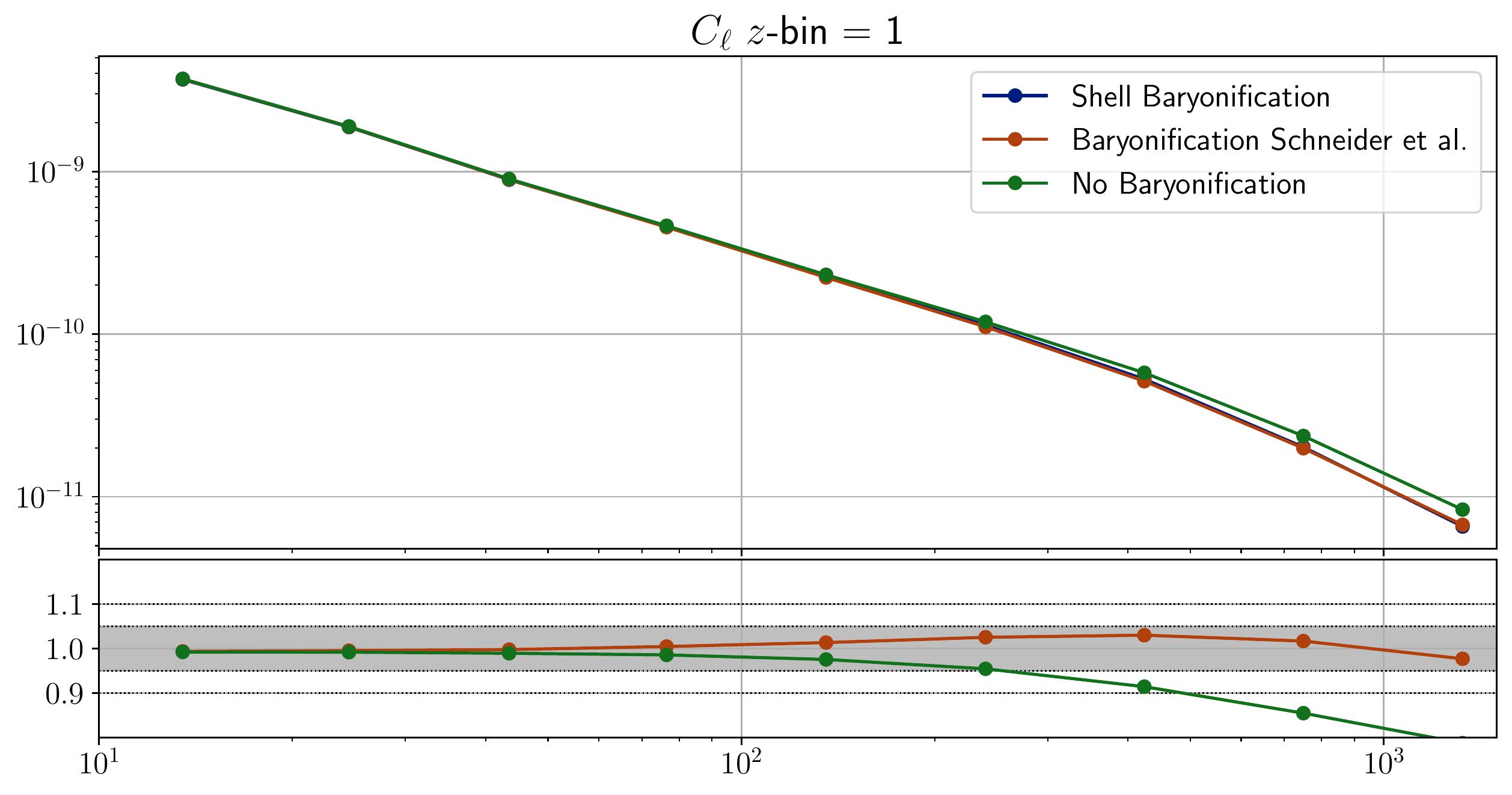}
     \includegraphics[width=0.47\textwidth]{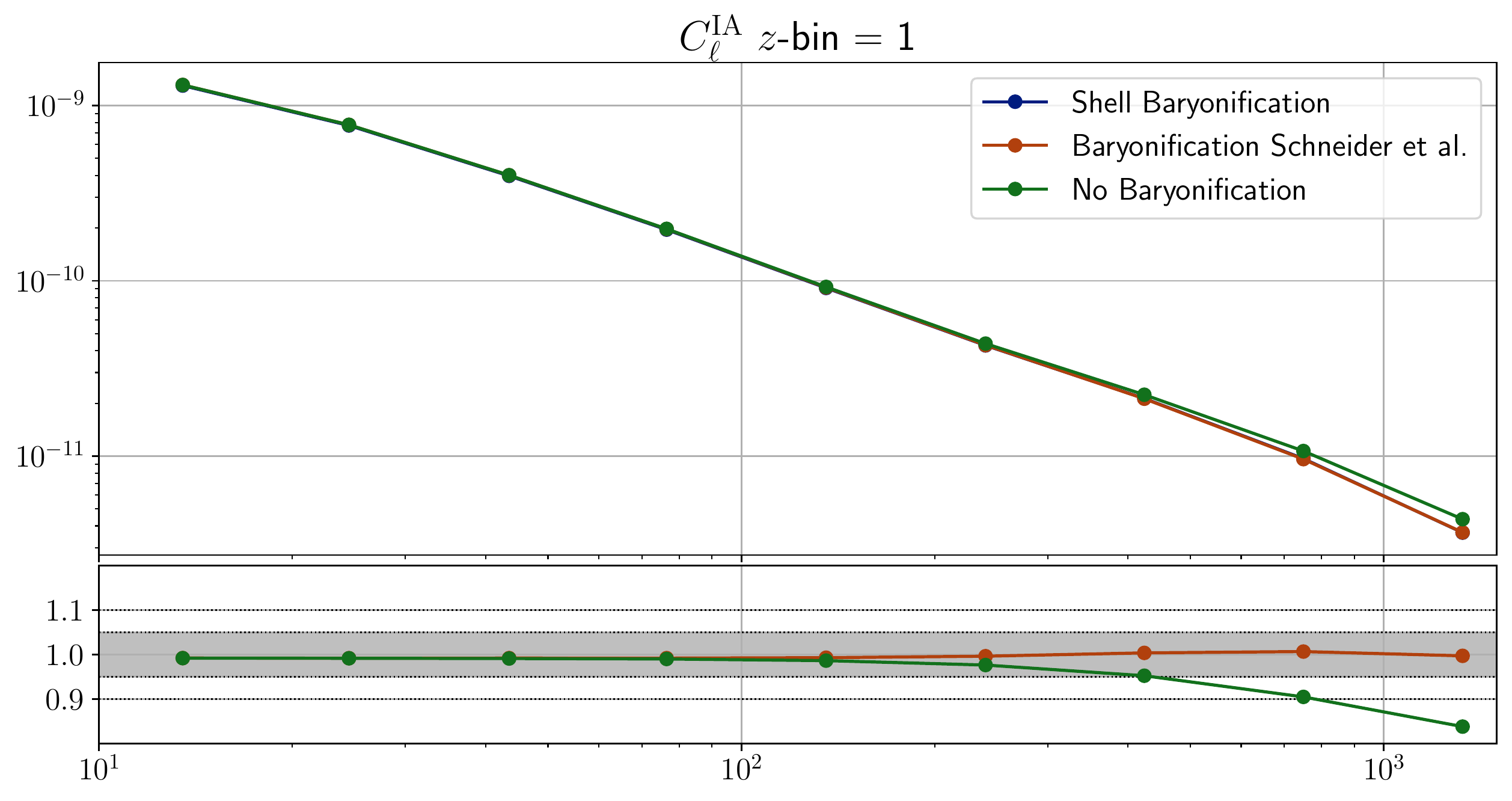}
     \includegraphics[width=0.47\textwidth]{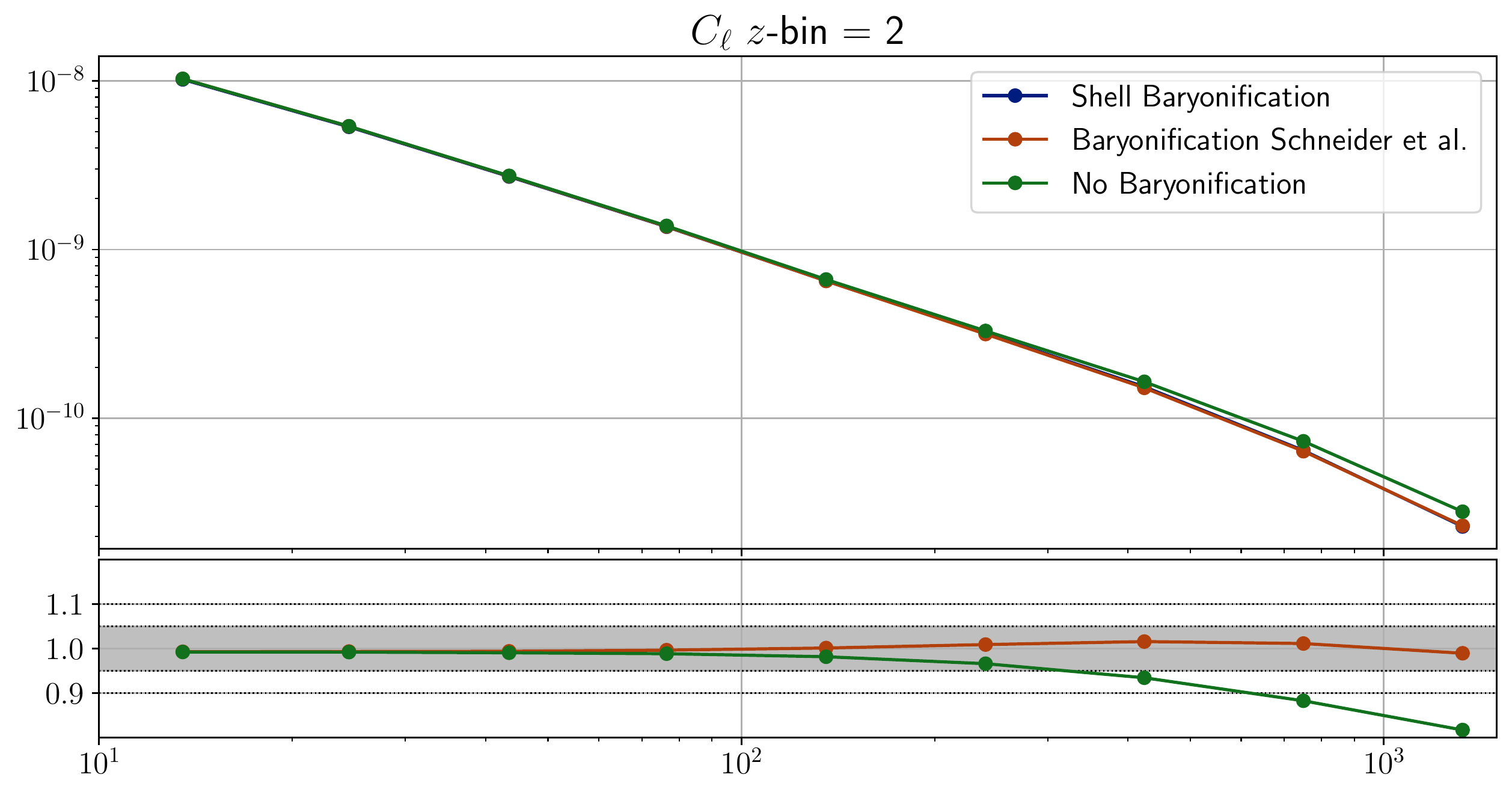}
     \includegraphics[width=0.47\textwidth]{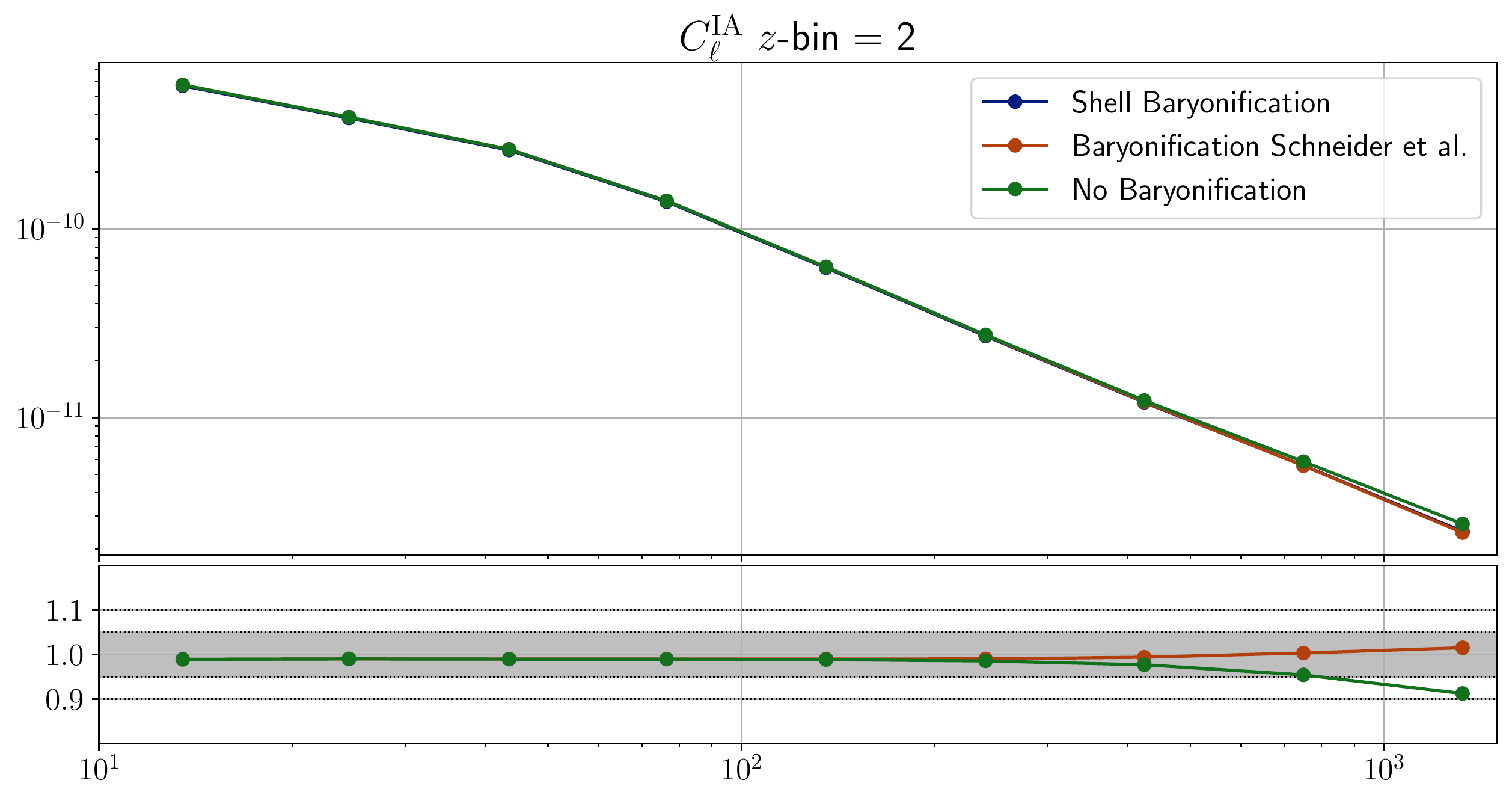}
     \includegraphics[width=0.47\textwidth]{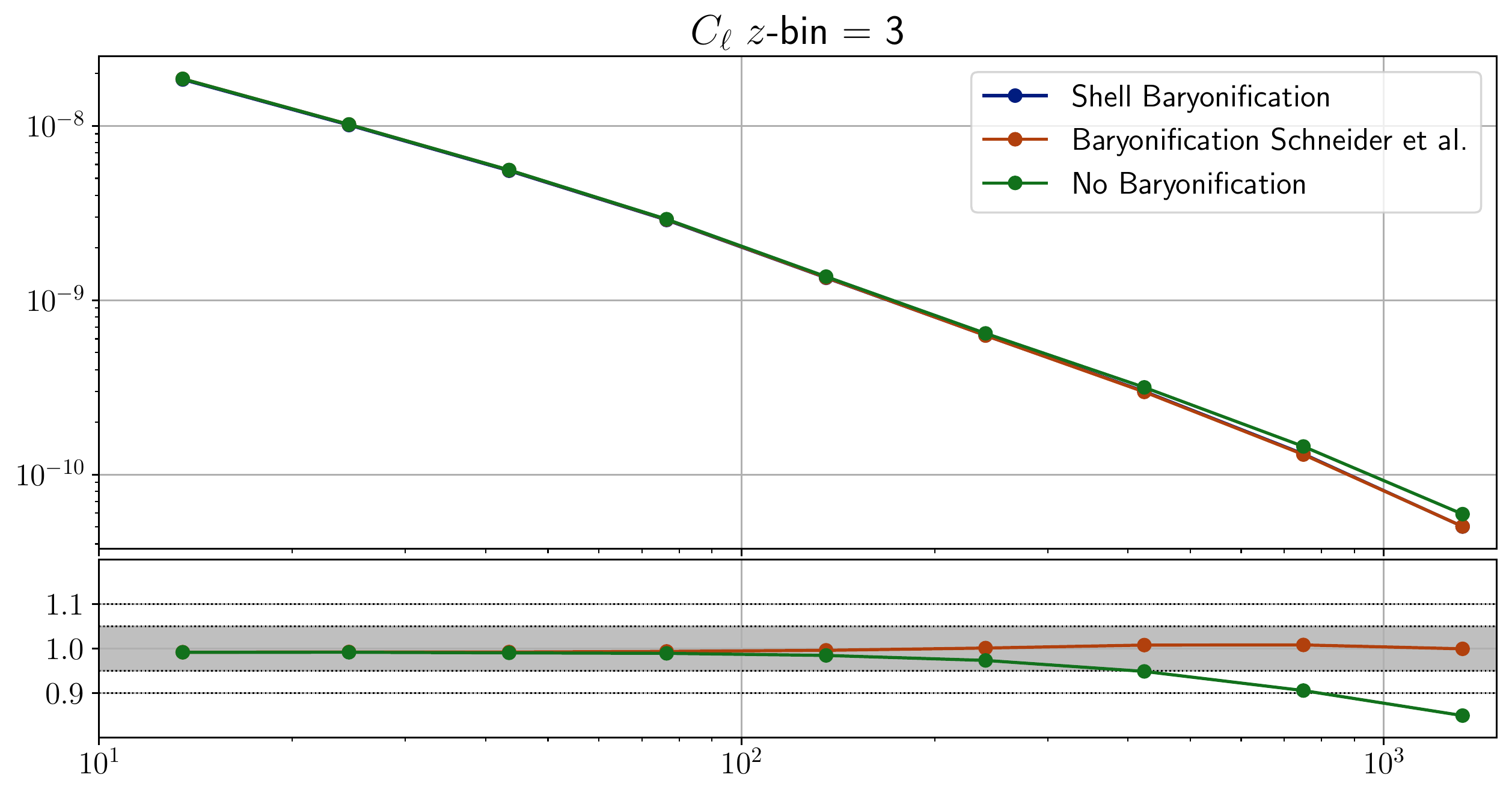}
     \includegraphics[width=0.47\textwidth]{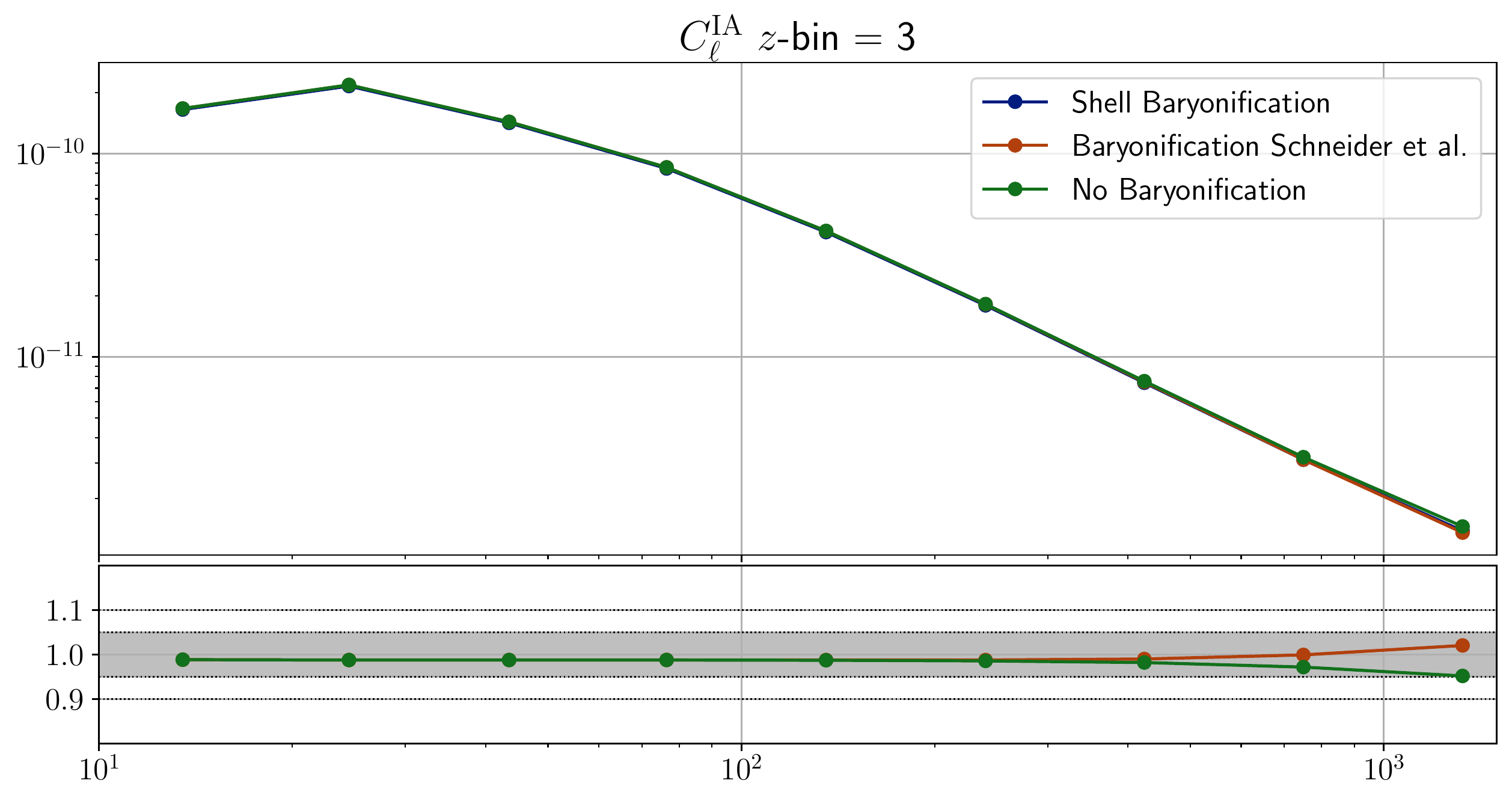}
     \includegraphics[width=0.47\textwidth]{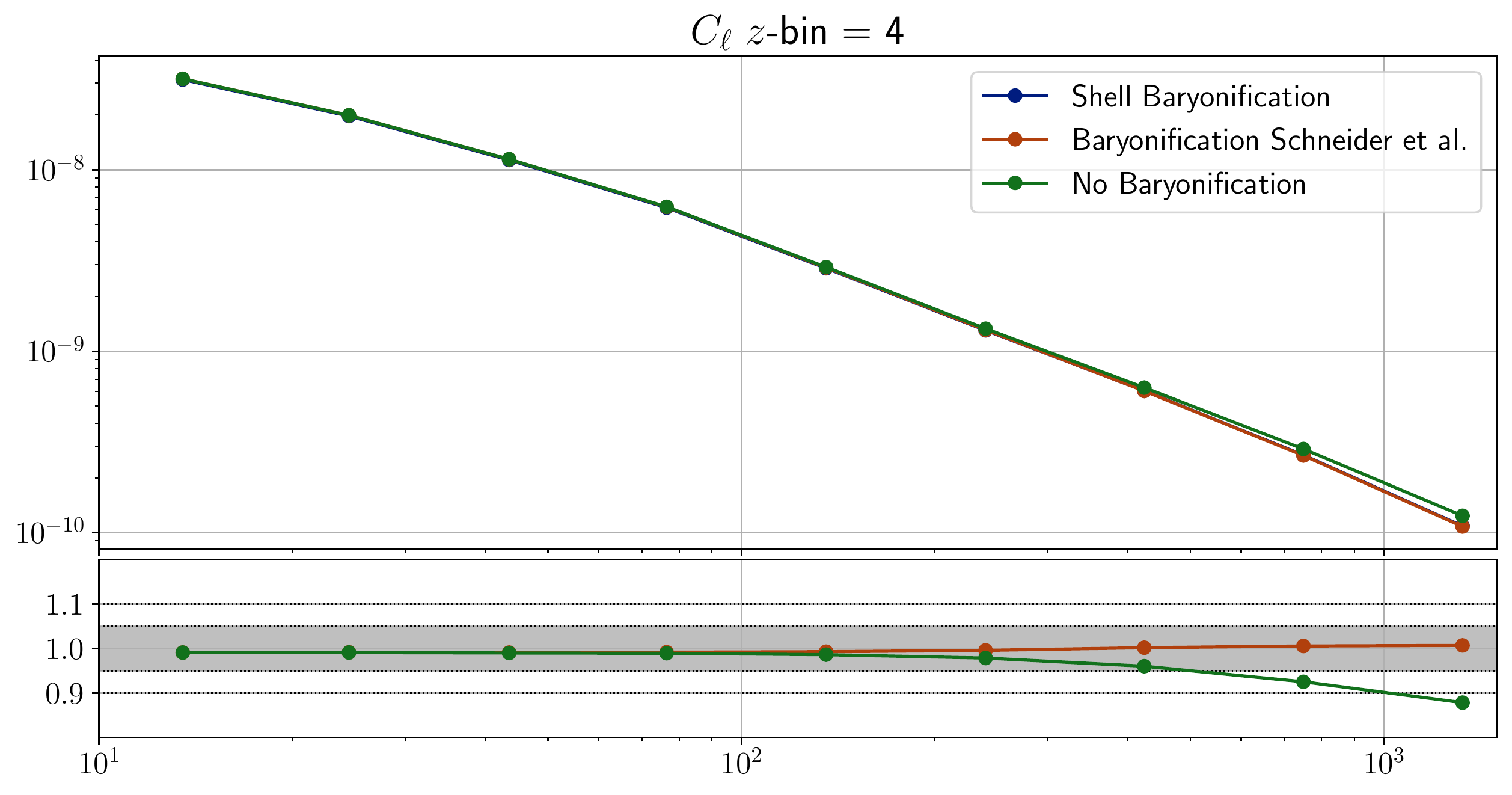}
     \includegraphics[width=0.47\textwidth]{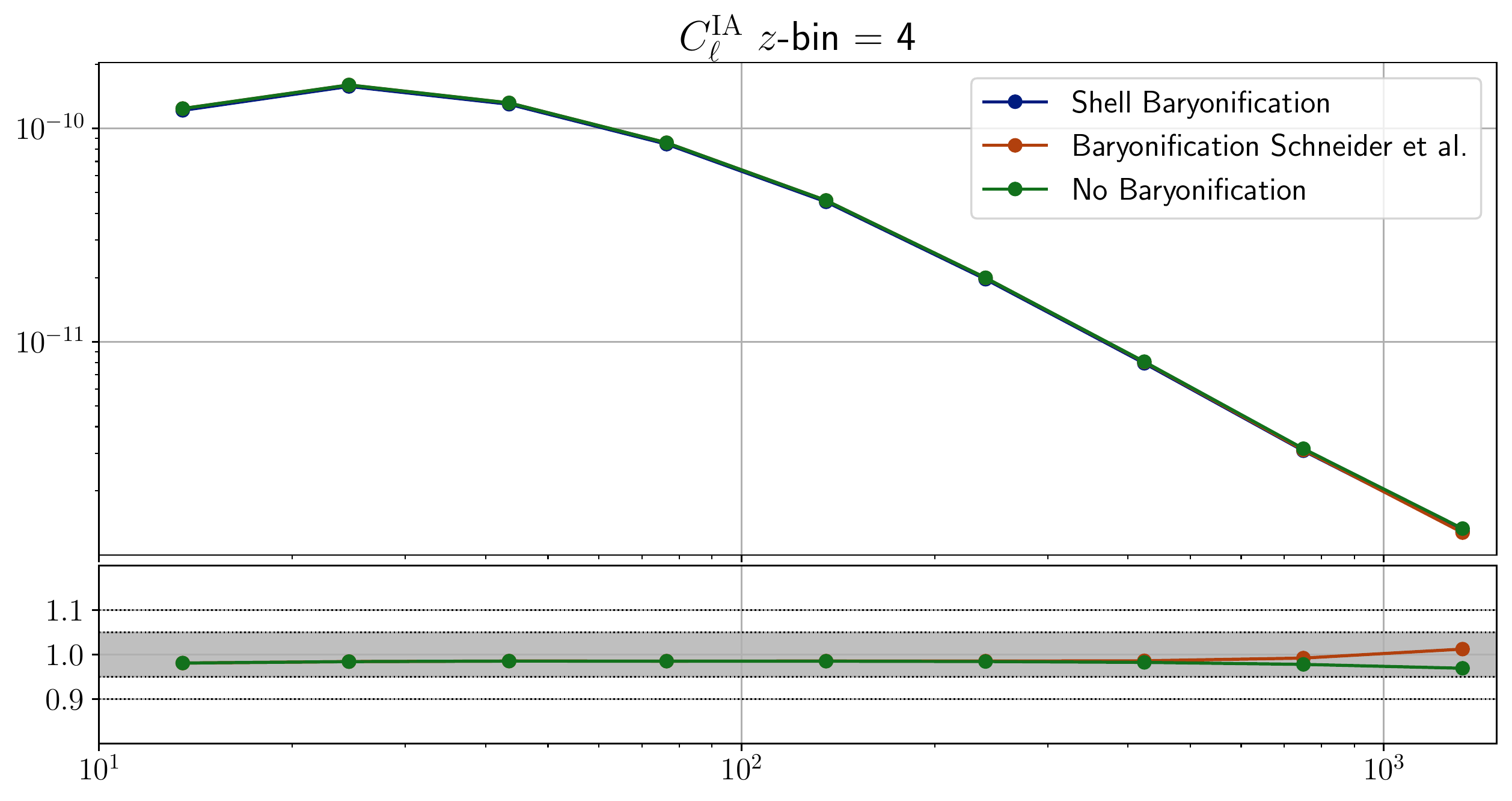}
     \includegraphics[width=0.47\textwidth]{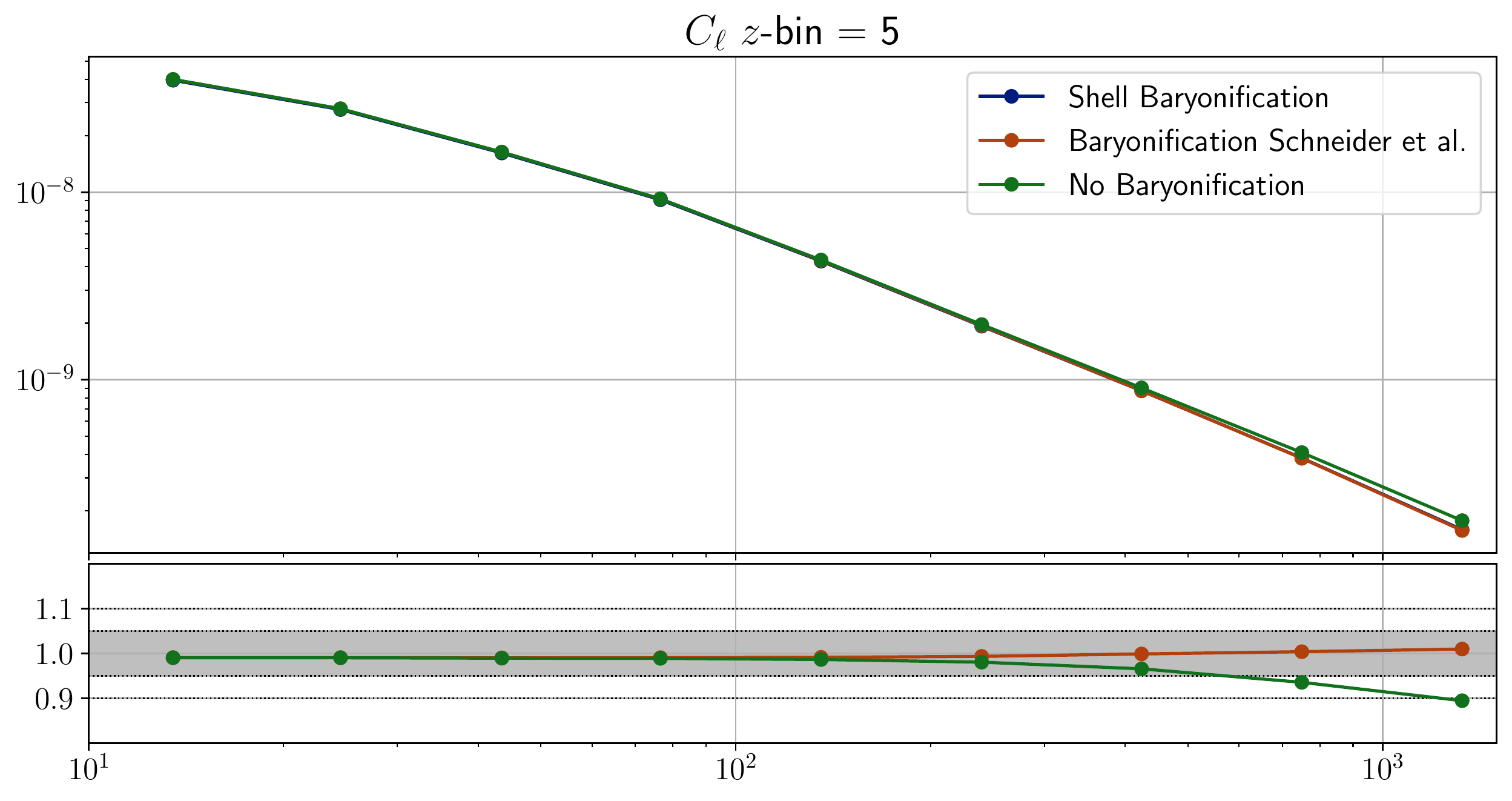}
     \includegraphics[width=0.47\textwidth]{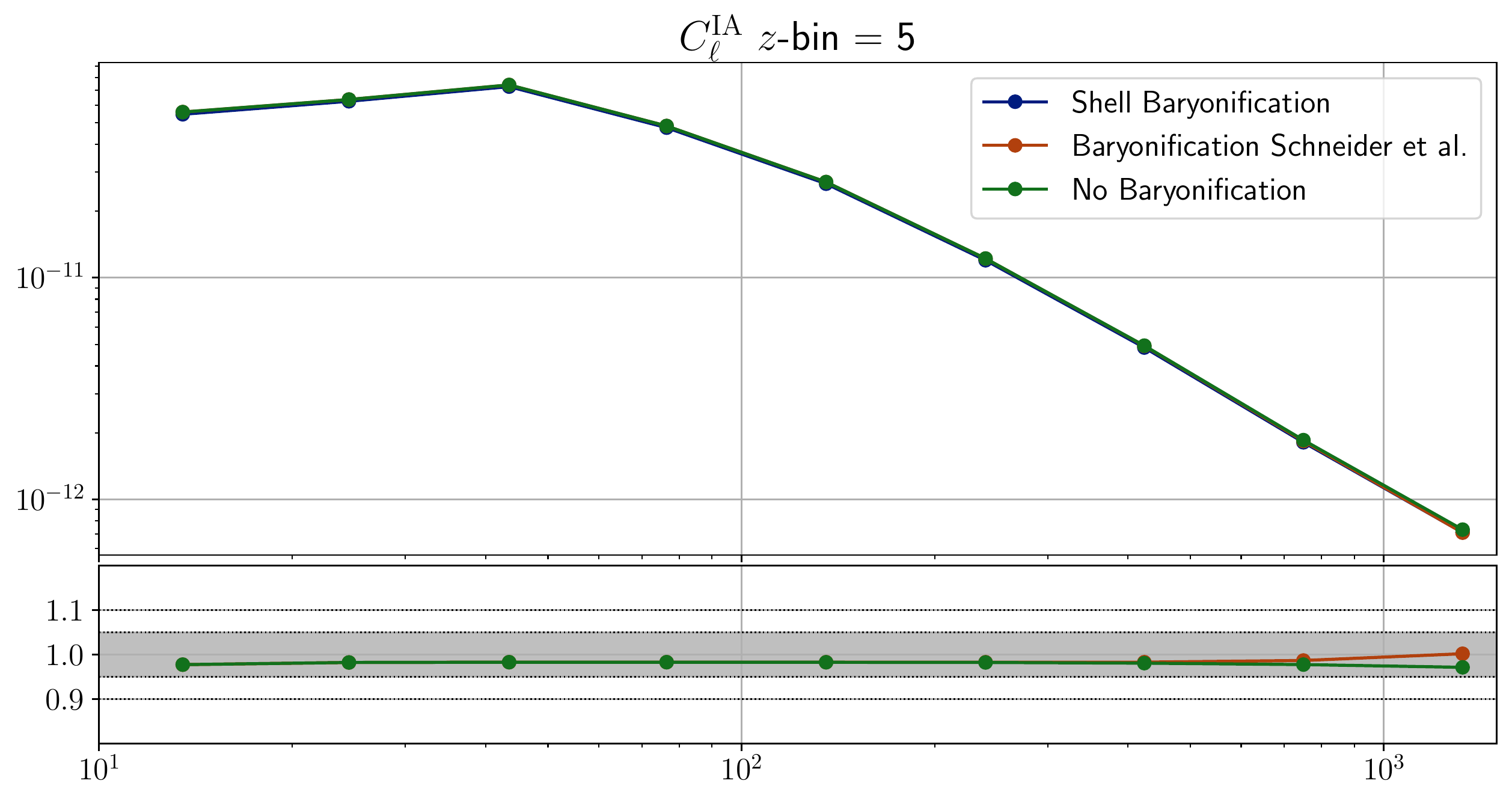}
     \caption{Power spectra of the shell baryonification, the original baryonification from~\cite{aurel_bray2}, and the dark-matter-only-case  using a fiducial simulation. The left-hand column shows the power spectra of the signal for each redshift bin and the right-hand column for the intrinsic alignment maps (II).
      \label{fig:bary_spec_acc}}
     
 \end{figure*}

We compare the power spectra of maps generated with the original baryon correction model and our shell baryonification in Figure~\ref{fig:bary_spec_acc}. The shell baryonification was performed as explained above using a random fiducial simulation and our fiducial baryonic correction parameters (see section~\ref{sec:barys} and table~\ref{tab:grid}). The original baryonification model was performed in the same way as in~\citetalias{kids_net}, which is the normal procedure as described in~\cite{aurel_bray2}. We reran the simulation with the identical initial condition in snapshot mode. Afterwards, we used the \amiga~halo finder~\cite{amiga1,amiga2} to identify all relevant halos and performed the normal baryon corrections inside the generated snapshots. Finally, we generate the shells of the past lightcone using the replication scheme of the original simulations and project them with \ufalcon~and the fiducial redshift distributions. The power spectra of the resulting maps is shown in Figure~\ref{fig:bary_spec_acc}. Both approaches agree very well over all considered scales. Additionally to the two baryonic feedback models, we also plot the power spectrum of the maps without any corrections. One can clearly see that there is a significant drop of the power on small scales (high $\ell$-modes) for the baryon-corrected maps.

 \begin{figure*}
     \centering
     \includegraphics[width=1.0\textwidth]{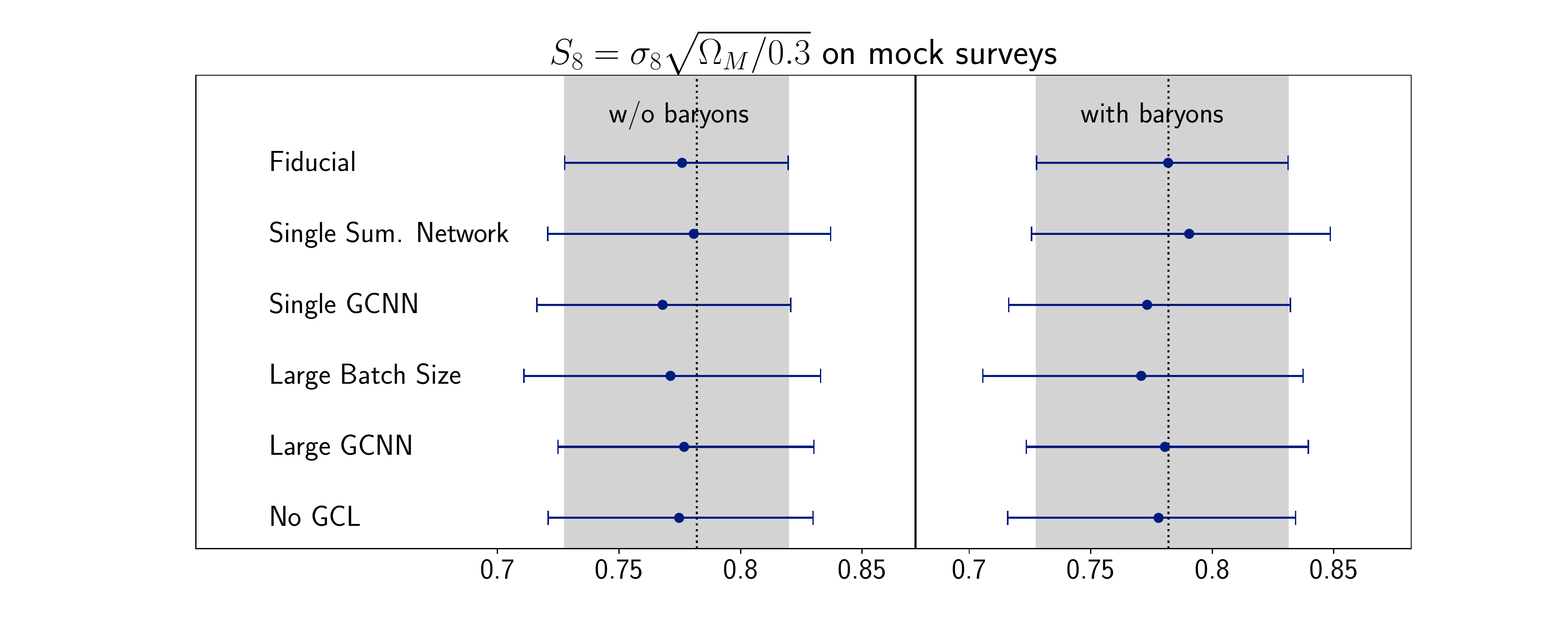}
     \caption{Constraints of various different settings using the fiducial benchmark simulations as mock observation. The first constraint ``Fiducial" is equivalent to the fiducial network constraints of Figure~\ref{fig:s8_mocks} and serves as reference point. The left side of the plot shows the constraints without baryonic correction, while the right side includes these corrections. \label{fig:s8_bench}}
 \end{figure*}
 \begin{table*}[t!]
     \centering
     \renewcommand{\arraystretch}{2}
     \begin{tabular}{llcccccccc}
Type & Baryons &  $\Omega_M$ & $\sigma_8$ & $h_0$ & $\Omega_b$ & $n_s$ & $w_0$ & $A_\mathrm{IA}$ & $S_8$  \\
\hline
\hline
GCNN & no &  $0.262_{-0.162}^{+0.048}$ & $0.884_{-0.277}^{+0.136}$ & $0.725_{-0.085}^{+0.095}$ & $0.045_{-0.015}^{+0.015}$ & $0.962_{-0.092}^{+0.031}$ & $-0.930_{-0.285}^{+0.316}$ & $0.514_{-0.663}^{+0.670}$ & $0.774_{-0.048}^{+0.043}$  \\
GCNN & yes &  $0.259_{-0.159}^{+0.046}$ & $0.903_{-0.286}^{+0.149}$ & $0.726_{-0.086}^{+0.094}$ & $0.045_{-0.015}^{+0.015}$ & $0.966_{-0.096}^{+0.104}$ & $-0.891_{-0.297}^{+0.335}$ & $0.457_{-0.665}^{+0.673}$ & $0.786_{-0.055}^{+0.046}$  \\
GCNN Optimized & no &  $0.251_{-0.151}^{+0.041}$ & $0.903_{-0.268}^{+0.144}$ & $0.721_{-0.081}^{+0.028}$ & $0.046_{-0.016}^{+0.014}$ & $0.958_{-0.088}^{+0.028}$ & $-0.923_{-0.268}^{+0.289}$ & $0.479_{-0.542}^{+0.542}$ & $0.776_{-0.044}^{+0.037}$  \\
GCNN Optimized & yes &  $0.239_{-0.130}^{+0.037}$ & $0.940_{-0.228}^{+0.188}$ & $0.722_{-0.082}^{+0.028}$ & $0.045_{-0.015}^{+0.015}$ & $0.966_{-0.068}^{+0.054}$ & $-0.854_{-0.279}^{+0.304}$ & $0.423_{-0.488}^{+0.492}$ & $0.793_{-0.051}^{+0.039}$  \\
Power Spectrum & no &  $0.259_{-0.154}^{+0.047}$ & $0.874_{-0.253}^{+0.150}$ & $0.726_{-0.086}^{+0.094}$ & $0.045_{-0.015}^{+0.015}$ & $0.967_{-0.097}^{+0.103}$ & $-0.884_{-0.298}^{+0.329}$ & $0.774_{-0.705}^{+0.856}$ & $0.765_{-0.052}^{+0.053}$  \\
Power Spectrum & yes &  $0.261_{-0.145}^{+0.055}$ & $0.886_{-0.265}^{+0.156}$ & $0.728_{-0.088}^{+0.092}$ & $0.045_{-0.015}^{+0.015}$ & $0.969_{-0.099}^{+0.101}$ & $-0.843_{-0.243}^{+0.410}$ & $0.734_{-0.700}^{+0.883}$ & $0.779_{-0.059}^{+0.061}$  \\
     \end{tabular}
     \caption{Full parameter constraints of our analysis. We present the constraints of all setting presented in section~\ref{sec:results} with and without the treatment of baryon corrections. \label{tab:full_res}}
 \end{table*}
 
 \section{Additional Network Benchmarks}
 \label{ap:more_nets}
 
 In this appendix we provide more details about the benchmark networks. These networks are used to test the robustness of the pipeline with respect to certain hyperparamters of the used networks.
 
 \subsection{Architectures \& Training}
 
 Besides the fiducial architecture presented in section~\ref{sec:network}, we train three different networks. The first network has the identical architecture as the fiducial networks, however, it was trained on 16 GPUs instead of eight, leading to a total batch size that is twice as big as the fiducial one. For the second benchmark model we doubled the number of residual layers. And for the last benchmark model we replaced the graph convolutional layers (GCL) inside the residual layers with standard 1D convolutional layers. The GCL of \deepsphere~have the major advantage that they are approximately rotational equivariant. However, a potential disadvantage is that they only connect pixels in a certain neighborhood. This means that the disconnected patches of the KiDS-1000 data might not be correlated if one applies only GCL layers. This does not happen in our networks since we also apply normalization layers that affect all pixels, however, it might still decrease the performance. Standard 1D convolution are not rotational equivariant, however, they automatically connect the disconnected patches of the KiDS-1000 data. We apply 1D convolutions on the \healpix~pixel arrays with a filer size of 16. Besides the mentioned changes, all other parameters, e.g. initial layers or optimizer settings remained unchanged. In the next section we will show the impact of these choices on the constraints of the degeneracy parameter.
 
 \subsection{Constraints}
 
 We present the constraints on the degeneracy parameter for various settings in Figure~\ref{fig:s8_bench}. All constraints are generated as described in sections~\ref{sec:inference} and~\ref{sec:mock_tests}, using a kernel scale of $h=0.4$ and we used the fiducial benchmark simulations as mock observations. It can be seen that the constraints of all settings are very consistent. The fiducial constraints from section~\ref{sec:mock_tests} are the tightest because they represent the combination of six networks. Besides that, all other settings have a very similar performance, indicating that our results are fairly robust with respect to the hyperparameter choices.

\section{Full Parameter Constraints}
 \label{ap:full_cons}
 We present the constraints on all cosmoligical parameters in table~\ref{tab:full_res}. It is important to note that most of the constraints are dominated by their prior distribution (see table~\ref{tab:grid}), which is why we did not include them in the main paper.  The only parameters with meaningful constraints are the degeneracy parameter $S_8$ and the intrinsic alignment amplitude $A_\mathrm{IA}$. Nevertheless, all results are consistent with each other.

\end{document}